\documentclass[12pt, a4paper]{article}

\pdfoutput=1

\usepackage{hyperref}
\usepackage{enumerate}
\usepackage{color}
\usepackage{float}
\usepackage{slashed}
\usepackage{bbold}
\usepackage{mathtools}
\usepackage{amsmath}
\usepackage{amssymb}
\usepackage{multirow}
\usepackage{arydshln}
\usepackage[usenames,dvipsnames]{xcolor}
\usepackage{caption}
\usepackage{graphicx}
\usepackage{inputenc}
\usepackage{tikz}
\usepackage{cite}
\usepackage{dsfont}

\usetikzlibrary{matrix}
\usetikzlibrary{calc,fit}

\DeclareFontEncoding{U}{}{}
\DeclareFontFamily{U}{bbold}{}
\DeclareFontShape{U}{bbold}{m}{n}
{  <5> <6> <7> <8> <9> gen * bbold
	<10> <10.95> bbold10
	<12> <14.4> bbold12
	<17.28> <20.74> <24.88> bbold17
}{}
\DeclareSymbolFont{bbold}{U}{bbold}{m}{n}
\DeclareSymbolFontAlphabet{\mathbbold}{bbold}

\newcommand{\nn}{\nonumber}

\newcommand{\bi}{\begin{itemize}}
	\newcommand{\ei}{\end{itemize}}
\newcommand{\ben}{\begin{enumerate}}
	\newcommand{\een}{\end{enumerate}}
\newcommand{\be}{\begin{equation}}
\newcommand{\ee}{\end{equation}}
\newcommand{\bea}{\begin{eqnarray}}
\newcommand{\eea}{\end{eqnarray}}
\newcommand{\eq}[1]{Eq.~(\ref{#1})}

\newcommand{\eps}{\epsilon}

\newcommand{\tev}{{\rm TeV}}

\newcommand{\bfr}{\begin{mdframed}[backgroundcolor=gray!20] }
	\newcommand{\efr}{\end{mdframed}}

\newcommand{\op}{\mathcal{O}}

\newcount\hour \newcount\minute
\hour=\time \divide \hour by 60
\minute=\time
\def\dgz{\delta g_{1z}}

\def\dka{\delta\kappa_\gamma}
\def\lz{\lambda_z}

\def\eps{\epsilon}

\renewcommand{\thefootnote}{\fnsymbol{footnote}}

\begin{document}

	\thispagestyle{empty}
	\begin{titlepage}
	
\begin{flushright}
DESY 17-231 \\
\end{flushright}
		

		\vspace{1cm}
		\begin{center}
			{\Large\bf\color{black} Diboson at the LHC vs LEP} 
			\bigskip\color{black}
			\vspace{1cm}\\
			{{\large  Christophe Grojean$^{a,b,}$\footnote{christophe.grojean@desy.de}, Marc Montull$^{a,}$\footnote{marc.montull@desy.de}, Marc Riembau$^{a,c,d,}$\footnote{marc.riembau@unige.ch}}
				\vspace{0.3cm}
			} \\[7mm]
			{\it $^a$ {DESY, Notkestrasse 85, 22607 Hamburg, Germany}} \\
			{\it $^b$ {Institut f\"ur Physik, Humboldt-Universit\"at zu Berlin, 12489 Berlin, Germany}} \\
			{\it {$^c$\, IFAE and BIST, Universitat Aut\`{o}noma de Barcelona, 08193 Bellaterra, Barcelona, Spain}}\\
			{\it {$^d$\, D\'epartment de Physique Th\'eorique, Universit\'e de Gen\`eve, Gen\`eve, Switzerland}}\\
		\end{center}
		\bigskip
		
		\begin{abstract}
			
		We use the current CMS and ATLAS data for the leptonic $pp \to WW, WZ$ channels to show that diboson production is, for a broad class of flavour models, already competitive with LEP-1 measurements for setting bounds on the dimension six operators parametrising the anomalous couplings between the quarks and the electroweak gauge bosons, at least under the assumption  that any new particle is heavier than a few TeV.  We also make an estimate of the HL-LHC reach with $3$\,ab$^{-1}$. We comment on possible BSM interpretations of the bounds, and show the interplay with other searches for a simplified model with vector triplets. We further study the effect of modified $Z$-quark-quark couplings on the anomalous triple gauge coupling bounds. We find that their impact is already significant and that it could modify the constraints on $\dgz$ and $\dka$ by as much as a factor two at the end of HL-LHC ($\lz$ is only marginally affected), requiring a global fit to extract robust bounds. We stress the role of flavour assumptions and study explicitly flavour universal and minimal flavour violation scenarios, illustrating the differences with results obtained for universal theories.

		\end{abstract}
		\bigskip
		
	\end{titlepage}
	
	\tableofcontents 
	
	\newpage

\renewcommand{\thefootnote}{\arabic{footnote}}

\section{Introduction}
\label{sec:intro}
\setcounter{footnote}{0}

The Large Hadron Collider (LHC) is probing the Standard Model (SM) at higher energies than ever before, reaching new regions never explored so far. For this reason, we must take the chance to learn as much as possible from it. With the discovery of a scalar particle consistent with the Higgs boson, the SM can in principle be consistent up to the Planck scale. Nonetheless in many UV completions predicting a light Higgs, e.g. supersymmetric or composite Higgs models, one requires other new particles with masses around the electroweak scale unless there is some fine tuning. So far though, the LHC has not seen any robust hints of new physics, which indicates that any new particles must be either too weakly coupled to the SM or heavy enough to not have been seen. In the first case, one may expect to see direct effects, like for example a resonance showing up once enough luminosity is collected. In the second case, one expects to see the effects of the new particles indirectly, for example, by modifying the differential cross sections of particular processes with respect to the SM prediction.

The study of diboson production, $pp \to WV, V=W,Z$, offers a way to probe physics scenarios of the second class. The interest in such channels both at lepton and hadron colliders is not new~\cite{Hagiwara:1986vm, Gounaris:1996rz,Campbell:2011bn} but it has recently received renewed attention, see in particular Refs.~\cite{Frye:2015rba, Butter:2016cvz,Falkowski:2016cxu,Green:2016trm, Panico:2017frx,Azatov:2017kzw, Zhang:2016zsp, Baglio:2017bfe,Franceschini:2017xkh,Ellis:2018gqa,Alves:2018nof}. This is  first due to the fact that together with $pp \to Vh$~\cite{Biekoetter:2014jwa,Banerjee:2018bio}, diboson production directly probes the interactions of the Goldstone bosons via the gauge boson longitudinal polarizations, and therefore is one of the first places where to expect signs of new physics related to the electroweak symmetry breaking. Furthermore, in  Refs.~\cite{Falkowski:2016cxu, Franceschini:2017xkh, Baglio:2017bfe} it has been shown that at high energy, the leading amplitudes for $pp \to VV, Vh$ grow with the center of mass energy faster than the SM ones and therefore diboson production can benefit from the higher energy probed at the LHC to reveal sign of new physics. See Refs.~\cite{Contino:2016jqw, Azatov:2016xik,Azatov:2017kzw, Dror:2015nkp, Biekoetter:2014jwa, Farina:2016rws, Alioli:2017jdo, Franceschini:2017xkh, Alioli:2017nzr, Liu:2018pkg} for studies using this high energy behaviour to increase the sensitivity to $d=6$ operators; notice that in some cases the new LHC bounds can improve on the LEP-1 and LEP-2 bounds. As shown in Refs.~\cite{Falkowski:2016cxu, Franceschini:2017xkh, Baglio:2017bfe, Gupta:2014rxa}, in the Higgs basis and at the dimension-six level~\cite{Falkowski:LHCHXSWG-INT-2015-001,deFlorian:2016spz},  there are a priori seven Beyond the Standard Model (BSM) coefficients that modify the diboson amplitude $pp \to WV$ at high energies. These are three anomalous triple gauge couplings (aTGC), traditionally parametrised by $\dka, \, \dgz, \, \lambda_\gamma$, and four anomalous couplings between the light quarks and the $Z$ gauge boson, $\delta g_L^{Zu}$, $\delta g_R^{Zu}$, $\delta g_L^{Zd}$, $\delta g_R^{Zd}$ ($\delta V\bar{q}q$ hereinafter), that will be introduced later.  Interestingly, the  $pp \to VV$ and $pp \to Vh$ amplitudes at high energy become equal, as expected by the Goldstone equivalence theorem, and actually only depend on five combinations of the $d=6$ operators~\cite{Falkowski:2016cxu, Franceschini:2017xkh, Baglio:2017bfe}. There has not been yet a complete global analysis establishing the future bounds on these five independent so-called High Energy Parameters, but some first results have been obtained in the $WZ$~\cite{Franceschini:2017xkh} and $Zh$~\cite{Banerjee:2018bio} channels, showing some nice complementarity. Combining with LEP constraints on the Z couplings to fermions, one could in principle univocally derive bounds on the aTGCs. The purpose of our work is to stress that, if this strategy is perfectly fine for universal theories, the  aTGC bounds  obtained that way do not directly apply when other flavour assumptions are considered and one needs to perform a global fit to derive bounds on both aTGCs and $V\bar{q}q$ couplings. 
 \medskip
 
In this work, we study the constraining power of diboson data to set bounds on the anomalous couplings between the $W$ and $Z$ gauge bosons and the light quarks. In particular, we make use of the differential distributions reported by the experimental collaborations. We find that due to the enhanced sensitivity at high energies, $pp \to WV$ can already be competitive or even surpass LEP-1 on setting bounds on $\delta V \bar{q}q$, at least under the assumptions that these anomalous couplings are generated by new particles with masses equal or greater than a few TeV to ensure the validity of the EFT, see section~\ref{sec:validity}. We refer the reader to Refs.~\cite{Frye:2015rba,Franceschini:2017xkh, Azatov:2017kzw, Panico:2017frx,Liu:2018pkg} where new differential distributions and experimental searches are proposed in order to increase the sensitivity to the effective field theory (EFT) operators entering diboson production. If these are implemented by the experiments, the increase of sensitivity could allow diboson production at HL-LHC to set much stringent bounds on the BSM amplitudes, reaching the point where they are smaller than the SM, and therefore can start constraining BSM scenarios with a characteristic coupling  smaller than a typical SM gauge coupling.

We rely on the differential distributions from the up-to-date diboson measurements performed by ATLAS and CMS with up to 20\,fb${}^{-1}$ of data at 8\,TeV and 13\,fb${}^{-1}$ of data at 13\,TeV, see Table~\ref{tab:experimental} for details. We also estimate the sensitivity expected at the high-luminosity run of the LHC (HL-LHC) with an anticipated  total of 3\,ab$^{-1}$ of data. We consider two  general flavour structures of the higher dimensional operators: 
\textit{i)} Flavour Universality (FU), where the EFT operators satisfy a $U(3)^5$ family symmetry, which, in the Higgs basis, corresponds to $[\delta g_{L,R}^{Zu,d}]_{ij} = A_{L,R}^{u,d} \  \delta_{ij}$, and 
\textit{ii)} Minimal Flavour Violation (MFV) where this symmetry is broken only by spurions of the Yukawa couplings, leaving $[\delta g_{L,R}^{Zu,d}]_{ij} \simeq \left( A_{L,R}^{u,d}+B_{L,R}^{u,d} \ \frac{m_i^2}{m_3^2} \right)\delta_{ij}\,$.~\footnote{When simulating the diboson production, we only modify the couplings to the $u$ and $d$ quark since the BSM effects from the heavier quarks are PDF-negligible, below 1\%, within the flavour assumptions considered. Therefore the diboson analysis presented in this paper does not distinguish MVF and $U(2)^5$-flavour symmetric setups.} It would also be interesting to combine and compare the LEP-1 bounds on the $\delta V \bar{q}q$ couplings with other flavour scenarios, e.g. the anarchic case (see Ref.~\cite{Efrati:2015eaa}) or the diagonal one, i.e. diag$({\delta g_{L,R}^{Zu,d}}_{11}, {\delta g_{L,R}^{Zu,d}}_{22}, \, {\delta g_{L,R}^{Zu,d}}_{33})$. We leave these analysis for future work since the non-Gaussianity of the fit makes it non-trivial to go from a more general case to a more restrictive one.
Diboson production at hadron colliders is insensitive to these assumptions since the cross section is dominated by the light quarks, while the constrains from LEP-1 can change by an order of magnitude, see the results of Ref.~\cite{Efrati:2015eaa} that we summarise in  Appendix~\ref{app:lepconstraints}. Another interesting UV assumption is that of universal theories~\cite{Peskin:1991sw,Barbieri:2004qk,Wells:2015uba}. These can be defined as those theories whose EFT can be fully described by bosonic operators and deviations of the light quark couplings can be written in terms of the gauge boson oblique parameters. Given that the LEP-1 bounds for these types of theories is one or two orders of magnitude stronger than those for MFV and FU, we found that with the current experimental searches diboson production is not competitive with LEP-1 for unviersal theories.~\footnote{See Appendix~\ref{app:lepconstraints} for the bounds on universal theories expressed in the Higgs basis.}
\medskip

Due to the larger systematics at the LHC, the conclusion that it can surpass LEP-1 and LEP-2 in setting bounds on the EFT operators may come as a surprise, but it follows from the fact that some BSM amplitudes can grow with the characteristic scale of the hard process probed at the LHC. 
However, it is important to keep in mind, as stressed in Refs.~\cite{Contino:2016jqw,Biekoetter:2014jwa, Farina:2016rws, Alioli:2017jdo,Franceschini:2017xkh}, that the larger systematics also imply in many cases that the new LHC bounds are valid only when the BSM contribution is larger than the SM one, limiting in some cases the generality of these bounds to  subsets of possible UV theories. 
We comment more on the EFT interpretation in section~\ref{sec:interpretation}. Nonetheless, given that the LHC is running and we do not know what new physics may lie ahead, it is still important to make sure that all the regions of the EFT parameter space are explored in the most model independent way as possible.
\medskip

Besides studying the bounds on $\delta V\bar{q}q$, we also look at the impact of non-vanishing $\delta V\bar{q}q$ in the aTGC determination under the different flavour assumptions considered. Looking at this effect was first mentioned and motivated in Refs.~\cite{Falkowski:2014tna,Zhang:2016zsp} and checked explicitly in Ref.~\cite{Baglio:2017bfe} using the channel $pp \to WW$ at 8\,TeV by ATLAS. We extend this analysis by first performing a global fit to the present data for all the channels in Table~\ref{tab:experimental} and also by studying the impact of different flavour assumptions. We also estimate the sensitivity that one can hope to reach at  HL-LHC, concluding that the effect of $\delta V \bar{q}q$ will be more and more important in the future. It should be noted that the analysis we provide is done at leading order (LO). We expect that the NLO effects are most relevant for amplitudes with final transverse polarizations due to the non-interference effects shown in Ref.~\cite{Azatov:2016sqh,Baglio:2017bfe}. The NLO effects can also be relevant in certain regions of the phase space; for instance in the amplitude of $pp \to WZ$ which nearly vanishes for the $\pm \mp$ and $\pm0$ polarizations when the polar scattering angle is $\theta \simeq \pi/2$  \cite{Franceschini:2017xkh}. Since in our study we are mostly interested in  the cases where the two gauge bosons have longitudinal polarizations, we do not expect much difference in our conclusions even though it would be interesting to study in more detail the NLO effects. 
\medskip

We briefly comment on  possible interpretations of the  EFT bounds derived. To gain perspective and a sense of the usefulness of the constraints coming from diboson production, we study a simplified model of heavy vector triplets and compare the diboson bounds to the ones from other searches like dijets, resonant diboson or Higgs coupling measurements, finding that diboson can be complementary to other processes in exploring the parameter space of the model, and can be the leading probe in important regions of parameters.
\medskip

The paper is organized as follows. In section~\ref{sec:dibsonHE}, we give the conventions and review  the high energy behavior of diboson production at the LHC. In particular, we note that the high-energy diboson amplitudes in the Higgs basis are controlled by seven independent parameters in FU and MFV setups as opposed to five parameters only for universal theories. In section~\ref{sec:EFTbounds}, we present the bounds on the $\delta V \bar{q}q$ and the effect of allowing these to be non-zero in the aTGC exclusion plots. 
In section~\ref{sec:HL-LHC}, we estimate the $\delta V \bar{q}q$ bounds that can be expected by the end of HL-LHC  and we quantify the effect of letting $\delta V \bar{q}q$ and aTGC float in global EFT fit. In section~\ref{sec:interpretation}, we briefly review the validity of the EFT approach in presence of non-negligible contributions from the dimension six BSM quadratic amplitudes, and we review various UV scenarios and power counting rules which motivate the various assumptions on the values of the parameters used through the paper. We also study a toy model with heavy triplets as a concrete example. And we compare our HL-LHC bounds for this toy model with those coming from Higgs coupling measurements and dijet searches. We conclude in section~\ref{sec:conclusionsDB}. Four appendices provide further technical details and cross-checks.

\section{Theoretical framework}
\label{sec:dibsonHE}
	
We work in the so called Higgs basis~\cite{Gupta:2014rxa, Falkowski:LHCHXSWG-INT-2015-001, deFlorian:2016spz}, and follow the conventions of Ref.~\cite{Falkowski:LHCHXSWG-INT-2015-001} where $\alpha, G_F$ and $m_Z$ are taken as the input parameters. The Higgs basis parametrizes the $d=6$ EFT operators as modifications to the SM vertices, and where the fields are in the mass eigenstates and in the unitary gauge. In this basis and considering only operators with $d \le 6$, the relevant terms for $pp \to WV$ production are: 
	\be
	{\cal L}_{\rm diboson} \supset {\cal L}_{\rm TGC} + {\cal L}_{V\bar{q}q} \,.
	\label{eq:intEFT}
	\ee
The first term contains the SM interactions between the electroweak gauge bosons together with the $d=6$ aTGC deformations,
	\begin{eqnarray} 
	\label{eq:TGC}
	{\cal L}_{\rm TGC}  &= &  
	i  e    \left ( W_{\mu \nu}^+ W_\mu^-  -  W_{\mu \nu}^- W_\mu^+ \right ) A_\nu   
	+ 
	i  e  \left [  (1 + \delta \kappa_\gamma )  A_{\mu\nu}\,W_\mu^+W_\nu^-   
	\right ]
	\nonumber \\  &  + & i g \, c_W  \left [  (1 + \delta g_{1,z})   \left ( W_{\mu \nu}^+ W_\mu^-  -  W_{\mu \nu}^- W_\mu^+ \right ) Z_\nu 
	+ (1 +  \delta \kappa_z) \, Z_{\mu\nu}\,W_\mu^+W_\nu^-    \right ] 
	\nonumber \\  & + &    
	i {e  \over m_W^2 }  \lambda_\gamma W_{\mu \nu}^+W_{\nu \rho}^- A_{\rho \mu}  
	+  i {g  \, c_W \over m_W^2}   \lambda_z W_{\mu \nu}^+W_{\nu \rho}^- Z_{\rho \mu}   \,.
	\end{eqnarray} 
The second term in the Lagrangian~(\ref{eq:intEFT}) contains the SM contribution and deviations to the couplings between the up and down quarks to the $W, \, Z$, gauge bosons,
	\begin{eqnarray} 
	{\cal L}_{V\bar{q}q} &= & \sqrt{g^2 + g'{}^2} Z_\mu   \left [  
	\sum_{f \in u, d}  \bar f_L  \gamma_\mu \left (T^3_f - s_W^2 Q_f +   \delta g^{Zf}_L  \right)  f_L 
	+    \sum_{f \in u, d}  \bar f_R  \gamma_\mu \left ( - s_W^2 Q_f  +   \delta g^{Zf}_R  \right)  f_R   \right ] \nonumber\\
	&+& {g \over \sqrt 2}  \left (W_\mu^+ \bar u_L \gamma_\mu  \left (I_3 +  \delta g^{Wq}_L  \right)  d_L + {\mathrm h.c.} \right ) \,.
	\label{eq:vertices}
	\end{eqnarray} 
	\medskip
Since at dimension six the following relations are satisfied (see for instance Ref.~\cite{Falkowski:LHCHXSWG-INT-2015-001}):
	\be
	\delta \kappa_z = \delta g_1^z - \tan^2 \theta \, \delta \kappa_\gamma\, , \hspace{1.3cm} \lambda_z = \lambda_\gamma \, ,  \hspace{1.3cm} \delta g^{Wq}_L = \delta g^{Zu}_L  -  \delta g^{Zd}_L\,,
	\ee
	the deviations of the schematic form $\sim g_{SM} \, (1 + \delta$) can be parametrized by two independent aTGC (which we choose to be $\delta \kappa_{\gamma}$, $\dgz$), and four independent corrections to $Z \bar{q}q$ vertices (which we choose to be $\delta g_L^{Zu}$, $\delta g_R^{Zu}$, $\delta g_L^{Zd}$, $\delta g_R^{Zd}$). Notice that the aTGC parametrized by $\lambda_\gamma = \lambda_z$ introduces a new type of coupling non-existent in the SM. In total there are, in the Higgs basis, seven parameters  that contribute to the leading deformations to diboson production (three aTGC and four $\delta V \bar{q}q$).
	\medskip
	
		In the Lagrangian~(\ref{eq:vertices}) we have not included right-handed charged currents nor dipole contributions since under FU and MFV they are either zero, or are suppressed by the Yukawas of the light quarks. We also ignore the deformations in the lepton sector since their bounds from LEP-1 data are an order of magnitude better than those on the quark sector~\cite{Efrati:2015eaa}. Finally we also ignored the shift to the $W$ mass, $\delta_m$, since its current existing bound is such that it numerically gives in diboson production an effect ten times smaller than a modified quark coupling.

\subsection{High energy behaviour and correlations}
\label{subsec:SMhighE}

\medskip

\begin{figure}[!h]
		\centering
		\includegraphics[scale=0.21]{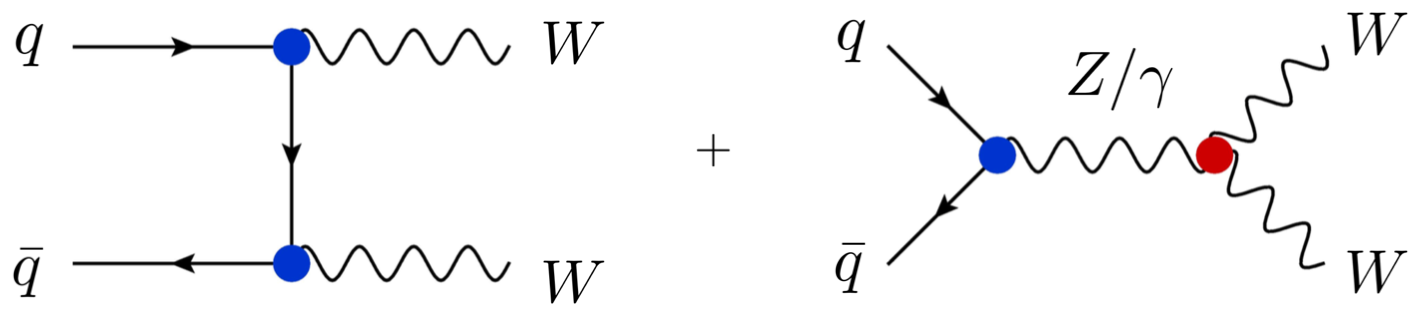}
		\caption{Representative contributions to diboson production.
			The sensitivity of the measurements, already  with  $\mathcal{O}(20)\,\textrm{fb}{}^{-1}$ of data and certainly even more at HL-LHC, is such that they can improve the LEP-1 constraints on the quark couplings to gauge bosons (blue). This also implies that the LEP-1 bounds are no longer stringent enough to make these parameters negligible when setting bounds on the  anomalous triple gauge couplings (red).}
		\label{fig:diagrams}
\end{figure}

In the Higgs basis, the energy growth of the amplitudes that interfere with the SM in the high energy limit can be understood as follows. At tree level in the SM, and in the unitary gauge, the leading amplitude for $q\bar{q}'\to WW\,(WZ)$ is given by the sum of three diagrams, consisting of an s-channel exchange of the $\gamma, \, Z$ bosons ($W$ boson), and a t-channel contribution. Taking as an example the case of $\bar{q} q \to WW$, where $\bar{q}q= \bar{u}u, \, \bar{d}d$, one finds that the tree level SM amplitude is given by the Feynman diagrams of Fig.~\ref{fig:diagrams}. One can check that at large center of mass energy, $\hat{s} \gg m_W^2$, the total amplitude for $q \bar{q}\to W^+_0 W^-_0$ is given by~\cite{Hagiwara:1993ck}
	\be
	\mathcal{M}_\gamma + \mathcal{M}_Z + \mathcal{M}_t = i\,  \hat{s} \left[ - \frac{e^2 \,\sin\theta}{2m_W^2} Q_q\,-\frac{e^2 \,\sin\theta}{2m_W^2}\frac{1}{s_W^2}(T_q^3- s_W^2 Q_q)\ +\frac{e^2\,\sin\theta}{2m_W^2}\frac{T_q^3}{s_W^2}\right] + \cdots 
	\label{eq:SMamp}
	\ee
	where $W^{\pm}_0$ stand for the longitudinal polarizations of the $W^\pm$ gauge bosons, $\hat{s}$ is the squared center of mass energy, the dots denote  sub-leading contributions at high energy, $Q_q$ and $T^3_q$ are the electric charge and $SU(2)$ weak isospin of the initial quarks and $\theta$ is the angle between $W^+$ and the beam axis. ~\footnote{There is another term that grows with energy but we neglected it since it is proportional to the quark masses. Its energy growth is canceled with the diagram including the Higgs, however, in our energy range it is negligible, and  as a first approximation one can think of the quarks to be massless.} 
	
	The key point of \eq{eq:SMamp} is to notice that while each of the individual sub-amplitudes grows with $\hat{s}$, the sum does not. Therefore, any shift to the SM couplings, shown in blue and red in Fig.~\ref{fig:diagrams}, will spoil the cancellation of the different pieces in \eq{eq:SMamp}, and therefore the resulting amplitude will be proportional to $\hat{s}$. In the Higgs basis it is especially clear to see that all the coefficients modifying diboson production  with a shift to the SM couplings will generically induce an amplitude that grows with $\hat{s}$. 
	
	Notice that the interaction given by $\lambda_\gamma W_{\mu \nu}^+W_{\nu \rho}^-(s_W A_{\rho\mu}+c_W Z_{\rho\mu})$ in \eq{eq:vertices} is  not present in the SM. In this case one cannot use the spoiling of the SM amplitude cancellation of \eq{eq:SMamp} to see whether its effect asymptotically  grows with the center of mass energy. 
Nonetheless, one can see by direct calculation that the amplitude induced by this operator actually grows with $\hat{s}$ as a consequence of the presence of extra derivatives in the interaction.
\medskip 

\subsection{Helicity amplitudes at high energy and correlations between aTGC and $\delta V \bar{q}q$}
	 To estimate which operators or combinations of operators will be the most constrained by diboson production at the LHC, one can study each helicity amplitude as done in Refs.~\cite{Falkowski:2016cxu, Franceschini:2017xkh}. In the limit where $\hat{s} \gg m_W^2$, the leading helicity amplitudes for the partonic scattering $q\bar{q} \to WW$ are given by~\footnote{The $L,R$ stand for the initial helicities of the quarks, while  $\pm$ and $0$ stand for the transverse and longitudinal polarizations of the final electroweak bosons respectively. We computed these amplitudes using FeynCalc~\cite{Shtabovenko:2016sxi} using the BSMC package~\cite{Falkowski:2015wza} for FeynRules~\cite{Alloul:2013bka}, finding agreement with the expressions presented in Ref.~\cite{Falkowski:2016cxu}, which also was a cross check for the .ufo file used in the \texttt{Madgraph5} simulations.} 
	\bea\nonumber
	\mathcal{M}(LL;00) &=& i \frac{\hat{s}}{m_W^2}\frac{e^2 \,\sin\theta}{2s_W^2}  \left[ (2T_q^3)\, \delta g_L^{Wq} - \delta g_L^{Zq} -\delta g_{1z} (T_q^3-s_W^2 Q_q)+\delta \kappa_\gamma t_W^2(T_q^3-Q_q)\right]  \\\nonumber
	\mathcal{M}(RR;00) &=& i \frac{\hat{s}}{m_W^2}\frac{e^2 \,\sin\theta}{2s_W^2}  \left[    \delta g_R^{Zq} -\delta g_{1z} \, s_W^2  \, Q_q + \delta \kappa_\gamma \, t_W^2\,  Q_q  \right] \nonumber \\
	\mathcal{M}(LL;\pm\pm)  &=&  i \frac{\hat{s}}{m_W^2}\frac{e^2 \,\sin\theta}{2s_W^2}  \,   T_q^3 \lambda_\gamma \,,
	\label{eq:AmpWW}
	\eea
	where $\delta g^{Wq}_L = \delta g^{Zu}_L  -  \delta g^{Zd}_L$ and $\delta g_{L,R}^{Zq}$ corresponds to the anomalous vertex of the incoming quark $q$, defined in \eq{eq:vertices}.
	For $qq' \to W Z$, the energy growing amplitudes  are
	\bea\nonumber
	\mathcal{M}(LL;00) &=& -i \frac{\hat{s}}{m_W^2} \frac{e^2\,\sin\theta}{2\sqrt{2}s_W^2c_W}\left[\delta g_L^{Zu}-\delta g_L^{Zd} - \delta g_{1z}c_W^2\right]\\
	\mathcal{M}(LL;\pm\pm) &=& -i\frac{\hat{s}}{m_W^2}\frac{e^2\,\sin\theta}{2\sqrt{2}s_W^2c_W}\, \lambda_z.
	\label{eq:AmpWZ}
	\eea
We can see, as pointed out in Refs.~\cite{Falkowski:2016cxu, Franceschini:2017xkh}, that in the asymptotic high energy regime there are only five independent combinations of parameters entering $pp \to WV$ since 
	\be
	\mathcal{M}(u_L \bar{d}_{L} \to W_0 Z_0) \, =  {1 \over \sqrt{2} \, c_W} \, \left(\mathcal{M}(\bar{d}_L d_L \to W_0 W_0) - \mathcal{M}(\bar{u}_L u_L \to W_0 W_0)\right) + {\cal O}(\hat{s}^0)\,.
	\ee
Therefore, there are only four relevant independent combinations for the longitudinal polarizations and one for the transverse ones that can be probed in the high energy limit.~\footnote{This counting may change for other flavour assumptions, since the right-handed charged current, that could be present away from the FU/MFV setups, gives rise also to an energy-growing amplitude  for $pp \to WZ$: $\mathcal{M}(RR;00) = -i\frac{e^2\,\hat{s}\,\sin\theta}{2\sqrt{2}m_W^2s_W^2c_W}\,\delta g_R^{Wq}\,+\,\op(\hat{s}^0)$.}  These four directions for the longitudinal polarizations are the so-called High Energy Parameters (HEPs)  introduced in Ref.~\cite{Franceschini:2017xkh}, see Table~2 in this reference.  For completeness, in Appendix~\ref{sec:Comp}, \eq{eq:HEPparameters},  these four HEPs are written explicitly in terms of the Higgs basis~\cite{Falkowski:LHCHXSWG-INT-2015-001} parameters. In Appendix~\ref{sec:Warsaw} we express the amplitudes shown in Eqs.~(\ref{eq:AmpWW}), (\ref{eq:AmpWZ}) in the Warsaw basis.

Notice that if the experimental sensitivity is low such that the quadratic BSM squared amplitudes dominate the cross section, all the channels above  show a similar behaviour at high-energy. On the other hand, when the experimental sensitivity is getting good enough to probe BSM deformations subdominant to the SM, the channels that feature an interference between SM and BSM will be of better use to bound anomalous couplings. As shown in Refs.~\cite{ Azatov:2016sqh,Franceschini:2017xkh}, this selects the production  of two longitudinally gauge bosons as the preferred channel. It is nonetheless possible to also use the production of transversally polarized gauge bosons when relying on specific kinematic observables to \textit{resurrect} the interference~\cite{Azatov:2017kzw, Panico:2017frx}.  In our analysis, which uses the current experimental observables, we find that the high energy bins are the most important in setting constrains. For these, we observe that the quadratic pieces are in general equal or larger than the interference parts, and therefore the question of the BSM/SM interference is not so relevant in our analysis.
\medskip

	From Eqs.~(\ref{eq:AmpWW}) and~(\ref{eq:AmpWZ}),   there are a total of seven coefficients parameterizing the five directions growing as $\hat{s}$ in the processes $pp \to WV$. Hence, in the asymptotic high energy limit,  two completely flat directions  are anticipated among the Higgs basis coefficients. 
A simple way to see the flat directions explicitly is by noting that any deviation of $\dgz$ and $\dka$ in  Eqs.~(\ref{eq:AmpWW}), (\ref{eq:AmpWZ}) can be compensated by a modification of the vertex corrections $\delta V \bar{q}q$. Naively, if one assumes that the largest sensitivity comes from the high energy bins, the characteristic energy of diboson production is $\sqrt{\hat{s}} \sim$ TeV. At these energies, one expects that the subleading amplitudes, which grow with $\sqrt{\hat{s}}/m_W$ instead of $\hat{s}/m_W^2$, can set bounds that are worse by a factor $\sqrt{\hat{s}}/m_W \sim 10$ (as long as the BSM squared amplitudes dominate the cross section, which, as we will see,  is the case in our analysis). These subleading amplitudes involve a longitudinal and a transverse vector boson in the final states. For $pp\to WW$, they are given by
	\bea\nonumber
	\mathcal{M}(LL;0\pm) &=& -\frac{e^2 \sqrt{\hat{s}}\,\cos^2\frac{\theta}{2}}{\sqrt{2}m_Ws_W^2}\bigg[(2T^3_q)\delta g_L^{Wq}-\delta g_L^{Zq} -\frac{1}{c_W^2} \delta g_{1z}(T^3_q-s_W^2Q_q)-T_q^3(\delta\kappa_\gamma+\lambda_\gamma)\bigg]\,,\\
	\mathcal{M}(RR;0\pm) &=& -\frac{e^2 \sqrt{\hat{s}}\,\sin^2\frac{\theta}{2}}{\sqrt{2}m_Ws_W^2}\left[t_W^2Q_q\delta g_{1z}-2\delta g_R^{Zq}\right],
	\label{eq:subWW}
	\eea
while for $pp\to WZ$, one has
	\bea\nonumber
	\mathcal{M}(LL;\pm0) &=& -\frac{e^2 \sqrt{\hat{s}}}{2 m_Ws_W^2c_W}\bigg[  
	\delta g_L^{Zu}+\delta g_L^{Zd} + (	\delta g_L^{Zu}-\delta g_L^{Zd})\cos\theta \,-\,c_W^2(2\delta g_{1z}+\lambda_z)\sin^2\frac{\theta}{2} 
	\bigg],\\
	\label{eq:subWZ}
	\eea
up to subleading terms suppressed by $\sim 1/\sqrt{\hat{s}}$. One can check that the combination of coefficients entering in the subleading amplitudes cannot be obtained as a linear combination of the directions appearing in the leading $\hat{s}/m_W^2$ amplitudes. Hence, one naively expects to find some directions in the EFT space that are ${\cal{O}}(10)$ times less constrained than the five directions given by the amplitudes leading at high energy. We confirm this naive estimate later in section~\ref{sec:corr} where we study the correlations among the different constraints in the Higgs basis.
		
To conclude this section, it should be noted that the previous counting is different for universal theories. As discussed in Appendix~\ref{app:lepconstraints}, the high-energy diboson amplitudes depend only on five independent parameters and no flat direction is expected in the global fit to diboson data. Anticipating the results that will be presented in the rest of the paper, one should be aware nonetheless that LHC diboson data will not be competitive to LEP-1 to constrain the $Z\bar{q}q$ couplings in universal theories, at least by using only the current leptonic experimental distributions.
		
\section{Results with current LHC data}
\label{sec:EFTbounds}
	
\subsection{Data used and statistical analysis}
\label{sec:MCandStat}
		
To get the bounds on the different BSM parameters of Eqs.~(\ref{eq:TGC})--(\ref{eq:vertices}), we have used all the leptonic channels of the $pp \to WW, \, WZ$ channels reported by CMS and ATLAS, see Table~\ref{tab:experimental}. We indicate in each case the differential distribution used to perform the combined fit. We limited the analysis to purely leptonic decays due to their high sensitivity and the ease with which one can reproduce the experimental analyses. See Ref.~\cite{CMSATLASTGC} for a summary of the ATLAS and CMS constraints. There are nonetheless other channels that would be interesting to add, e.g. two quarks and two leptons in the final state~\cite{Sirunyan:2017bey}, since they can set even tighter constrains than the purely leptonic channels.~\footnote{See Ref.~\cite{Liu:2018pkg} where projections for the semi-leptonic channels at HL-LHC are studied in detail, and new experimental observables are proposed.}
\medskip
			\begin{table}[t]
				\begin{center}
					\begin{tabular}{|crllc|c|}
						\hline
						Experiment & $\mathcal{L}$[fb$^{-1}$] & $\sqrt{s}$ & Process & Obs. & Ref. \\
						\hline
						ATLAS & 4.6 & 7\,TeV & $WW\to\ell\nu\ell\nu$ & $p_{T\ell}^{(1)}$ &~\cite{ATLAS:2012mec}, Fig.~7  \\
						ATLAS & 20.3 & 8\,TeV & $WW\to\ell\nu\ell\nu$ & $p_{T\ell}^{(1)}$ & $\, \ $~\cite{Aad:2016wpd}, Fig.~11\\
						CMS & 19.4 & 8\,TeV & $WW\to\ell\nu\ell\nu$ & $m_{\ell\ell}$ &~\cite{Khachatryan:2015sga}, Fig.~4\\
						\hline
						ATLAS & 20.3 & 8\,TeV & $WZ\to\ell\nu\ell\ell$ & $p_{TZ}$ &~\cite{Aad:2016ett}, Fig.~5\\
						CMS & 19.6 & 8\,TeV & $WZ\to\ell\nu\ell\ell$ & $p_{TZ}$ &~\cite{Khachatryan:2016poo}, Fig.~7\\
						ATLAS & 13.3 & 13\,TeV & $WZ\to \ell\nu \ell\ell$ & $m_{WZ}$ &~\cite{ATLAS:2016qzn}, Fig.~3\\
						\hline
					\end{tabular}
					\caption{Data used to extract the current LHC bounds.}
					\label{tab:experimental}
				\end{center}
			\end{table}
			
To perform the fit, we calculate the  BSM cross sections at tree level with \texttt{MadGraph5}~\cite{Alwall:2014hca}, while using \texttt{FeynRules 2.0}~\cite{Alloul:2013bka} to generate the .ufo file for the BSMC model~\cite{Falkowski:2015wza}. This procedure gives the cross section in terms of the seven BSM parameters $\delta g_{L,R}^{Zu,d}$ and $\dgz, \, \dka, \, \lambda_\gamma$. We perform a simulation to get the cross section for each bin for every differential distribution shown in Table~\ref{tab:experimental}, and then perform the cuts as described by the experimental collaborations in each case.~\footnote{In some cases, like $WW \to \nu \ell \nu \ell$, the cuts performed by the experiments for some sub-chanels are performed using a Boosted Decision Tree and not just a cut and count approach. In this case we only generate the subchannel for which we can easily reproduce the cuts, i.e. $WW \to \nu_e e \nu_\mu \mu$ and then fit to the total combination assuming that it does not depend on the lepton flavour.} To get the BSM cross section, we have generated for each bin several simulations corresponding to different values of the BSM coefficients and then we have fitted them to a general quadratic polynomial of the seven BSM coefficients  $\delta g_{L,R}^{Zu,d}$ and $\dgz, \, \dka, \, \lambda_\gamma$ which we schematically call $\delta_i$. In other words, we write~\footnote{When simulating the BSM cross sections, we modify the four $Z\bar{q}q$ couplings $\delta g_L^{Zu}$, $\delta g_L^{Zd}$, $\delta g_R^{Zu}$, $\delta g_R^{Zd}$,  for all the quark generations at the same time, as one would do in the FU case, see \eq{eq:fubounds}. Nonetheless, due to the proton's PDF, the contribution of the light quarks $u,d$ is more than a factor ten greater than the one of $c,s$, so one can safely assume that the modifications of $Zqq$ for second and third generation give negligible contributions to diboson production. We expect that the results we get for the diboson fit on the $Z\bar{q}q$ couplings for the FU case also apply for the $Z\bar{q}q$ couplings for the first two generations of the MFV case, since in the MFV case $[\delta g_{L,R}^{Zu,d}]_{11}\simeq [\delta g_{L,R}^{Zu,d}]_{22}$.} 
	\be
		\label{eq:s_SMBSM}
	\sigma_{SM+BSM} (\delta_1,..., \delta_n)= \sigma_{SM} + a_i \, \delta_i + b_{ij} \,  \delta_i \delta_j \,,
	\ee
where the indices $i,j$ go from $i,j=1,...,7$, where $\sigma_{SM}$ corresponds to the SM contribution, and  $a_i$ and $b_{ij}$ are numerical coefficients that characterize the BSM contribution which we determine by varying the BSM parameters $\delta_i, \delta_j$ in the \texttt{MadGraph5} simulation.  For $n$ number of BSM parameters one has $n$ independent $a_i$ coefficients and $n+n (n-1)/2$ independent $b_{ij}$ coefficients. So, for $n=7$, $35$ coefficients in total have to be fitted to obtain the full expression (\ref{eq:s_SMBSM}). We then built the ratio $\delta \mu$ defined as
	\be
	\mu(\vec{\delta} ) \,= \, \frac{\sigma_{SM+BSM}(\vec{\delta})}{\sigma_{SM}} = 1 + \frac{\sigma_{BSM}(\vec{\delta}  )}{\sigma_{SM}} = 1 + \delta \mu(\vec{\delta} ) \, .
	\label{eq:signalstrength}
	\ee
Currently, the fully leptonic $WW$ and $WZ$ cross sections have been computed at NNLO in QCD taking into account both on-shell and off-shell contributions~\cite{Grazzini:2016swo,Grazzini:2016ctr,Grazzini:2017ckn} and at NLO in EW but only on-shell~\cite{Bierweiler:2013dja,Baglio:2013toa}. If, in the fiducial phase space considered, the effects of taking into account the NLO corrections can be encapsulated by an overall $k$-factor, ~\footnote{This would not be the case if the LO amplitude is highly suppressed. This is actually what is happening for the WZ production channel as emphasized in Ref.~\cite{Franceschini:2017xkh}: in the central region of the detector the $\pm0$ and $\pm \mp$ LO amplitudes exactly vanish. We thank G. Panico for pointing this out to us.} the higher order corrections will mostly cancel in the ratios $\delta \mu$, and that is why we use them to perform the global fit which, as mentioned, is done at LO. This might not hold for transverse polarizations as a result of the non-interfering effects pointed out in Ref.~\cite{Azatov:2017kzw}, and an analysis can be found in Ref.~\cite{Baglio:2017bfe}. Finally, we build a $\chi^2$ function
	\be
	\chi^2 = \sum_{I\in \text{channels}}\,\sum_{i\in \text{bins}} \frac{(\tilde{\sigma}_{SM}^\textit{bkg}+ \mu \, \tilde{\sigma}_{SM}^\textit{signal} - \sigma_\textit{measured})^2_{I,i} }{(\Delta_\textit{syst})_{I,i}^2 +(\Delta_\textit{stat})_{I,i}^2}, \,
	\label{eq:chi2raw}
	\ee
where the first sum runs through all the channels under study, and the second  sum runs over each bin for the chosen differential distribution, $\sigma_{measured}$ is the measured cross section including signal and background, $\tilde{\sigma}_{SM}^{bkg}$  and $\tilde{\sigma}_{SM}^\textit{signal}$ correspond to the simulated cross sections for the signal and background done by the experimental collaborations, $\Delta_\textit{syst}$ is the theoretical uncertainty given by the experimental collaborations on the predicted SM cross sections, $\tilde{\sigma}_{SM}^\textit{bkg}$ and $\tilde{\sigma}_{SM}^\textit{signal}$, and finally $\Delta_\textit{stat}$ is the statistical error. When needed, we multiply and divide \eq{eq:chi2raw} by the integrated luminosity squared and compute the $\chi^2$ function using the number of events shown in the figures referred to in Table~\ref{tab:experimental}. 
\medskip

From the correlation matrices, central values and errors given in Ref.~\cite{Efrati:2015eaa}~\footnote{We thank the authors of Ref.~\cite{Efrati:2015eaa} for providing the Mathematica code with all the aforementioned quantities that had more precision than in the paper, and allowed to get a more reliable $\chi^2$ for LEP-1.}, we build a $\chi^2$ function for the LEP-1 measurements at the $Z$-pole. To perform the global fits to get the aTGC bounds, we combine the two $\chi^2$ for diboson at the LHC and LEP-1 as $\chi^2 = \chi^2_{LHC} + \chi^2_{LEP-1}$.

\subsection{Correlations among the Higgs basis parameters}
\label{sec:corr}

When performing a $\chi^2$ fit, in the Gaussian limit,  one can easily find the correlation between two parameters by looking at the entries of the correlation matrix. In our case, given that the $\chi^2$ function is not Gaussian due to the non-negligible size of the $d=6$ BSM quadratic amplitudes, we cannot easily extract a correlation matrix. Therefore, to get a sense of the correlations among the different BSM coefficients, we perform a global fit and look at the two dimensional plots for each pair of coefficients profiling over all others. We show all these correlations in Appendix~\ref{sec:AppCorr}. 

\begin{figure}[t]
\centerline{\includegraphics[scale=0.23]{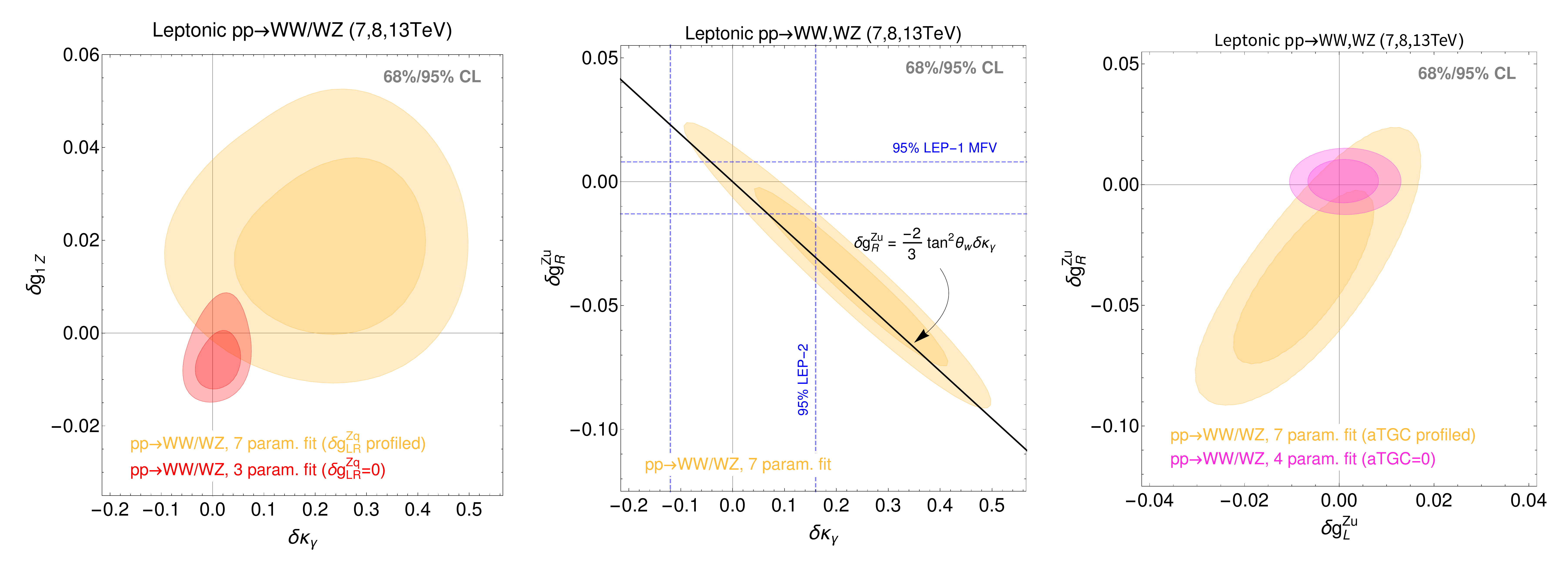}}
\caption{$68\%$ (dark shaded) and $95\%$ (light shaded) CL regions using the LHC diboson data reported in Table~\ref{tab:experimental}. {\bf Center:} fit to $\delta \kappa_\gamma$ and $\delta g_R^{Zu}$ profiling over all other five parameters. The line shows the expected flat direction in the $\hat{s} \rightarrow \infty$ limit that can be deduced from Eq.~(\ref{eq:AmpWW}). {\bf Left (Right):} in yellow the fit to aTGC (vertex corrections) marginalising over all other parameters, and in red (pink) the fit when the four $\delta V\bar{q}q$ (three aTGC) are set to zero.}
\label{fig:flatdirection}
\end{figure}

As an example of the correlations among the different parameters, in the center of Fig.~\ref{fig:flatdirection}, we show the projection of the $\chi^2$ function onto the two dimensional plane $\left (\dka, \, \delta g_R^{Zu} \right)$. The least constrained direction in this plot follows the slope given by the combination appearing in the amplitude $\mathcal{M}(RR;00)$ of \eq{eq:AmpWW}. The high energy flat direction is about ten times less constrained than orthogonal direction, in agreement with the naive estimate made in section~\ref{subsec:SMhighE}.

The large correlation shown in the center of Fig.~\ref{fig:flatdirection} makes $\delta g_R^{Zu}$ and $\dka$ very sensitive to each other. For reference, we show in horizontal blue dashed lines the allowed 95\% CL bounds set by LEP-1 on $\delta g_R^{Zu}$ and in vertical the 95\% CL bounds set by LEP-2 on $\dka$. 

From this plot one can intuitively see that if $\delta g_R^{Zu}$ is not set to zero but can vary within the range allowed by LEP-1, one may modify the bounds on $\dka$ in a non-negligible way. This indicates that the bounds on the aTGCs should include the $\delta V \bar{q}q$ deformations if a FU or a MFV scenario is assumed. Also, one can see that the assumptions on $\dka$ will have a large impact on the sensitivity of diboson production to $\delta g_R^{Zu}$. We see that the sensitivity of diboson to the different parameters is ultimately limited by the correlations, making a global combination crucial. 

\medskip

Fortunately, in a broad class of models, the parameter $\dka$ is expected to be generated only via loops, and, parametrically smaller than the other parameters, it can be neglected when setting constraints. The same holds true for $\lambda_\gamma$ which is also typically loop suppressed. This is because both $\dka$ and $\lambda_\gamma$ modify the magnetic moment and electric quadrupole moment of the $W$ which are only generated at one loop in minimally coupled theories~\cite{Hagiwara:1993ck, Giudice:2007fh}. Because of the large correlations, setting them to zero can greatly increase the accuracy of the fit to the various $\delta V \bar{q}q$.

\subsection{ $\delta V \bar{q}q$: LHC bounds vs LEP-1 constraints}
\label{subsec:LHCboundsvsLEP1}
		
\begin{figure}[t]
\centerline{\includegraphics[scale=0.37]{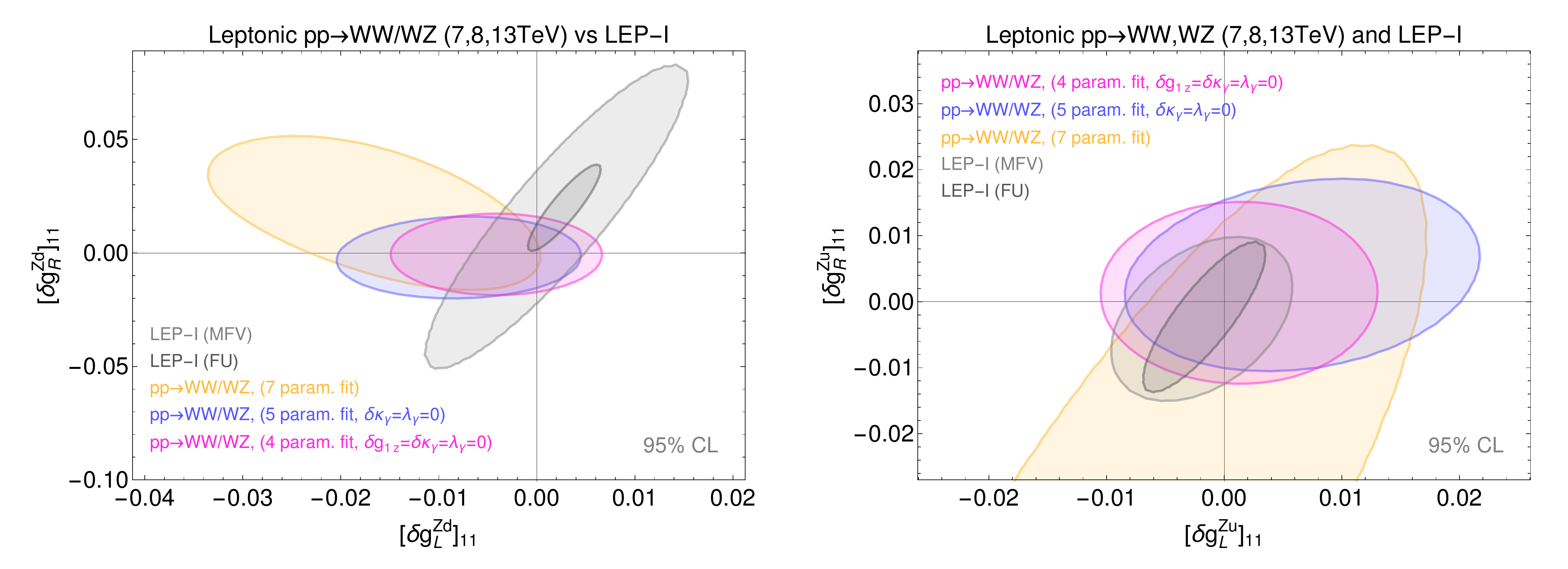}}
\caption{95\% CL regions for the anomalous couplings between the light quarks and the electroweak bosons. In light (dark) gray, the LEP-1 constraints assuming MFV (FU). In yellow, the diboson bounds after profiling over the remaining five parameters. In blue (pink) the same but setting $\delta\kappa_\gamma=\lambda_\gamma=0$ ($\delta\kappa_\gamma=\lambda_\gamma=\dgz=0$).}
\label{fig:vertex_all}
\end{figure}

	In Fig.~\ref{fig:vertex_all}, we show the allowed 95\% CL regions for the BSM coefficients $\delta g_{L}^{Zu}$, $\delta g_{R}^{Zu}$, $\delta g_{L}^{Zd}$, $\delta g_{R}^{Zd}$ defined in Eqs.~(\ref{eq:TGC}) and~(\ref{eq:vertices}), assuming \textit{i)}~that the aTGC are not negligible (yellow), \textit{ii)}~that $\lambda_\gamma = \dka = 0$ (blue) and 
\textit{iii)}~that $\lambda_\gamma = \dka = \dgz= 0$ (pink). In gray we show the bounds extracted from the LEP-1 fit of Ref.~\cite{Efrati:2015eaa}, assuming that the EFT obeys either a MFV (light gray) or a FU (dark gray) flavour structure. To avoid confusion, we remind that when extracting the diboson bounds, we do not differentiate the cases of MFV and FU since diboson production is mostly insensitive to possible differences between the light generations and the third generation that could appear in the MFV case; the only difference is a matter of interpretation, i.e. if one assumes FU the diboson bounds on the $Z\bar{q}q$ anomalous couplings apply to all the three quark generations, while if one assumes MFV they only apply  to $u,\, d, \,c$ and $s$ quarks. 
	
	We find that even for the most general case that includes all the seven BSM parameters (yellow), the diboson bounds for the down-type couplings are already competitive with those from LEP-1 one under the MFV scheme.  The LHC bound on $\delta g_R^{Zd}$ 
is better than the LEP-1 under the MFV hypothesis and it remains competitive under the FU assumption. On the contrary, for the up type quarks, we find that the LHC bounds are still significantly worse than those from LEP-1, even under the MFV assumption.
	
	Assuming that $\lambda_\gamma = \dka =0$ (blue), we find a big improvement on the diboson fit with respect to the seven parameter fit (yellow). The most striking difference being that for the up-type quark couplings, $\delta g_L^{Zu}$ and $\delta g_R^{Zu}$, the diboson bounds become of the same order of magnitude as those from LEP-1; from these two couplings, it is $\delta g_R^{Zu}$ that benefits the most from setting $\lambda_\gamma = \dka =0$. Notice that the improvement to $\delta g_R^{Zu}$ is due to setting $\dka=0$, since as shown in Fig. \ref{fig:flatdirection} they are strongly correlated. On the other hand $\delta g_R^{Zu}$ is insensitive to $\lambda_\gamma$.  For the down type couplings, we also find an improvement of about a factor two when setting the  two aTGC to zero. With these improvements, the current LHC diboson data set constraints on  $\delta g_R^{Zd}$ that are of same order as those derived from LEP-1 data in a FU setup. For MFV scenarios, the LHC bounds significantly outperform the LEP-1 ones.
	
In pink, we report the constraints for scenarios in which all the three aTGC are negligible compared to $\delta V\bar{q}q$. 
Actually, letting $\dgz$ float or not does not significantly change the conclusion: we see that the left handed couplings get a significant improvement with respect to the blue region, while the right handed ones are almost insensitive to this extra assumption on $\dgz$. 
This, again, can be understood from the ``correlation matrix" of Fig.~\ref{fig:gridplot} that shows that $\delta g_{1z}$ is mostly correlated with the left handed couplings.
\medskip
	
Note that the correlations among the left and right couplings from LEP-1 measurement are not aligned with the correlation appearing in $pp \to WV$ data, which gives some synergetic value to the combination of the two sets of data. This can be seen for example in Table~\ref{tab:nolooptgcs} that gives the individual constraints from diboson and LEP-1 and their combination when $\delta\kappa_\gamma=\lambda_\gamma=0$. Notice also that while the LEP-1 data for down quarks has a two sigma excess (driven by the $Zb\bar{b}$ asymmetry) in an analysis in a FU context, the LHC diboson data presents a two sigma excess as well, but in the opposite direction. So the combination alleviates the tension with the SM.	
\medskip

One should remember that the bounds from $pp \to WV$ in Fig.~\ref{fig:vertex_all} only constrain BSM theories where the new particles are above few TeV (see the discussion on the validity of EFT analysis in section~\ref{sec:interpretation}), while those from LEP-1 apply to theories where the new particles can be as light as $\gtrsim {\cal O}(100)$\,GeV. 

To conclude this section, we note that in the fits of the $pp \to WV$ data, the quadratic amplitudes appear to be non-negligible, modifying the constraints by a factor $\sim1.5-2$ when $\delta \kappa_\gamma$ is neglected, and by a larger factor when $\delta \kappa_\gamma$ is taken into account, as a result of the correlations identified earlier.
 We comment on what it means for the EFT interpretation and possible BSM models in section~\ref{sec:interpretation}. 

\subsection{LHC bounds on aTGC and interplay with $\delta V \bar{q}q$}
\label{sec:aTGCvsZqq}
	
\begin{figure}[t]
\centerline{\includegraphics[scale=0.23]{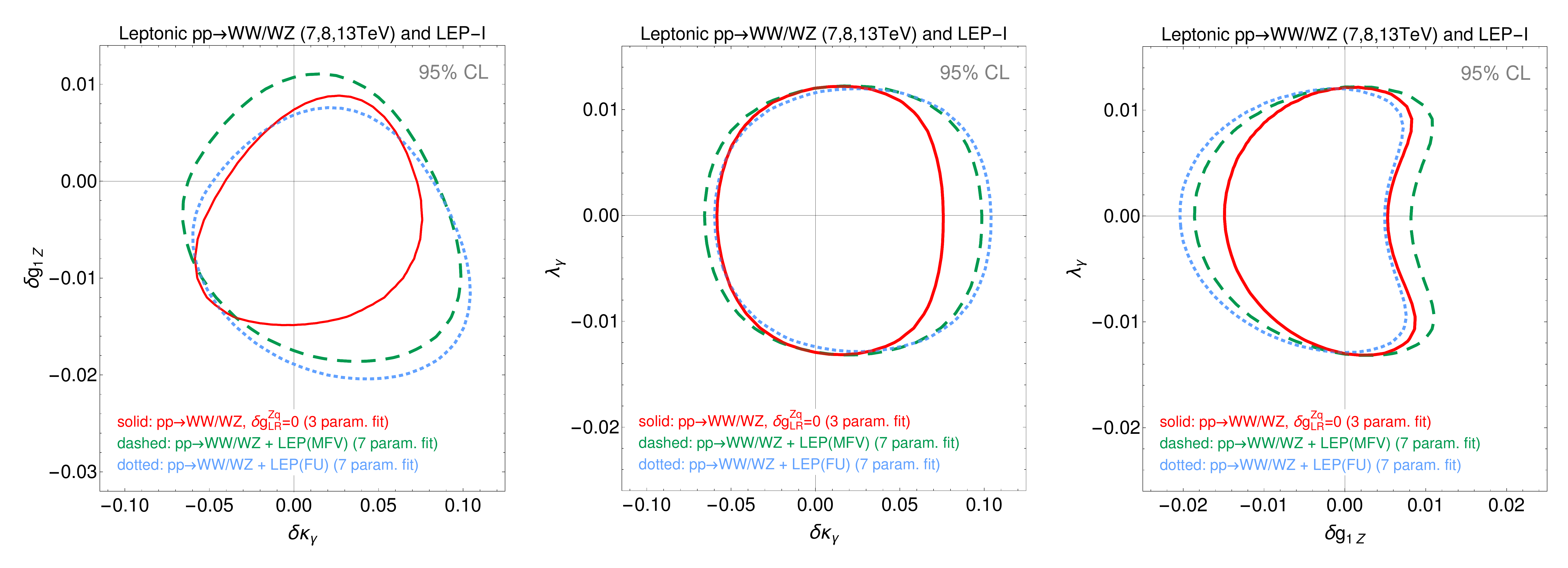}}
\caption{95\% CL regions for the aTGCs. The red curves show the bounds from a  three parameter fit of the current LHC diboson data, imposing $\delta V \bar{q}q=0$ and  profiling over the aTGC not shown on the plots.  The green and blue curves show the bounds from a  seven parameter global fit using the current LHC diboson data as well as the LEP-1 data under the MFV and FU assumptions respectively; the parameters not shown on the plots are profiled over.}
\label{fig:aTGCs_wVertex}
\end{figure}
	
Figure~\ref{fig:aTGCs_wVertex} presents the 95\% CL regions for the three aTGC parametrized by $\dgz, \, \dka, \, \lambda_\gamma$. In red, we show a fit to the three aTGC setting $\delta g_{L}^{Zu}=\delta g_{R}^{Zu}=\delta g_{L}^{Zd}=\delta g_{R}^{Zd}=0$ and profiling over the one aTGC not appearing in the plot. In this case we only use the LHC data from Table~\ref{tab:experimental}. In dashed green and dotted blue, we make a fit to the seven BSM parameters, the three aTGC $\dgz, \, \dka, \, \lambda_\gamma$ and the four $\delta g_{L,R}^{Zu,d}$, and profile over those not appearing in the plot; in this case we use $\chi^2 = \chi^2_{diboson} + \chi^2_\textit{LEP-1}$, assuming FU (dashed green) and MFV (dotted blue).
	
	From  Fig.~\ref{fig:aTGCs_wVertex}, we see that the effect of not neglecting the $\delta V\bar{q}q$ is the largest in the $(\dka, \delta g_{1z})$ plane, where the constrained area in parameter space varies around $50\%$ from one assumption to the other. This points to a large correlation between $\dka$ and $\dgz$ on the $\delta V \bar{q}q$ parameters, which is to be expected since they appear in the same high energy amplitudes as seen in Eqs.~(\ref{eq:AmpWW}) and~(\ref{eq:AmpWZ}). The determination of $\lambda_\gamma$ is insensitive to the different assumptions, as expected from the fact that it is the only parameter appearing in the amplitudes that grow with $\hat{s}$ and have final polarizations $\pm \pm$.
	\medskip
		
\begin{figure}[t]\centering
		\includegraphics[scale=0.316]{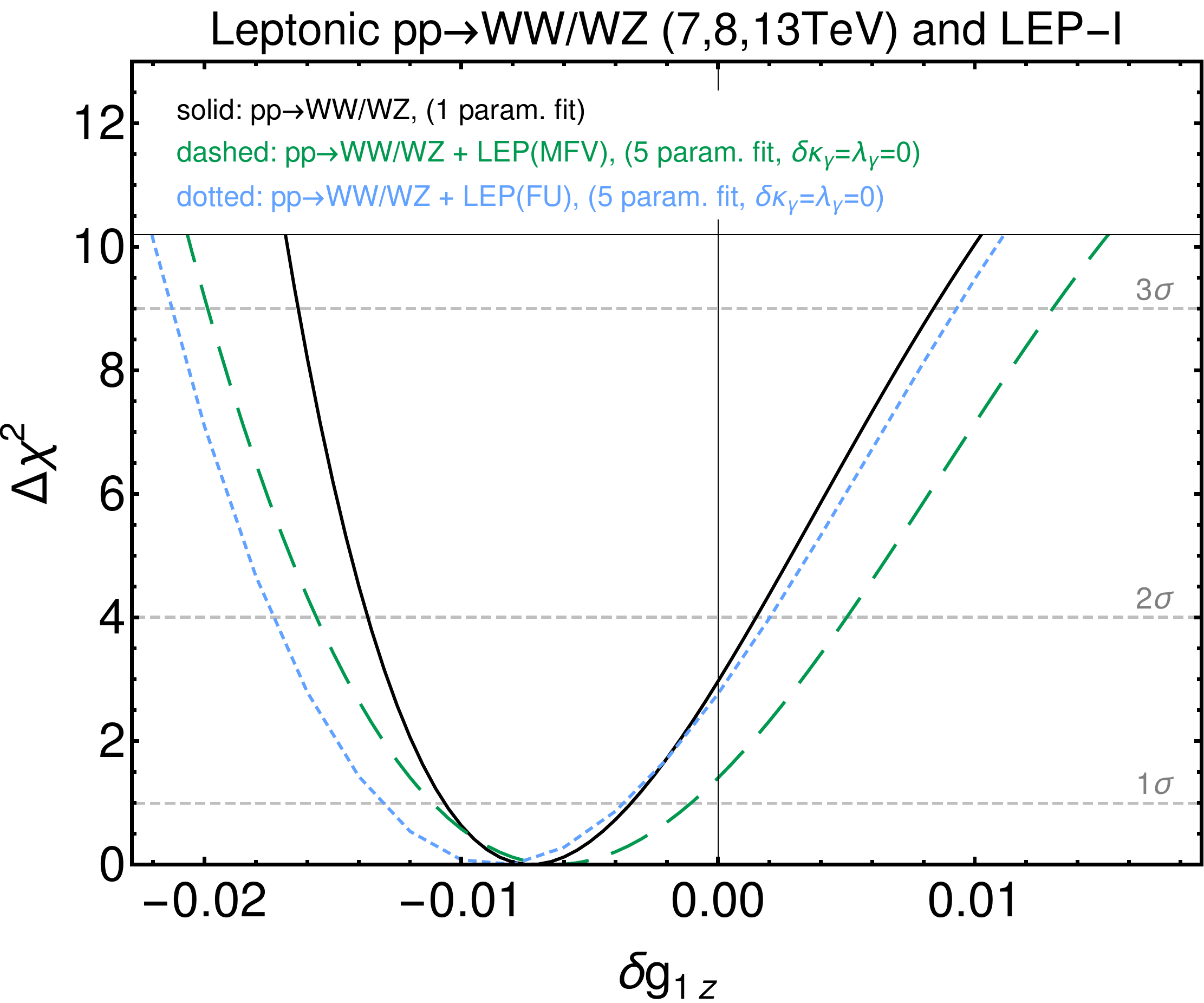} \quad\quad
		\includegraphics[scale=0.32]{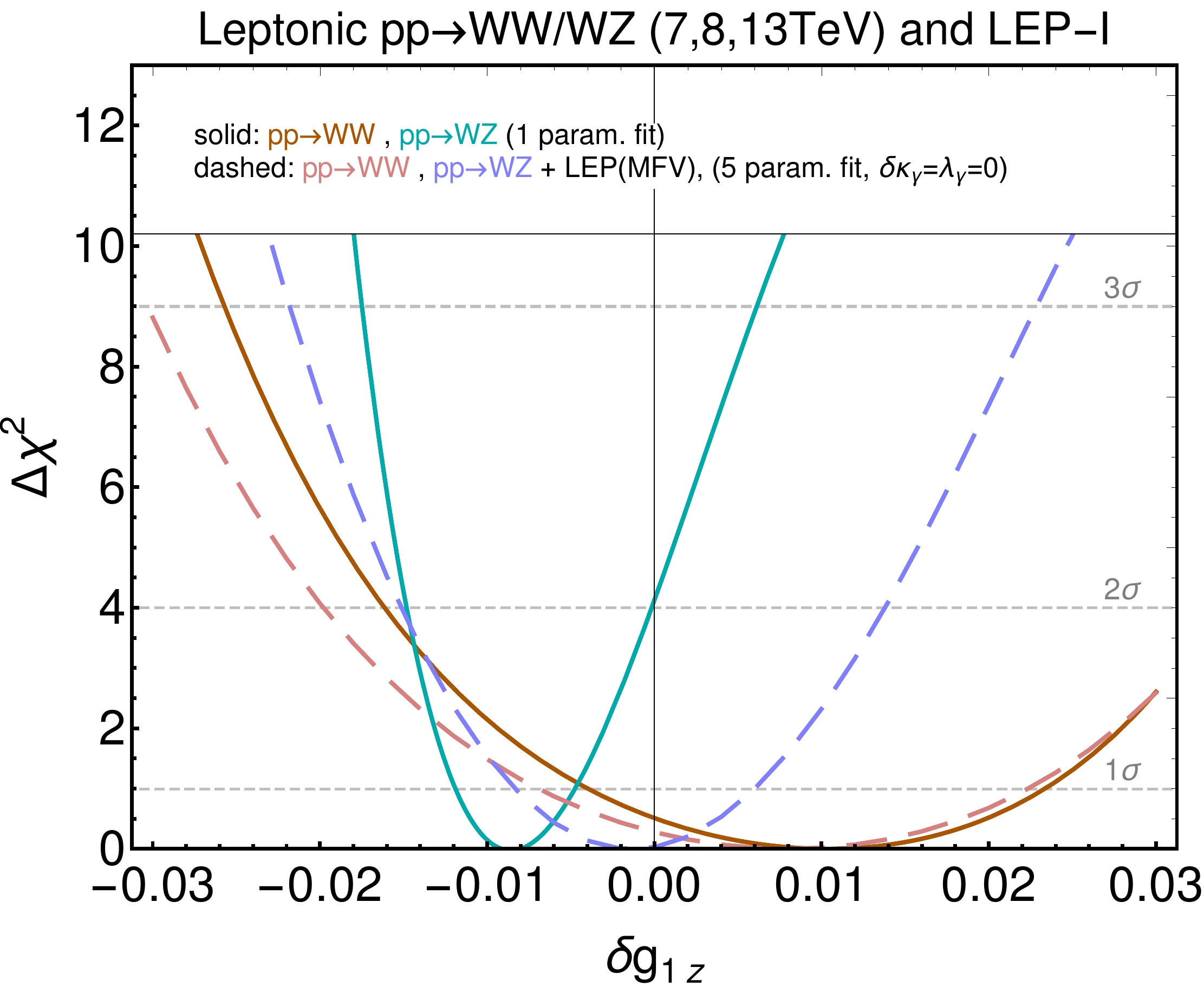}
		\caption{ Constraints on $\delta g_{1z}$ assuming $\delta \kappa_\gamma$ and $\lambda_\gamma$ to be loop suppressed, i.e. negligible. In solid, exclusive fits where only $\delta g_{1z}$ is taken into account. In dashed and dotted, fits profiling over the fermion-electroweak boson vertices under the two different MFV and FU assumptions respectively. {\bf Left:} $\Delta \chi^2$ combining all the channels shown in Table \ref{tab:experimental}. {\bf Right:} Showing the $\Delta \chi^2$ for the $WW$ and $WZ$ channels separately.}
		\label{fig:dg1z_onedimensional_noloop}		
	\end{figure}

	Given that in many BSM models $\delta \kappa_\gamma$ and $\lambda_\gamma$ are assumed to be loop induced and therefore parametrically smaller than $\delta g_{1z}$, we also study the effect of profiling over $\delta V\bar{q}q$ when $\dgz$ is the only aTGC modifying the diboson production. Since the global fit is non-Gaussian, this particular case with $\dka=\lz=0$ cannot be obtained simply from the general case.   On the left plot of Fig.~\ref{fig:dg1z_onedimensional_noloop}, we show in solid black the one parameter exclusive fit to $\delta g_{1z}$, setting all the other parameters to zero. In dashed green and dotted blue, we allow $\delta g_{L}^{Zu}, \, \delta g_{R}^{Zu}, \, \delta g_{L}^{Zd}, \, \delta g_{R}^{Zd}$ to be different than zero and perform a global fit.  On the right plot of Fig. ~\ref{fig:dg1z_onedimensional_noloop} we perform the analysis separating the $WW$ and $WZ$ channels, and find that currently the $WZ$ channels dominate the total $\Delta \chi^2$.
	\medskip

	Similarly to Fig.~\ref{fig:aTGCs_wVertex}, Fig.~\ref{fig:dg1z_onedimensional_noloop} tells us that profiling $\delta V \bar{q}q$ in the fit changes the current constraints on $\delta g_{1z}$ by a factor of about $25\%$. Also, we find that once the $\delta V \bar{q}q$ are introduced, the FU and MFV assumptions which modify $\chi^2_\textit{LEP-1}$ yield qualitatively similar size effects to $\dgz$ but still with at least 10\% differences between the two.
	\medskip
	
In this section, we have presented an analysis using all the current leptonic diboson data to set constraints to both aTGCs and $\delta V\bar{q}q$ vertices, and study the correlations under different flavour schemes. The LEP-1 constraints for $\delta V \bar{q}q$ in Fig.~\ref{fig:vertex_all} could lead to the conclusion that  for a MFV setup larger deviations could be obtained. However, this is not the case because the cross correlations among LEP-1 and LHC diboson data make the global fit more constraining than both sets of data alone. The largest correlation in the fit appears to be between $\dka$ and $\delta g_R^{Zu}$ as established in section~\ref{sec:corr}.

\section{Projected bounds for  HL-LHC}
\label{sec:HL-LHC}
\subsection{Data used and assumptions for HL-LHC}

To estimate the bounds at HL-LHC, as a first step and for simplicity, we  simulated the channels  $pp \to WW \to \nu \ell \nu \ell$ and $pp \to WZ \to \ell \nu \ell \ell$. We build a $\chi^2$ function with the same form as in \eq{eq:chi2raw} and we inject the SM signal, i.e., we assume that the measured number of events will be the same as in the SM prediction, so that $\sigma_{measured} = \tilde{\sigma}_{SM}^{bkg} + \tilde{\sigma}_{SM}^\textit{signal}$ defined after \eq{eq:chi2raw}. Therefore the $\chi^2$ can be written as:
\be
	\chi^2 = \sum_{I\in \text{channels}}\,\sum_{i\in \text{bins}} \frac{[\, \delta \mu(\vec{\delta} \, )]^2}{(\delta_{syst})_{I,i}^2 +(\delta_{stat})_{I,i}^2  } \,
\label{eq:chi2HLLHC}
\ee	
where we define $\delta_{syst} = \sqrt{b+({\cal L} \, \Delta_{syst})^2}/s$, with $\Delta_{syst}$ being the absolute systematics error in the cross section, ${\cal L}$ being the integrated luminosity, and $\delta_{stat} = 1/\sqrt{s}$. $s$ and $b$ stand for the number of simulated SM signal and background events, and  $\delta \mu(\vec{\delta} \, )$ is defined in \eq{eq:signalstrength}.  As usual, events with misidentified particles, such as misidentified leptons in processes with $W+\text{jets}$ or top production (see e.g. section 5 of Ref.~\cite{Khachatryan:2015sga}), are included within the background. 

There has  been no extensive study of the systematic uncertainties and the expected background for $pp \to WV$, especially in the high energy bins. A 5\% of systematic uncertainties is claimed to be possible in Ref.~\cite{Franceschini:2017xkh} in the fully leptonic $WZ$ channel within the fiducial region used for their analysis, and is used as a benchmark in Ref.~\cite{Liu:2018pkg} for the semileptonic $WV$ and $Wh$ channels. In Ref.~\cite{Frye:2015rba}, instead, it is claimed that this accuracy can only be reached by measuring ratios of cross sections. We take a pragmatic approach and consider two scenarios for the uncertainties at HL-LHC: a pessimistic one where  $\delta_{syst} = 30\%$ is assumed for all the bins, which corresponds to an extrapolation of the uncertainty in the overflow bins of the experimental analysis, and a more aggressive scenario where one assumes $\delta_{syst} =  5\%$ for all bins.

For the  $WW$ channel, we consider the $m_{\ell\ell}$ distribution  and for the $WZ$ channel, we consider the $m_T^{WZ}$ distribution. In both cases we have chosen the observables and cuts followed by the experimental collaborations in \cite{Khachatryan:2015sga,ATLAS:2016qzn}. The binning used in our analysis also follows the experimental collaborations for low transverse masses, while we add more bins at higher transverse masses to increase the sensitivity. This corresponds to the following binning,
\bea
m_T^{WZ} &\in&\{140,180, 250, 450, 600, 750, 900, 1100\} \ \text{GeV}\,, \nonumber \\
m_{\ell \ell} &\in& \{50, 125, 200, 300, 500, 700, 900, 1100, 1300, 1500, 1700\} \ \text{GeV} \,,
\eea
where in each case the overflow bin is chosen to contain at least ten events. As a small cross check, we compared our estimated bounds on the aTGC at HL-LHC with 3\,ab$^{-1}$, shown in red in Fig.~\ref{fig:aTGCs_wVertex_HLLHC}, with those in Fig.~3 of Ref.~\cite{Gianotti:2002xx}. There the channels $W\gamma \to \ell \nu \gamma$ and $WZ \to \ell \nu \ell \ell$ were considered and bounds on the aTGC were derived for a run at 14\,TeV with a total accumulated luminosity of up to 1\,ab$^{-1}$. Our bounds, assuming $\delta_{syst} = 5\%$ in the leptonic $WW$ and $WZ$ channels, turn out to be more conservative than those in Ref.~\cite{Gianotti:2002xx} but overall of the same order. So, our simple assumptions are in line with the existing literature and should give  a reliable and conservative estimate of the HL-LHC reach. Note that there are several ways to improve the diboson analysis: \textit{i)} the semileptonic channels can be considered on top of the purely leptonic ones, \textit{ii)} more refined observables like those presented in Refs.~\cite{Franceschini:2017xkh, Azatov:2017kzw, Panico:2017frx} can be studied.  Therefore even with $\delta_{syst} = 5\%$, our estimates on the diboson reach at HL-LHC are probably on the conservative side. To compare the traditional experimental analysis with new proposals, in section~\ref{sec:vectortriplets} we compare the HL-LHC reach of leptonic $WZ$ estimated in Ref.~\cite{Franceschini:2017xkh} with our combination of the leptonic $WW$ and $WZ$ using the $m_{\ell\ell}$ and $p_T^Z$ differential distributions.
\medskip

\subsection{HL-LHC projections on $\delta V \bar{q}q$ vs LEP-1}
\label{subsec:HLLHCboundsvsLEP1}

\begin{figure}[t]\centering
\centerline{\includegraphics[scale=0.37]{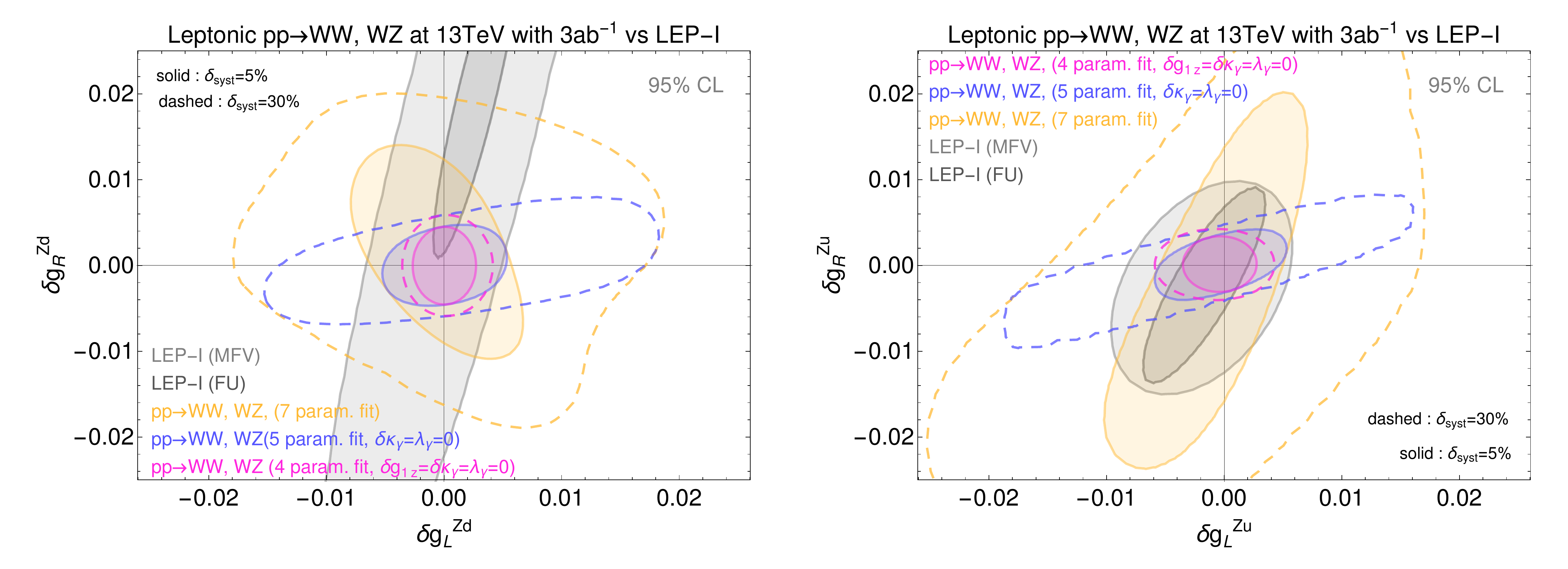}}
\caption{Estimated 95\% CL bounds at HL-LHC on the anomalous couplings between the light quarks and the electroweak bosons. In yellow,  diboson bounds after profiling over the remaining five parameters. In blue (pink),  same but setting also $\delta\kappa_\gamma=\lambda_\gamma=0$ ($\delta\kappa_\gamma=\lambda_\gamma=\dgz=0$). Solid and dashed stand for an assumed $\delta_{syst}=5\%$ and $\delta_{syst} = 30\%$ respectively.  Light (dark) gray regions correspond to the LEP-1 bounds assuming MFV (FU).}
\label{fig:VertexHLLHC}
\end{figure}	

Figure~\ref{fig:VertexHLLHC} shows the allowed 95\% CL regions for $\delta g_{L}^{Zu}$, $\delta g_{R}^{Zu}$, $\delta g_{L}^{Zd}$, $\delta g_{R}^{Zd}$  in the three different scenarios: \textit{i)} the three aTGC, $\lambda_\gamma, \dka, \dgz$, are kept as floating parameters in the fit (yellow), \textit{ii)} $\lambda_\gamma$ and   $\dka$ are set to zero (blue), and \textit{iii)}  the three aTGC  are set to zero (pink).  A total accumulated luminosity of  3\,ab$^{-1}$ is assumed. In order to appreciate the improvement compared to LEP,  the gray regions report the bounds extracted from the LEP-1 data  under the MFV (light gray) and FU (dark gray)  assumptions. Clearly, for low enough systematics, HL-LHC will  surpass the LEP-1 bounds for any new physics scenario with a built in MFV structure  that does not generate anomalously large aTGC, i.e. scenarios for which $\dka=\lambda_\gamma=0$ (blue) is a good approximation. Under the FU assumption, the HL-LHC bounds on $\delta g_{R}^{Zu}$ and $\delta g_{R}^{Zd}$ vastly surpass the LEP-1 bounds whenever $\dka=\lambda_\gamma=0$, while the bounds on $\delta g_{L}^{Zu}$ and $\delta g_{L}^{Zd}$ are only slightly better.
In any case, it should be noted that the blue and pink bounds improve by one order of magnitude at HL-LHC compared to the current bounds. 
As long as the systematics remain low enough, the seven parameter FU fit  also improves by about a factor three the bounds for all the $\delta V \bar{q}q$ with respect to the current bounds shown in Fig.~\ref{fig:vertex_all}. The seven parameter FU fit equals or surpasses the LEP-1 constraints for $\delta g_L^{Zd}, \, \delta g_L^{Zu}$ and $\delta g_R^{Zd}$.
On the other hand, with higher systematic uncertainties, $\delta_{syst} = 30\%$, the improvement from the seven parameter and five parameter fits with respect to the current constraints will be limited and mostly concern the right handed couplings. Only for $\delta g_R^{Zd}$, the HL-LHC will show an improvement over LEP-1 in all the cases, both for MFV and FU structures.

\subsection{HL-LHC projections on aTGC and interplay with $\delta V \bar{q}q$}

\begin{figure}[t]
\centerline{\includegraphics[scale=0.235]{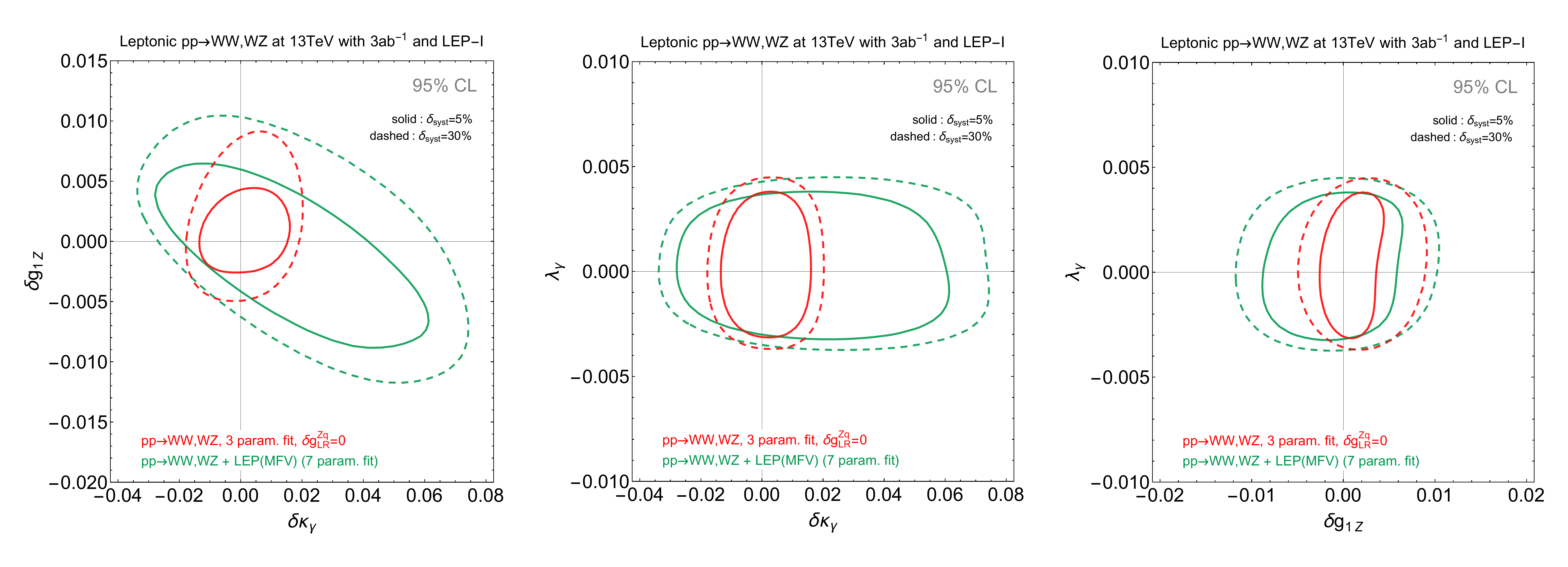}}
\caption{Estimated 95\% CL bounds on the aTGC at HL-LHC. Solid and dashed stand for $\delta_{syst}=5\%$ and $\delta_{syst}=30\%$ respectively. The bounds on $\dka$, $\dgz$ change by a factor two to three between the 3-parameter and the 7-parameter fits, while the bound on $\lambda_\gamma$ remains unaffected. The aTGC bounds in 3-parameter fit agrees well with the ones obtained for universal theories, see Fig.~\ref{fig:compPlot}.}
\label{fig:aTGCs_wVertex_HLLHC}
\end{figure}

Figure~\ref{fig:aTGCs_wVertex_HLLHC} shows the allowed 95\% CL regions for the three aTGC parametrized by $\dgz, \, \dka, \, \lambda_\gamma$. In red, we show a fit to the three aTGC setting $\delta g_{L}^{Zu}=\delta g_{R}^{Zu}=\delta g_{L}^{Zd}=\delta g_{R}^{Zd}=0$ and profiling over the one aTGC not appearing in the plot.  In green, we make a fit to the seven BSM parameters, namely the three aTGC $\dgz, \, \dka, \, \lambda_\gamma$ and the four $\delta g_{L,R}^{Zu,d}$ and we profile over those not appearing in the plot. We use the HL-LHC projections to build $\chi^2_{diboson}$ while the $\chi^2_\textit{LEP-1}$ is built from the global fits performed in Ref.~\cite{Efrati:2015eaa}. We find that at HL-LHC the differences between assuming MFV or FU for $\chi^2_\textit{LEP-1}$ are negligible when performing a combined global fit of LEP-1 and LHC data. For this reason in this section we only present results with the FU hypothesis for the LEP-1 fit.
\medskip

At HL-LHC, the aTGC bounds shown in Fig.~\ref{fig:aTGCs_wVertex_HLLHC} are qualitatively similar to those of Fig.~\ref{fig:aTGCs_wVertex} obtained with the current data. The main difference between the two is that the features found with the current data regarding the impact of $\delta V \bar{q}q$ are accentuated at HL-LHC. 
This is particularly true for $\delta \kappa_\lambda$ and $\delta g_{1z}$:
the bounds on the $\dka$, $\dgz$ vary by more than $100\%$ if instead of setting $\delta V \bar{q}q =0 $ they are included in a global fit combining the LEP-1 data in the context of FU or MFV scenarios. 
On the other hand $\lambda_\gamma$ will remain mostly unaffected, as anticipated from Eqs.~(\ref{eq:AmpWW}) and~(\ref{eq:AmpWZ}). 
\medskip

\begin{figure}
	\centering
	\includegraphics[width=0.45\linewidth]{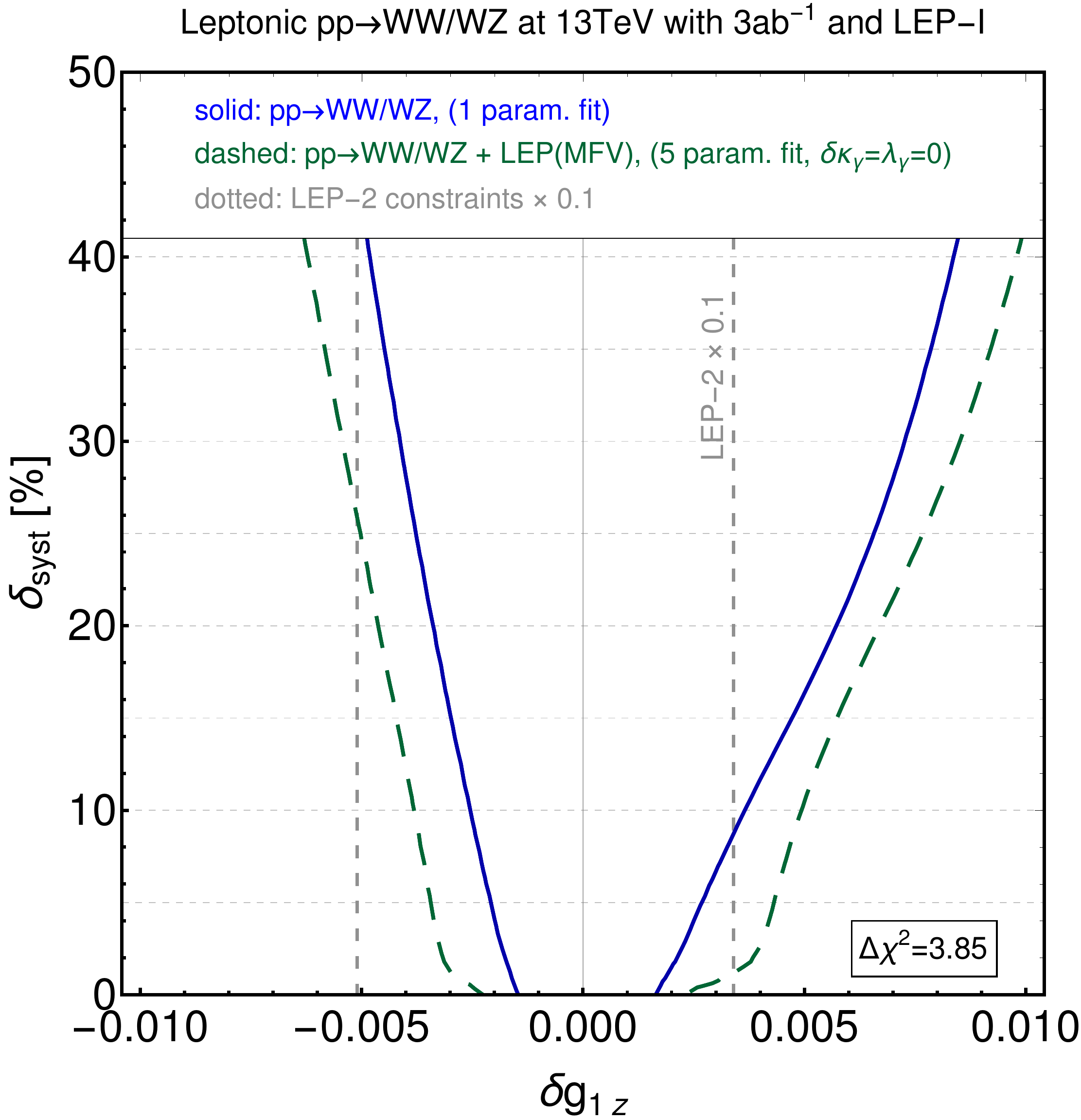} \quad\quad
	\includegraphics[width=0.45\linewidth]{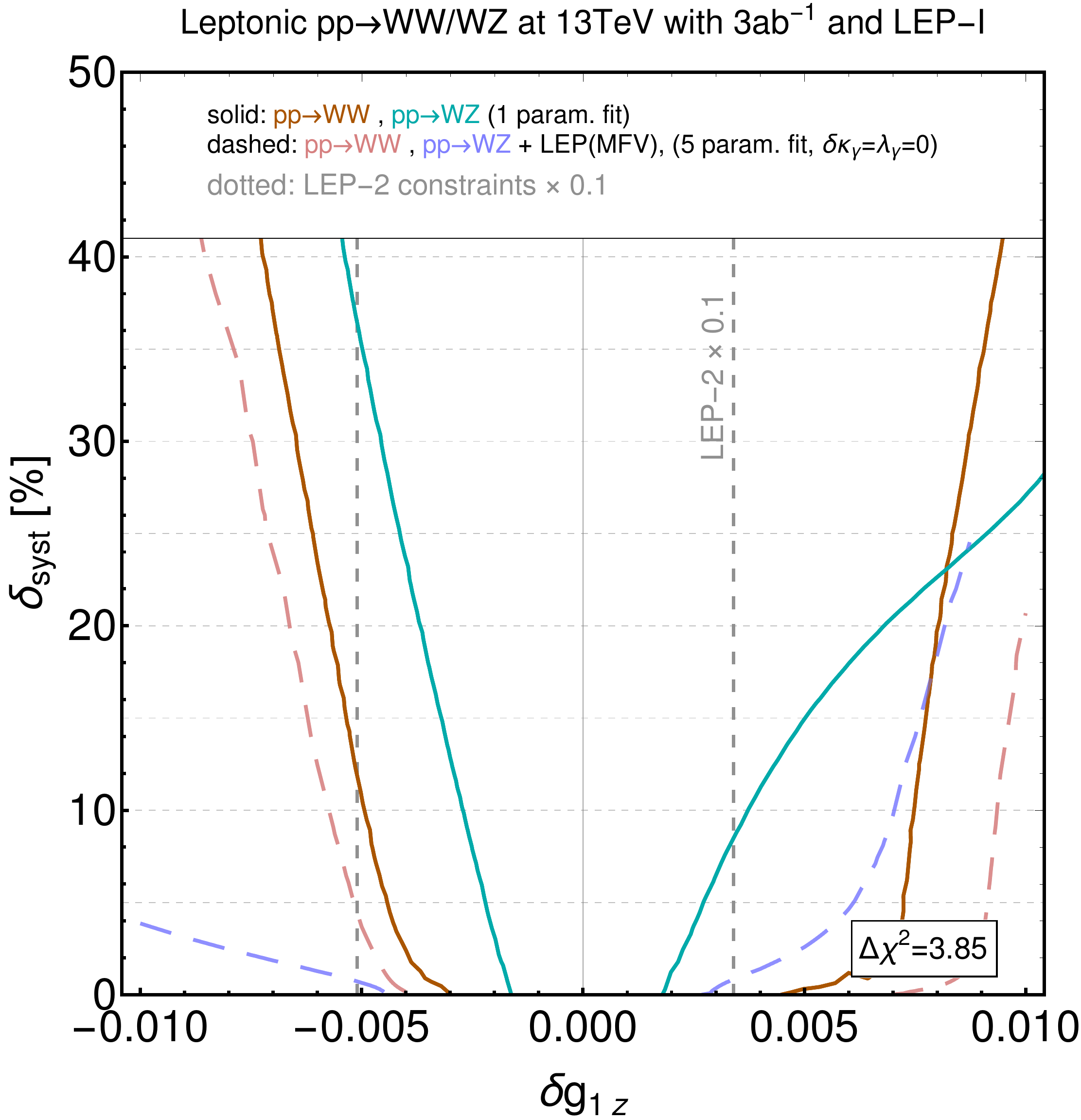}
	\caption{95 \%CL  bound at HL-LHC on $\dgz$ as a function of the assumed systematic uncertainty  for the case where the light quark vertices are neglected and for the case where we take them into account and combine the fit with LEP-1 data. For comparison, in light gray we show the LEP-2 constraints, rescaled by a factor 0.1.  {\bf Left:} $\Delta \chi^2$ combining the projections for $WW$ and $WZ$. {\bf Right:} Showing the projected $\Delta \chi^2$ for the $WW$ and $WZ$ channels separately.}
	\label{fig:dg1z_onedimensional_MFV_syst}
\end{figure}

Figure~\ref{fig:dg1z_onedimensional_MFV_syst}  shows the 95\%CL bound on $\dgz$ when setting $\dka = \lambda_\gamma = 0$ as a function of the assumed systematic uncertainty. 
Two cases are considered: \textit{i)} all deviations in the light quark vertices are  neglected and set to zero, and \textit{ii)} the diboson data are combined with the LEP-1 data and the light quark vertices are profiled over. The bound on $\dgz$ is rather robust and does not show a strong dependence on the assumed systematic uncertainty, changing by a factor two between when the systematics vary from 0\% to 50\%  (the statistical uncertainty is of course kept). The HL-LHC bound will be of the order of 0.1\%, an order of magnitude better than the current existing bound. And further improvement can be anticipated, e.g. by relying on the new analyses proposed in Ref.~\cite{Franceschini:2017xkh}.

\section{Interpretation of the constraints}
\label{sec:interpretation}

In this work we have performed a global analysis of the diboson data at the LHC and inferred bounds on aTGCs as well as on anomalous couplings of the quarks to the EW gauge bosons. We found that in some cases these bounds surpass the LEP-1 and LEP-2 bounds. Nonetheless, it is important to stress that this is only so for certain regions of the parameter space.  As in any EFT analysis, the constraints on the Wilson coefficients are only valid when the characteristic energy of the processes remains smaller than the masses of the new particles. Furthermore,  both for the current LHC data and also for the HL-LHC ones, the quadratic terms of the BSM contributions to the diboson production cross section play a non-negligible role in settings bounds on the Wilson coefficients.  In that situation, further restrictions on the parameter space follow to ensure that the interference between the SM amplitude and the dimension-8 operators, formally of the same order as the square of the dimension-6 operator contributions, remains sub-dominant~\cite{Contino:2016jqw, Azatov:2016xik,Azatov:2017kzw, Dror:2015nkp, Biekoetter:2014jwa, Farina:2016rws, Alioli:2017jdo, Alioli:2017nzr, Butter:2016cvz,Falkowski:2016cxu,Panico:2017frx,Zhang:2016zsp, Baglio:2017bfe,Franceschini:2017xkh}. We comment on these two limitations, in the following, and also see how they appear in a concrete toy model with vector triplets.

\subsection{Quadratic BSM amplitudes}
\label{sec:validity}

 As already noted and extensively discussed in Refs.~\cite{Contino:2016jqw, Azatov:2016xik,Azatov:2017kzw, Dror:2015nkp, Biekoetter:2014jwa, Farina:2016rws, Alioli:2017jdo, Alioli:2017nzr, Butter:2016cvz,Falkowski:2016cxu,Panico:2017frx,Zhang:2016zsp, Baglio:2017bfe,Franceschini:2017xkh}, when setting bounds to the EFT coefficients, it may happen that these bounds only constrain BSM amplitudes that are larger than the SM one. This makes the quadratic dimension six BSM amplitudes to be non negligible. To get a sense of which BSM theories can be studied only using dimension six operators while neglecting those of dimension eight, it is useful to schematically write the ratio of amplitudes between the EFT and the SM. These estimates have already been discussed in Refs.~\cite{Azatov:2016sqh,Contino:2016jqw,Azatov:2017kzw}, here we only give a small review for convenience. Schematically, for the $WV$ channels with longitudinally polarized gauge fields, the ratio of EFT and SM amplitudes is given by
	\be
	\left| \frac{{\cal M_{\rm EFT}}}{{\cal M}_{\rm SM}} \right|^2 \,\sim \, 1+\frac{c_6}{g_{SM}^2}\frac{E^2}{\Lambda^2}+\left( \frac{c_6^2}{g_{SM}^4}+\frac{c_8}{g_{SM}^2}  \right)\frac{E^4}{\Lambda^4} + \cdots \,,
	\label{eq:EFTvalidity}
	\ee
where $c_6, \, c_8$ represent the coefficients in front of the $d=6,8$ operators. When the quadratic terms dominate, the following condition has to be fulfilled in order to be able to neglect the dimension-8 operators
	\be
	c_6^2 \gg g_{SM}^2 \, c_8\,.
	\label{eq:const1EFT}
	\ee	
For simple power counting rules such that $c_6 \sim c_8 \sim g_\star^2$, with $g_\star$ a charactestic coupling of the new physics degrees of freedom, the EFT validity condition simply requires that the BSM coupling must be larger than the SM one, $g_\star^2 \gtrsim g_{SM}^2$, which is nothing else than the condition that also ensures that the  quadratic BSM pieces dominate \eq{eq:EFTvalidity}.

In the channels with mixed longitudinal and transverse polarizations, for which the new physics amplitude only grows as $\sqrt{\hat{s}}$, the same conclusion applies. The channel with transverse polarizations  only is, however, slightly different. In that case, the linear/interference terms at the dimension-6 level is suppressed due to the necessity to go through a helicity flip~\cite{Azatov:2016sqh}:
	\be
	\left| \frac{{\cal M_{\rm EFT}}}{{\cal M}_{\rm SM}} \right|^2 \,\sim \, 1+\frac{c_6}{g_{SM}^2}\frac{m_W^2}{E^2}\frac{E^2}{\Lambda^2}+\left( \frac{c_6^2}{g_{SM}^4}+\frac{c_8}{g_{SM}^2} \right)\frac{E^4}{\Lambda^4} + \cdots \,.
	\label{eq:EFTvalidityTT}
	\ee
And the quadratic pieces can dominate the linear terms for smaller values of $g_\star$. One would then end up in a region of the parameter space where the EFT analysis would not be valid since the dimension-8 operators cannot be neglected. Recently, new observables have been proposed~\cite{Azatov:2017kzw, Panico:2017frx} to \textit{resurrect} the interference and then circumvent this (in)validity issue.

\subsection{Power countings and BSM interpretations}
\label{sec:powercountings}

Assessing the consistency of EFT interpretation requires some assumptions on the scaling of the Wilson coefficients of the higher dimensional operators. We present here different power counting and selection rules which inspired the particular choices of the BSM parameters kept in the fits presented  in sections~\ref{sec:EFTbounds} and~\ref{sec:HL-LHC}. They correspond to specific dynamics for the new physics above the weak scale. We follow the conventions of \cite{Falkowski:LHCHXSWG-INT-2015-001}, 
\be
\quad {\cal L}  = {\cal L}_{SM} + \sum_{d=6} \bar{c}_i \, {\cal O}_i + \dots \quad\quad \text{with} \quad\quad  \bar{c_i} = c_i \, v^2/\Lambda^2 = \hat{c}_i \, m_W^2/m_\star^2, 
\label{eq:conventionsOps}
\ee
where $[v]=[\Lambda]=\sqrt{\hbar}/L$, $[m_W]=[m_\star]=1/L$ and $c_i$ and $\hat{c}_i$ are dimensionless for all the operators except for ${\cal O}_{3W}$ which has dimensions of $1/L^4$ hence $c_{3W}$ and $\hat{c}_{3W}$ have dimensions of $[g^{-2}]=\hbar$. ~\footnote{Notice that we use the notation ${\cal O}$ instead of $O$ in \eq{eq:conventionsOps} with respect to the notation of \cite{Falkowski:LHCHXSWG-INT-2015-001}.}
\begin{itemize}
\item For generic weakly coupled and renormalizable BSM gauge theories, with new particles of spin at most equal to one and with interactions mediated by operators of mass dimension smaller or equal to 4, one can check that, after integrating the BSM sector the operators, defined in the SILH basis \cite{Falkowski:LHCHXSWG-INT-2015-001}, ${\cal O}_{HB}$, ${\cal O}_{HW}$ and ${\cal O}_{3W}$ can only be generated at one loop, while ${\cal O}_B$, ${\cal O}_W$, ${\cal O}_{2B}$, ${\cal O}_{2W}$ can be generated at tree level.  \footnote{See an example of this procedure in \cite{Arzt:1994gp}.}. Given that at dimension six
\be
	\dka \sim \bar{c}_{HB} + \bar{c}_{HW}\,, \hspace{1 cm}
	\lambda_\gamma \sim g^2 \, \bar{c}_{3W} \,, \hspace{1 cm} 
	\dgz \sim \bar{c}_W+\bar{c}_{2W} + \frac{g'^2}{g^2} \, \bar{c}_B + \bar{c}_{HW}  \ \dots\, \, ,
	\label{eq:TGCops}
	\ee
one expects $\dka, \lambda_\gamma$ to be loop suppressed with respect to $\dgz$ and therefore the leading effect is expected in $\dgz$.
\medskip

As shown in previous sections, diboson production can also be affected by the vertex corrections $\delta V \bar{q}q$. In weakly coupled gauge theories these can also be generated at tree level, either by contributions from the bosonic operators ${\cal O}_B$, ${\cal O}_W$, ${\cal O}_{2B}$, ${\cal O}_{2W}$ or via ${\cal O}_{HQ}^{(1)}$, ${\cal O}_{HQ}^{(3)}$, ${\cal O}_{Hu}$, ${\cal O}_{Hd}$, see for instance the model in section \ref{sec:vectortriplets}. Therefore, whether they can be neglected in the diboson process or not depends on the specific details of the UV theory. The flavour assumptions of the UV theory will also determine the strength of the LEP-1 bounds.

\item  Strongly-Interaction-Light-Higgs (SILH) models \cite{Giudice:2007fh} address the hierarchy problem by making the Higgs boson a pseudo Nambu--Goldstone boson associated with the symmetry breaking of a global symmetry of a new strong sector.  Hence, the Higgs boson is a composite particle belonging to the strong sector. On the other hand, the gauge bosons and fermions appearing in the SM are assumed to be elementary and external to the strong sector and they acquire their masses by mixing linearly to the resonances of the strong sector, a setup dubbed partial-compositness. Assuming that the strong sector can be characterised by one mass scale $m_\star$ and one coupling $g_\star$, it is possible to estimate the size of the various EFT coefficients appearing after integrating it out, see Ref.~\cite{Giudice:2007fh}. Focusing on the bosonic operators relevant to the aTGC and the electroweak precision tests, their estimated size is found to be of the order 
\bea
\hat{c}_{WH}, \, \hat{c}_{HB}, \sim \frac{g_\star^2}{16 \pi^2} \,, \quad \hat{c}_{3W}\sim \frac{1}{16 \pi^2},  \quad
\hat{c}_{2B} \sim \frac{g^2}{g_\star^2} \, , \quad \hat{c}_{2W} \sim  \frac{g'^2}{g_\star^2}\,, \quad \hat{c}_B, \hat{c}_W \sim {\cal O}(1) \,,
\eea
 which yields the following power countings for the aTGC's:
\be
	\delta g_{1z} \sim \frac{m_W^2}{m_\star^2}\,,\hspace{2cm}
	\dka \sim {g_\star^2 \over 16 \pi^2}\frac{m_W^2}{m_\star^2}\,, \hspace{2cm}
	\lambda_\gamma \sim {g_{SM}^2 \over 16 \pi^2}\frac{m_W^2}{m_\star^2} \,.
	\label{eq:TGCpc}
\ee
Therefore for SILH-like models,  $\delta \kappa_\gamma$ and $\lambda_\gamma$ are parametrically suppressed by one loop with respect to $\delta g_{1z}$. 
\medskip

Let us now asses the size of the vertex corrections $\delta V \bar{q}q$ that can also modify the diboson production. If the leading operators modifying the $W, \, Z$ couplings to the quarks are bosonic, one ends up with 
 \be
	 \delta g_{L,R}^{Zu,d} \sim \frac{m_W^2}{m_\star^2}.
	 \label{eq:elementaryquark}
	 \ee
which is comparable in size to the deviation expected for $\delta g_{1z}$. However in this case the LEP-1 bounds are very constraining (see subsection on universal theories in Appendix \ref{app:lepconstraints}) and the $\delta V \bar{q}q$ vertex corrections can be neglected in our analysis of diboson production. On the other hand, if one has non-bosonic operators modifying $\delta V \bar{q}q$, then, the LEP-1 bounds can be relaxed such that these vertex corrections affect the diboson production. This can happen when the light quarks have a high degree of compositeness $\eps_q$ due to linear mixings of the form  $\eps_q m_\star \bar{q}\Psi$, where $q$ stands for either a quark double or singlet and $\Psi$ a heavy resonance. In this case, it is possible to generate sizeable and uncorrelated coefficients for the operators ${\cal O}_{Hq}^{(3)}$, ${\cal O}_{Hq}^{(1)}$,  ${\cal O}_{Hu}$, ${\cal O}_{Hd}$ which yield vertex corrections of order:
	\be
	\delta g_{L,R}^{Zu,d} \sim \eps_q \frac{g_\star^2}{g_{SM}^2}\,\frac{m_W^2}{m_\star^2}.
	\label{eq:quarkcomposite}
	\ee
as shown in the example of section \ref{sec:vectortriplets}. In this scenario, the LEP-1 constrains are expected to be less stringent than in the universal case (when the EFT is described only by bosonic operators), and therefore the vertex corrections $\delta V \bar{q}q$ may affect diboson production at the LHC.

\item Strongly Coupled Multi-pole Interaction models \cite{Liu:2016idz} are hypothetic strongly coupled UV theories where the fermions and gauge bosons appearing in the SM can be resonances of a strong sector; nonetheless their EFT's can exhibit couplings at dimension four,  $g_{SM}$,  that are much smaller than the characteristic coupling of the strong sector $g_\star \lesssim 4 \pi$. On the other hand, the strong coupling $g_\star$ only manifests via higher derivative interactions coming from operators with $d>4$. The resulting power counting for these theories, when the SM gauge bosons are part of the strong sector, shows that $\lambda_\gamma$ and $\dka$ can be generated without the loop suppression. The three scenarios presented in Ref.~\cite{Liu:2016idz} {\it Pure Remedios}, {\it Remedios}+MCHM and {\it Remedios}+ISO(4) induce the following $d=6$ operators with sizes of order
\be
\hat{c}_{3W} \sim \frac{g_\star}{g_{SM}^3}\, , \hspace{2cm} \hat{c}_{2B}, \hat{c}_{2W} \sim {\cal O}(1)\, .
\ee
where $g_\star \gtrsim g_{SM}$. On the other hand only the {\it R}+MCHM and {\it R}+ISO(4) generate the operators ${\cal O}_{HB}$, ${\cal O}_{HW}$, ${\cal O}_{W}$ and ${\cal O}_{B}$ with sizes of
$
\hat{c}_{HB}, \hat{c}_{W}, \hat{c}_{B} \sim {\cal O}(1)\,,$
while $\hat{c}_{HW}$ is of order ${\cal O}(1)$ and $g_{\star}/g$ for {\it R}+MCHM and {\it R}+ISO(4) respectively. These power counting imply that for all these models, i.e. {\it Pure Remedios}, {\it R}+MCHM and {\it R}+ISO(4),
\be
\lambda_\gamma \sim \frac{g_\star}{g_{SM}} \frac{m_W^2}{m_\star^2} \,.
\ee
On the other hand, for the {\it Pure Remedios} case, $\dgz \sim \frac{m_W^2}{m_\star^2}$ while $\dka$ is not generated at dimension six. For the other two cases one has that
\be
\dgz \sim \dka \sim \frac{m_W^2}{m_\star^2} \quad (\text{{\it R}+MCHM})\,, \hspace{1cm} \dgz \sim \dka \sim  \frac{g_\star}{g_{SM}} \frac{m_W^2}{m_\star^2}  \quad (\text{{\it R}+ISO(4)})\,.
\ee
Hence, {\it Remedios}-type models show possible ways in which $\dka$ and $\lambda_\gamma$ could be not loop suppressed, giving further motivation for studying their bounds, while showing that neglecting these two aTGC may not always be appropriate.

\end{itemize}	
	
\subsection{Energy limitation}
\label{sec:EnergyLimitation}

An EFT has an intrinsic cutoff scale and for the analysis to be valid it should not use any event with a characteristic scale above this cutoff. Given that the center of mass energy of the interacting partons is not known at the LHC, it may be impossible to know the center of mass energy of a given process if the energy and momentum of the final states are not completely reconstructed. This is the case of the leptonic processes for $pp \to WV$ with one or two neutrinos in the final states. We set 3\,TeV as the energy for which the EFT stops to be valid following the analysis from Ref.~\cite{Falkowski:2016cxu} and checking that we did not get any events in the \texttt{Madgraph5} simulation above 3\,TeV. One could extend the EFT reach below the 3\,TeV mark without changing the experimental analysis by following the procedure explained in Ref.~\cite{Contino:2016jqw,Biekoetter:2014jwa,Racco:2015dxa,Falkowski:2015jaa}. This procedure is based on considering only the events with a characteristic energy below a pre-determined cutoff scale $E_{cut}$. The constraints obtained this way are, although not optimal since one is throwing away the events above the cutoff, totally consistent with the EFT expansion. 

To illustrate the effect of the energy limitation to stay within the validity region of an EFT analysis, one can consider the projections of the $\delta V \bar{q}q$ bounds onto the parameter space $(g_\star,m_\star)$ on models whose dynamics follows the power counting discussed above.
\begin{itemize}
\item For SILH-like models with elementary quarks, the scaling \eqref{eq:elementaryquark} naively leads to a simple 68\%CL lower bound on $m_\star$ independent of $g_\star$:
\begin{equation}
m_\star > (500, 900, 1300)\,\textrm{GeV at LEP-1, LHC, HL-LHC respectively}
\label{eq:boundsValidity1}
\end{equation}
(we used the constraints on $\delta g_R^{Zd}$ of Tables~\ref{tab:nolooptgcs} and~\ref{tab:projvertexs}, under the MFV assumption and setting the aTGC to zero).
However,  the bounds in \eq{eq:boundsValidity1} for the LHC and HL-LHC are not reliable since they fall outside the regime of validity of the EFT.  Hence, only the LEP-1 bound can be trusted for these types of models. Futhermore, as commented above, in this elementary quark SILH scenario, there exist some correlations among the four $\delta V\bar{q}q$ couplings and a more meaningful bound on $(m_\star, g_\star)$ should take into account these correlations.
For instance, the 95\%CL LEP-1 bound can be obtained from the universal theory fit, see Eq.~\ref{eq:Bosonicbounds}, leading to
\begin{equation}
m_\star>2.3\,\textrm{TeV}.
\end{equation}
In deriving this constraint, we have used $\delta g_R^{Zd} \sim g'^2  \hat{S} /(3g^2-3g'^2)$ and assumed the scaling $\hat{S}\sim m_W^2/m_\star^2$, where $\hat{S}$ is one of the oblique parameters relevant for universal theories.

\item For the SILH-like models with composite quarks, the situation is different and there the diboson channels at the LHC can be used to set  reliable constraints stronger than the ones derived at LEP. Indeed the scaling from \eq{eq:quarkcomposite} imposes at 68\%CL
\begin{equation}
g_\star/m_\star < (1.3, 0.7, 0.4)\,{\rm TeV^{-1}} \, \textrm{at LEP-1, LHC, HL-LHC respectively}.
\end{equation}
And this time, the validity constraint of $m_\star>3$\,TeV implies that at the LHC and HL-LHC, diboson data can reliably  constrain theories with $g_\star > 2.1$ and $1.2$ respectively, i.e., with a characteristic coupling slightly larger than the electroweak one.
\end{itemize}

\subsection{A model with triplets: diboson reach vs other searches}
\label{sec:vectortriplets}
		
	In this section we put the previous results in a global perspective, assessing the usefulness of diboson observables in a simple UV toy model where other types of searches are also constraining the parameter space. Our motivation stems from the fact that from an EFT point of view, non-universal corrections to the light quark vertices come from operators of the type $(\bar{f}\gamma_\mu f)(H^\dagger \overleftrightarrow{D}_\mu H)$, and in general grounds one expects to also generate the operators $(\bar{f}\gamma_\mu f)(\bar{f}\gamma_\mu f)$ and $(H^\dagger D_\mu H)(H^\dagger D_\mu H)$, which affect dijet processes and Higgs physics respectively. Considering a particular model allows one to compare these different searches and appreciate their complementarity.\footnote{There are other possible BSM scenarios with signals in diboson production but not in dijets, e.g., a model with vector-like quarks as  seen from Table~8 of Ref.~\cite{deBlas:2017xtg}. It may be interesting to see what diboson production can say about these types of scenarios.}

	We focus our attention to the general vector triplet models presented in Refs.~\cite{Contino:2011np,Biekoetter:2014jwa,Pappadopulo:2014qza}, which appear in various BSM scenarios, and can produce sizable and non-universal deviations to $\delta V \bar{q}q$ for the light quarks. We will see how the different searches are sensitive in the different limits of the parameter space. 
	
	For generality, we give the expressions for a model with custodial symmetry consisting on two vectorial resonances, $L_\mu$ and $R_\mu$, transforming respectively as $(1,3,1)$ and $(1,1,3)$ under $SU(3)_C\otimes SU(2)_L \otimes SU(2)_R$. At leading order,  these resonances couple to the SM currents as follows:
		\begin{multline}
		\mathcal{L}_\textit{int} = L^a_\mu \left(  \gamma_H  J_\mu^{Ha} \,+\, \gamma_V J_\mu^a + \sum_f \gamma_f J_\mu^{fa}  \right)
		\\
		+ R^0_\mu \left(  \delta_H J_\mu^H \,+\, \delta_V J_\mu + \sum_f \delta_f  J_\mu^{f}  \right)\,+\,\frac{1}{\sqrt{2}}(\delta_H R^+_\mu J_\mu^{-H}+h.c.)
		\label{eq:model}
		\end{multline}
where the SM currents are given by
		\begin{equation}
		\begin{split}
		& J_\mu^{Ha} = \frac{i}{2}H^\dagger \sigma^a\overleftrightarrow{D}_\mu H\,,
		\quad
		J_\mu^a = D^\nu W^a_{\nu\mu}\,,
		\quad
		J_\mu^{fa} = \bar{f}\gamma_\mu\sigma^a f\,,
		\quad
		J_\mu^{H} = \frac{i}{2}H^\dagger \overleftrightarrow{D}_\mu H\,,
		\\ 
		&
		J_\mu = \partial^\nu B_{\nu\mu}\,,
		\quad
		J_\mu^{f} = \bar{f}\gamma_\mu  f\,,
		\quad
		J_\mu^{H-} = \frac{i}{2} H^T \overleftrightarrow{D}_\mu H.
		\end{split}
		\end{equation}
The simplified UV model is fully characterized by the 11 arbitrary parameters $(\gamma_H, \gamma_V, \gamma_{f=Q_L,\ell_L}, $ $\delta_H, \delta_V, \delta_{f=Q_L, u_R, d_R, \ell_L, e_R})$. The couplings to each fermion also carry flavour indices. In the following, we will assume that they follow the MFV flavor scheme, with the two lighter generations having roughly the same $\gamma_f$ and $\delta_f$ couplings, and the third generation being different.
		
When both resonances have a mass  $m_\star \gg m_W$,  they can be integrated out to generate higher-dimensional interactions among the SM particles.  At order $1/m_\star^2$,  see Ref.~\cite{Biekoetter:2014jwa},  this yields
\begin{multline}
\mathcal{L}_\textit{tree}^{(6)}  \supset \bar{c}_W \, {\cal O}_W+ \bar{c}_B\, {\cal O}_B + \bar{c}_{2W} \, {\cal O}_{2W} + \bar{c}_{2B} \, {\cal O}_{2B}+\bar{c}_H\, {\cal O}_H \\[.2cm]
+  \sum_f (\bar{c}_{Hf} \, {\cal O}_{Hf} +\bar{c}^{(3)}_{Hf} \, {\cal O}^{(3)}_{Hf}) \,+\, \sum_{f,f'}  (\bar{c}_{ff'} \, {\cal O}_{ff'} + \bar{c}^{(3)}_{ff'}  {\cal O}^{(3)}_{ff'}) \,,
\label{eq:treeoperators}
\end{multline}
where the operators are defined as 
\begin{eqnarray}
& \displaystyle \nn
{\cal O}_B ={i g' \over 2 m_W^2} \left(H^\dagger D_\mu H\right) \partial_\nu B_{\mu \nu}\,, 
\quad
{\cal O}_W = {i g \over 2 m_W^2} \left(H^\dagger \sigma^i D_\mu H\right) \partial_\nu W^i_{\mu \nu}\,, 
\quad
{\cal O}_H = { 1 \over 2 v^2} \left(\partial_\mu |H|^2\right)^2\,,
\\
&  \displaystyle
{\cal O}_{2B} = {1 \over m_W^2} \left(\partial^\mu B_{\mu \nu}\right)^2\,, 
\quad 
{\cal O}_{2W} = {1 \over m_W^2} \left(\partial^\mu W^i_{\mu \nu}\right)^2\, , 
\quad
{\cal O}_{Hf} = {i \over v^2} \bar{f} \gamma_\mu f H^\dagger D_\mu H\,, 
\label{eq:toymodelSILH}
\\
&  \displaystyle \nn
{\cal O}_{Hf}^{(3)} =  {i \over v^2} \bar{f} \sigma^i  \gamma_\mu f H^\dagger \sigma^i D_\mu H\,, 
\quad
{\cal O}_{ff'}={1 \over v^2} \left(\bar{f} \gamma_\mu f\right)^2\,, 
\quad
{\cal O}_{ff'}^{(3)} = {1 \over v^2} \left(\bar{f} \sigma^i \gamma_\mu f\right)^2\,. 
\end{eqnarray}
At tree-level, the matching between the UV model and its EFT description leads to the following expression of the Wilson coefficients appearing in Eq.~(\ref{eq:toymodelSILH})
\bea
& \displaystyle \nn
\bar{c}_B = \frac{m_W^2}{m_\star^2}{ \delta_H \delta_V \over g^{\prime 2}} \,, 
\quad
\bar{c}_W =\frac{m_W^2}{m_\star^2}{\gamma_H \gamma_V \over g^2} \,, 
\quad 
\bar{c}_H = 3\frac{m_W^2}{m_\star^2}\frac{\delta_H^2+\gamma_H^2 }{g^2}\,,
\\
& \displaystyle \label{eq:OpCoef}
\bar{c}_{2B} = \frac{m_W^2}{4m_\star^2}\frac{\delta_V^2}{g^{\prime 2}}\,,
\quad
\bar{c}_{2W} = \frac{m_W^2}{4m_\star^2}\frac{\gamma_V^2}{g^2} \,, 
\quad 
\bar{c}_{Hf} =\frac{m_W^2}{m_\star^2}\frac{2}{g^2} \left(- \delta_H \delta_f + \delta_V \delta_f\right)\,, 
\\
& \displaystyle \nn
\bar{c}_{Hf}^{(3)} = \frac{m_W^2}{m_\star^2}\frac{2}{g^2} \left(-\gamma_H \gamma_f +   \gamma_V \gamma_f\right) \,, 
\quad
\bar{c}_{ff'} =- \frac{m_W^2}{2  m_\star^2} \frac{1}{g^2}   \delta_f \delta_{f'}\,,
\quad 
\bar{c}^{(3)}_{ff'} =- \frac{m_W^2}{2  m_\star^2} \frac{1}{g^2}  \gamma_f \gamma_{f'} \,.
\eea
		In \eq{eq:treeoperators}  the sums for $f, f'$ can run over $\{Q_L^i, u_R^i, d_R^i, \ell_L^i, e_R^i \}$ for the operators ${\cal O}_{Hf}, \, {\cal O}_{ff'}$, and over $\{Q_L^i, \ell_L^i \}$ for  ${\cal O}^{(3)}_{Hf}, \, {\cal O}^{(3)}_{ff'}$.
		\medskip
		
Among the seven parameters entering in fit to the diboson data, only five are generated at tree-level: the aTGC $\dgz$ and the four vertex corrections $\delta g_{L,R}^{Zu,d}$. The two other aTGCs $\dka$ and $\lambda_\gamma$ are generated at one loop by the operators ${\cal O}_{HB}, \, {\cal O}_{HW}$ and ${\cal O}_{3W}$ defined in Table~97 of Ref.~\cite{deFlorian:2016spz}. Specifically, we have\footnote{We assume the CKM matrix to be diagonal since the mixing effects are negligible in our analysis.}, 
\bea
&&\nn
\delta g_{1z} = -\frac{g^2 + g'^2}{g^2-g'^2}  \left(\bar{c}_W + \bar{c}_{2W}+\frac{g'^2}{g^2}(\bar{c}_B + \bar{c}_{2B}) \right) \,,
\\
&&\nn
\delta g_L^{Zu} = \frac{1}{2 } \left(- \bar{c}_{HQ} +\bar{c}_{HQ}^{(3)} + \bar{c}_{2W} + \bar{c}_{2B} \frac{g'^2}{g^2} - 
		\frac{2}{3}\frac{2 g'^2}{g^2 - g'^2} \left(\bar{c}_{2B} \frac{2 g^2 - g'^2}{g^2} + \bar{c}_{2W} + \bar{c}_B +\bar{c}_W \right) \right) \,,
\\
&&\nn
\delta g_L^{Zd} = \frac{1}{2 } \left( - \bar{c}_{HQ}-\bar{c}_{HQ}^{(3)}- \bar{c}_{2W} - \bar{c}_{2B} \frac{g'^2}{g^2} + \frac{1}{3}\frac{2 g'^2}{g^2 - g'^2} \left(\bar{c}_{2B} \frac{2 g^2 - g'^2}{g^2} + \bar{c}_{2W} + \bar{c}_B +\bar{c}_W \right) \right)\,,
 \\
&&\nn
\delta g_R^{Zu} =  \frac{1}{2 } \left( -\bar{c}_{Hu}- \frac{2}{3}\frac{2 g'^2}{g^2 - g'^2} \left( \bar{c}_{2B} \frac{2 g^2 - g'^2}{g^2} + \bar{c}_{2W} + \bar{c}_B + \bar{c}_W \right) \right)\,,
\\
&&
\delta g_R^{Zd} =  \frac{1}{2 } \left(-\bar{c}_{Hd}  + \frac{1}{3}\frac{2 g'^2}{g^2 - g'^2} \left( \bar{c}_{2B} \frac{2 g^2 - g'^2}{g^2} + \bar{c}_{2W} + \bar{c}_B + \bar{c}_W \right) \right)\,.
\label{eq:toycouplings}
\eea
The operators ${\cal O}_{ff'}$, ${\cal O}^{(3)}_{ff'}$ and ${\cal O}_{H}$ do not contribute to diboson production but they modify dijet and Higgs production which can then be used to set constraints on the parameters $\gamma_f, \, \delta_f, \gamma_H, \, \delta_H$ of the simplified UV model.

\subsubsection*{General models with only $L_\mu$}

In order to compare with experimental bounds and previous works, we will only consider the scenario where one has only the $L_\mu$ resonance. We further assume for simplicity that $\gamma_V\ll \gamma_H, \gamma_Q$. This minimal setup interpolates between the strongly and weakly coupled limits in the Higgs and fermionic sectors.
\medskip

\begin{figure}[t]
		\includegraphics[scale=0.36]{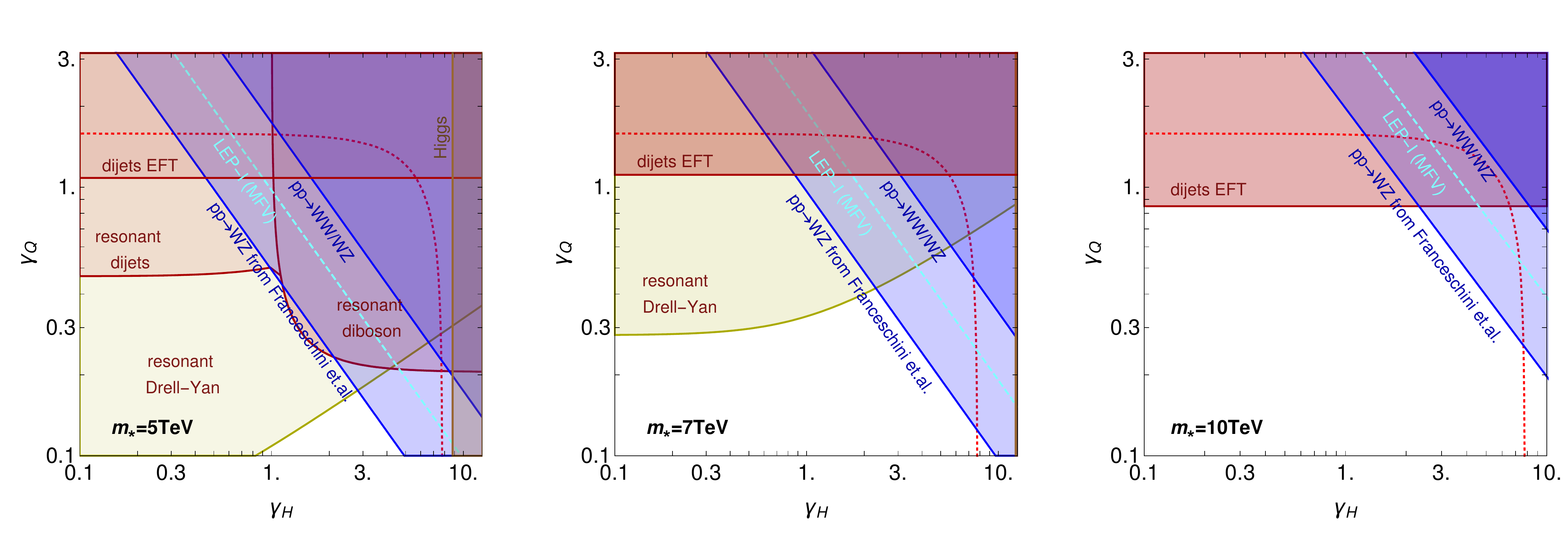}
		\caption{95\% CL exclusion regions in the $(\gamma_H,\gamma_Q)$ plane for various values of the resonance masses. In blue, the projected constraints from diboson data at HL-LHC, using our projections (dark shade) or the refined analysis strategy of Ref.~\cite{Franceschini:2017xkh}.  The other constraints come from a recast of the studies in Refs.~\cite{Alioli:2017jdo,Thamm:2015zwa, Chekanov:2017pnx, CMSfuturebounds, ATLASfuturebounds} and are commented in detail in the text.}
		\label{fig:UVmodel_constraints}
	\end{figure}

Figure~\ref{fig:UVmodel_constraints} shows, for fixed  $m_\star$,  the constraints on  $(\gamma_H,\gamma_Q)$  from various searches. The blue regions correspond to constraints imposed by the future diboson measurements at HL-LHC, using our projections (dark shade) or the refined the analysis strategy proposed in Ref.~\cite{Franceschini:2017xkh} (light shade). For comparison, the LEP-1 bound taken from Ref.~\cite{Efrati:2015eaa} under the MFV flavour scheme is indicated by the dashed light blue line. For this scenario,  HL-LHC hardly competes with LEP-1. The systematics uncertainty in diboson measurements would have to go significantly below 10\% to overcome the LEP-1 constraints.

For light resonance mass, $m_\star = 5 \, \tev$ (left plot), the parameter space is severely constrained by direct resonance searches in the leptonic channel (yellow), hadronic channel (red) or diboson channel (dark red).  The bounds have been obtained by recasting the projections of Refs.~\cite{Thamm:2015zwa, Chekanov:2017pnx}). To derive the bound from Drell--Yan searches,  we set $\gamma_\ell = \gamma_Q$. The bright red dotted line delineates the boundary between the regions in which $L_\mu$ has a width smaller or larger than 20\% its mass; this separates the regions where the direct searches may stop being sensitive to these resonances (at large $\gamma_Q$ and $\gamma_H$).  Finally, the sensitivity at HL-LHC  in the Higgs coupling measurements also cuts off the region with $\gamma_H>9$, corresponding to  $\xi = v^2 g_\star^2/m_\star^2 > 0.08$~\cite{Thamm:2015zwa, CMSfuturebounds, ATLASfuturebounds}.
			
At higher resonance masses, $m_\star = 7\, \tev$  (center plot) and $10 \, \tev$ (right plot), the direct resonance searches loose steam and  the diboson channels become more relevant in a larger portion of the parameter space. Already for $m_\star = 7\, \tev$,  the resonant dijet bound falls in the region where $\Gamma/m_\star > 20\%$, questioning its validity. The Higgs and resonant diboson constraints are too weak to set any constrain at these masses. For $m_\star = 10 \, \tev$, the constraints from resonant Drell--Yan searches fade away too.

Non-resonant dijet observables also impose severe constraints on the viable parameter space of our simplified model, the dijet EFT lines in Fig.~\ref{fig:UVmodel_constraints}. We used the results of Ref.~\cite{Alioli:2017jdo} that puts bound on coefficient $Z$ of the dimension-6 operator involving two gluon field strenghts
\be
\mathcal{L}\supset -\frac{Z}{2 m_W^2}D_\mu G^{A\, \mu\nu} D^\rho G^A_{\rho\nu} \,.
\label{eq:Zcoef}
\ee
Using the equations of motion, this operator can be rewritten in terms four-fermion operators.

The results presented in this section can be translated to the benchmark models A and B suggested in Ref.~\cite{Pappadopulo:2014qza}. The model A corresponds to a gauge bosons from an extended gauge symmetry and it features $\gamma_H\gamma_f\sim g^4/g_\star^2$ and $\gamma_H/\gamma_f\sim 1$.  Figure~\ref{fig:UVmodel_constraints} shows that this scenario is better probed via direct searches. The model B corresponds to a resonances from a composite sector, and it features  $\gamma_H\gamma_f\sim 1$ and $\gamma_H/\gamma_f\sim g_\star^2/g^2$, which projects in parameter space onto a line  parallel to the indirect diboson constraints in Fig.~\ref{fig:UVmodel_constraints}. The indirect probes can bring  information complementary to the direct constraints. 

In conclusion, the diboson channels give interesting constraints in regions where $\gamma_Q$ is small while $\gamma_H$ is large, which can be mapped to composite models with  heavy resonances strongly coupled to the Higgs boson but weakly coupled to the light quarks. In this section, we only studied the case with a left handed resonance $L_\mu$ in order to compare with the direct experimental searches and with previous phenomenological works. We find that, with an uncertainty $\delta_{syst}=10\%$,  the $pp \to WW, WZ$ leptonic channels at HL-LHC   will  access regions of the parameter space that remain blind to other searches. Nonetheless,  for this simplified scenario, LEP-1 is still slightly better than our HL-LHC projections. It would be interesting to study the case where more than one resonance is present, and therefore the various $\delta V \bar{q}q$ are less correlated. We expect, as shown in sections~\ref{subsec:LHCboundsvsLEP1} and~\ref{subsec:HLLHCboundsvsLEP1}, that in these cases diboson production will be significantly better than LEP-1 while being complementary to direct searches.

\section{Summary and outlook}
\label{sec:conclusionsDB}
	
The high energies accessible at the LHC open the possibility not only to directly produce new states, but also to enhance the sensitivity to new physics out of direct reach with effects that are encoded in higher dimensional operators involving the SM degrees of freedom. We offered a detailed analysis of diboson processes at LHC, which provides an interesting probe of some of these operators, in particular those that give rise to effects growing with the characteristic energy scale of the underlying hard process.
	\medskip
	
	Due to the expected increased sensitivity in the analyses, we reiterated that the interpretation of the diboson measurements in terms of anomalous triple gauge couplings has to be reconsidered. In particular, the effects of  anomalous couplings among the light quarks and the electroweak bosons can no longer be neglected a priori. On one hand, the current LHC diboson data already set stronger constraints than LEP-1 on the anomalous couplings $\delta V\bar{q}q$ for the down quark, at least under the hypothesis of MFV. On the other hand, both in the MFV and FU hypotheses, the aTGC fit is  found to be only marginally stable under profiling over the $\delta V\bar{q}q$ vertex corrections even when the LEP-1 constraints are imposed.
\medskip

		\begin{figure}[t]
	\centering
	\includegraphics[scale=0.48]{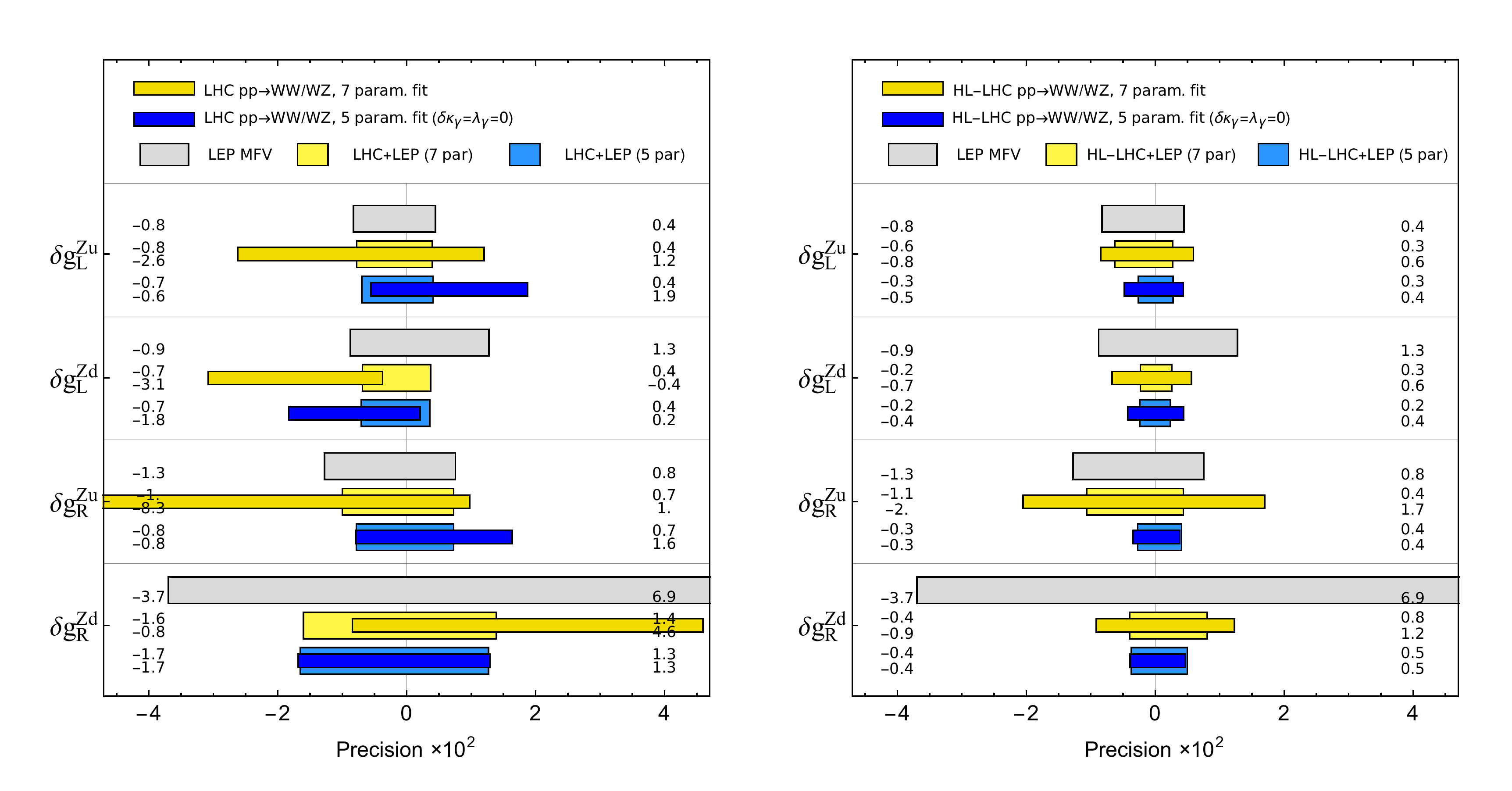}
	\caption{Current (\textbf{left}) and future (\textbf{right}) 95\% CL constraints on the anomalous vertices among light quarks and electroweak bosons. In gray, we show LEP-1 results assuming MFV. In yellow we use LHC diboson data and perform a global fit including the aTGCs. In blue, we only profile over $\delta g_{1z}$ and the vertex corrections. The thicker boxes combine LHC and LEP-1 data.}
	\label{fig:summaryVqq}
\end{figure}

	We did a simple estimate for the HL-LHC reach and found that the constraints will improve by a factor two to three. The different flavour assumptions on the vertices will have a seizable impact on the aTGC constraints. Quite remarkably, the precision on light quark couplings at HL-LHC will significantly surpass the LEP-1 constraints for both MFV and FU assumptions. And, as shown in Fig.~\ref{fig:dg1z_onedimensional_MFV_syst}, the HL-LHC may be able to set bounds on $\dgz$ of the order of $0.1\%$ in both FU and MVF scenarios. On the contrary, we checked that for universal theories which, as shown in Appendix~\ref{app:lepconstraints}, depend only on three aTGC and two $\delta V \bar{q}q$, the HL-LHC bounds are still far from reaching the LEP-1 precision. 	
\medskip

\begin{figure}[t]
\hspace*{-0.35in}
	\includegraphics[scale=0.45]{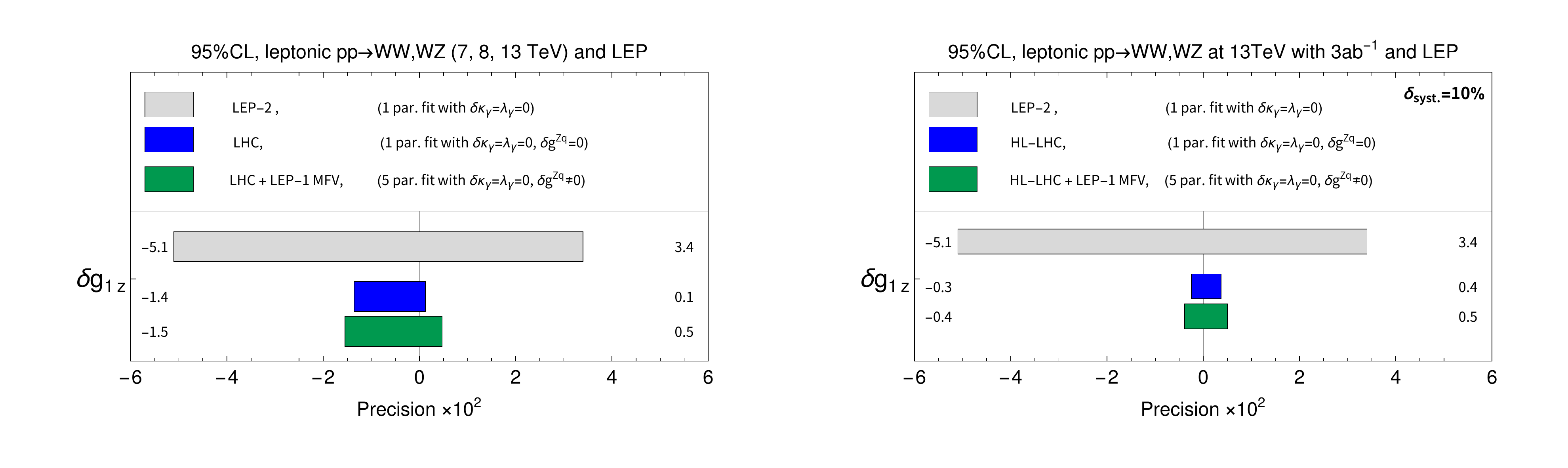} 
	\caption{Current (\textbf{left}) and future (\textbf{right}) 95\% CL constraints on $\delta g_{1z}$. In gray, LEP-2 results. In blue, constraint from LHC diboson data alone. In green, fit to LHC diboson data including the anomalous $V\bar{q}q$ vertices and profiling over them with the LEP-1 MFV constraints. In the projection of the HL-LHC bounds, a 10\%  systematic uncertainty in the channel $pp \to WW,WZ$ is assumed.}
	\label{fig:summaryaTGC}
\end{figure}

The left plot of Fig.~\ref{fig:summaryVqq} shows that the current leptonic diboson data can already set  bounds setting bounds on $\delta V \bar{q}q$ that are competitive with LEP-1, and they can also improve the bound on $\delta g_R^{Zd}$. The right plot shows that, by the end of  HL-LHC, leptonic $pp \to WV$ can be very competitive with the LEP-1 bounds or greatly surpass them if one assumes that $\dka = \lambda_\gamma=0$. Focusing on the aTGCs,  Fig.~\ref{fig:summaryaTGC} tells that, even when $\dka$ and $\lambda_\gamma$ are neglected, $\dgz$ is quite sensitive to the $\delta V \bar{q}q$ anomalous couplings: both with the current data and at  HL-LHC, the $\dgz$ bounds varying by about 30\% when $\delta V \bar{q}q$ are switch on and off in the global fit. 
\medskip

	We studied the interplay between the operators probed in diboson and the ones probed in other searches, as dijets or Higgs physics. This interplay can be intuitively understood by remembering  that the operators affecting diboson take the form $(\bar{f}\gamma_\mu f)(H^\dagger D_\mu H)$, and one can expect that generically might be accompanied by operators like  $(\bar{f}\gamma_\mu f)^2$ and $(H^\dagger D_\mu H)^2$ as well. As a concrete example, we presented a model in which all those deviations are indeed induced, showing that  measurements in diboson offer a complementarity exploration of the parameter space. It would be interesting to see how the direct and indirect bounds change for models with more resonances; we expect that diboson will fare better compared to the other searches when considering less simplified scenarios.
	\medskip 
	
	There are several interesting future directions.  Focusing on the current experimental searches, it would be interesting to study the semileptonic channels, which might benefit from fat jet techniques~\cite{Sirunyan:2017bey}, perhaps allowing to reach higher invariant masses than the leptonic ones.  Regarding new searches, one could follow the steps advocated in Refs.~\cite{Azatov:2017kzw, Franceschini:2017xkh, Liu:2018pkg, Panico:2017frx} and find new ways to increase the sensitivity to certain BSM physics allowing for more general interpretations of the bounds while also lowering the mass scale which one can probe. Regarding the results presented in this paper, we would like to encourage the experimental collaborations to use the current diboson searches to set bounds on the anomalous couplings between the light quarks and the $Z$ boson. It would also be very interesting to see how the degeneracy between aTGC and $\delta V\bar{q}q$ can be resolved by considering the production of a $Z$ in association with two jets by vector boson scattering.
In any case, we want to stress that, beyond the case of universal theories,  there exist flavour scenarios  for which robust bounds on the aTGCs can only follow from a global fit that include the effect of the $\delta V \bar{q}q$ anomalous couplings.

\section*{Acknowledgements}

We are particularly grateful to Zhengkang (Kevin) Zhang for collaborating in the initial stages of the project.
We also thank A.~Azatov, S.~Dawson, G.~Durieux, J.~Elias-Miro, A.~Falkowski, J.~Gu, R.S.~Gupta, G.~Panico, and A.~Pomarol  for helpful discussions and/or useful comments on the manuscript.
C.G. is supported by the Helmholtz Association through the recruitment initiative program. M.R. is supported by la Caixa--Severo Ochoa grant program, by the Spanish Ministry MEC under grants FPA2015-64041-C2-1-P, FPA2014-55613-P and FPA2011-25948, by the Generalitat de Catalunya grant 2014-SGR-1450 and by the Severo Ochoa excellence program of MINECO (grant SO-2012-0234).
We thank the Collaborative Research Center SFB676 of the Deutsche Forschungsgemeinschaft (DFG), ``Particles, Strings and the Early Universe", for support.

\newpage

\appendix
	
\section{Correlations in the Higgs basis}
\label{sec:AppCorr}

\begin{figure}[h!!]\centering
		\includegraphics[width=1\linewidth]{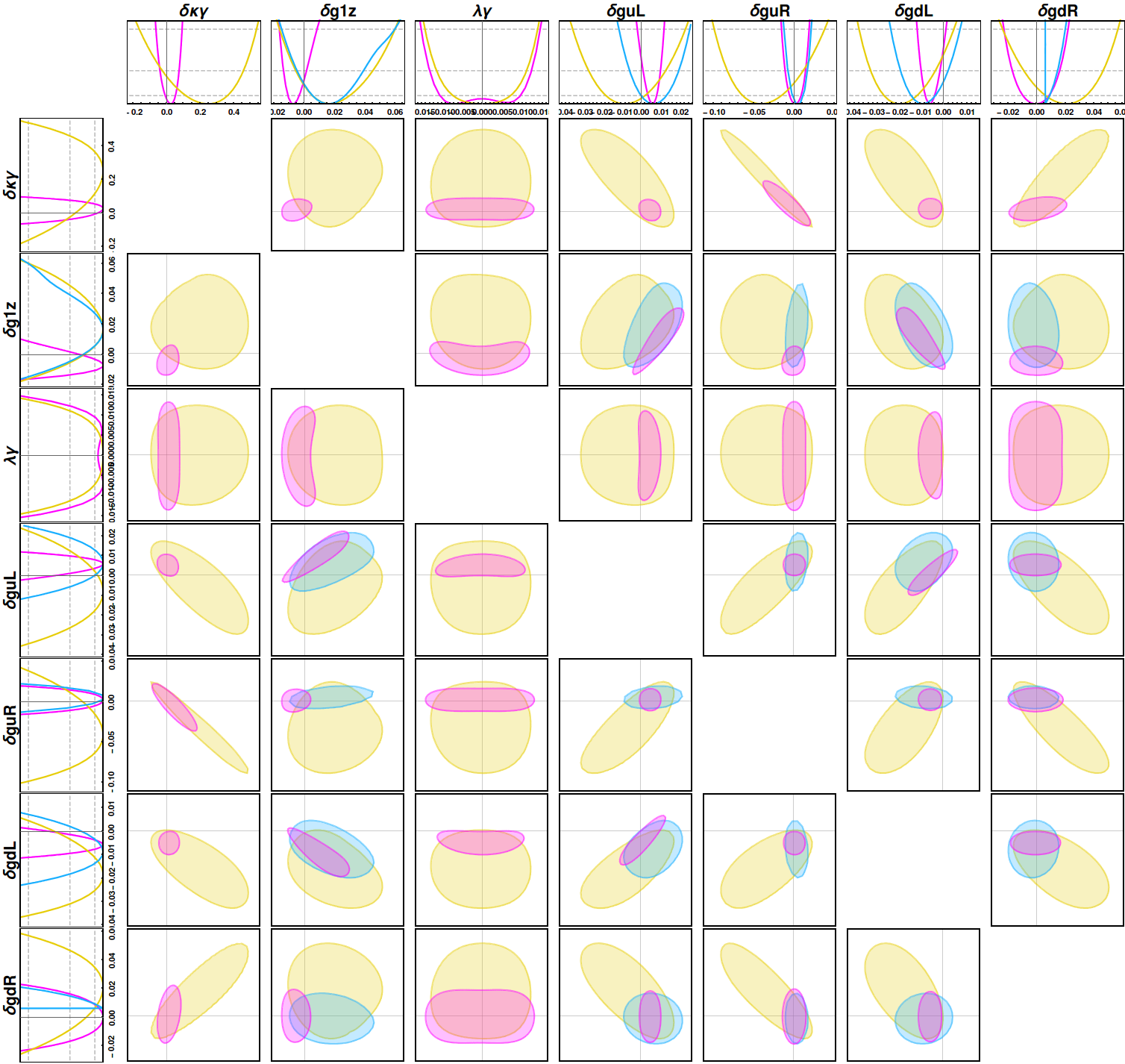} 
		\caption{One and two dimensional 95\% CL constraints for the seven parameters entering in diboson processes, using only the current LHC data in Table~\ref{tab:experimental}. In yellow,  all parameters are profiled. In blue, we profile over all parameters but setting $\delta \kappa_\gamma=\lambda_\gamma=0$. In pink, we do an exclusive fit setting to zero all parameters that do not appear in the plot labels.}
		\label{fig:gridplot}
	\end{figure}

\begin{figure}[h!!]\centering
		\includegraphics[width=1\linewidth]{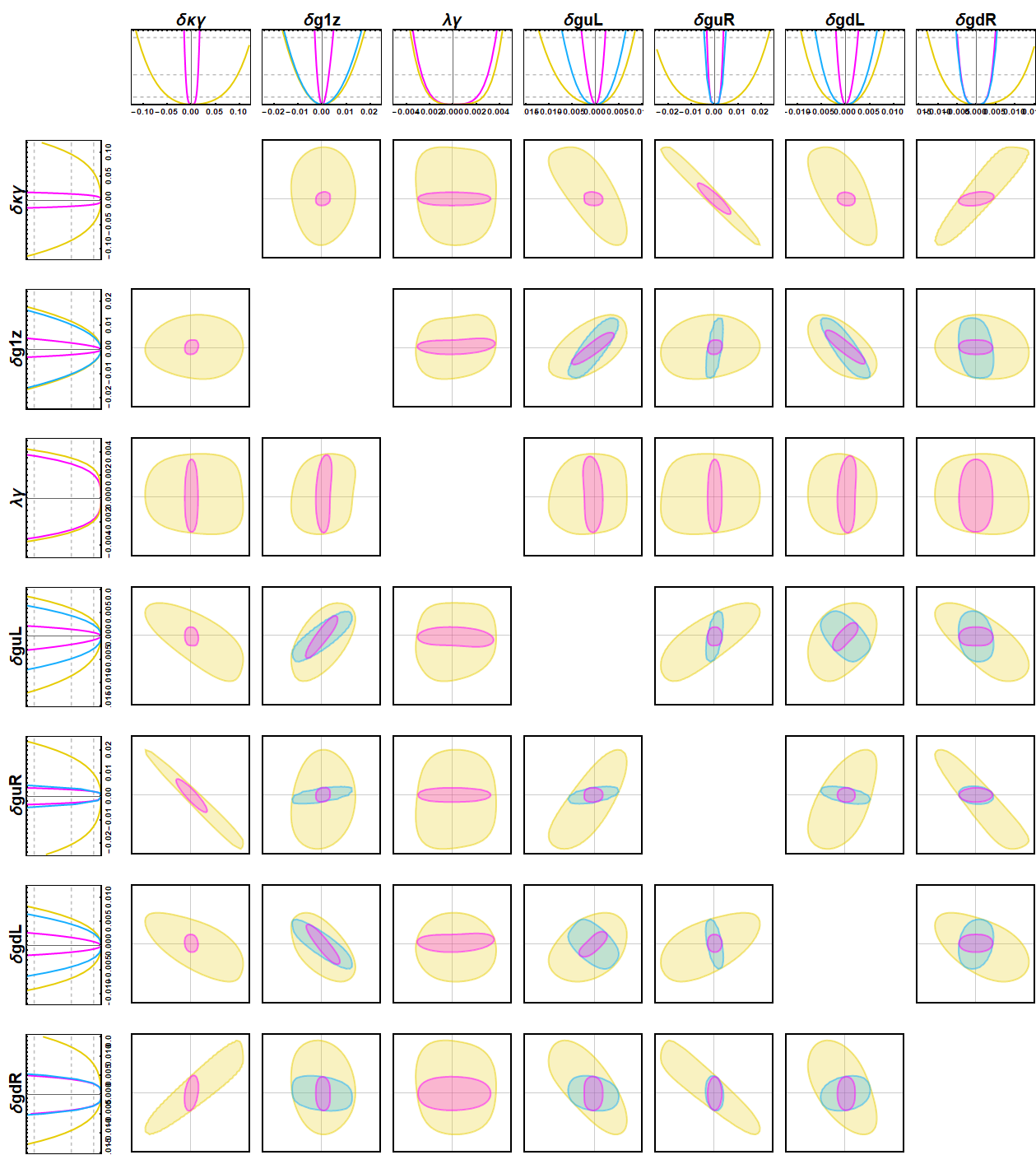} 
		\caption{One and two dimensional 95\% CL constraints for the seven parameters entering in diboson processes, using the $pp\to W^+W^-$ and $pp\to WZ$ projections for 13\,TeV with 3\,ab$^{-1}$ of integrated luminosity and assuming a 10\% systematic uncertainty. In yellow,  all parameters are profiled. In blue, we profile over all parameters but setting $\delta \kappa_\gamma=\lambda_\gamma=0$. In pink, we do an exclusive fit setting to zero all parameters not appearing in the plot labels.}
		\label{fig:gridploproj}
	\end{figure}

In Figs.~\ref{fig:gridplot} and~\ref{fig:gridploproj} we show the correlations between all the seven parameters relevant for diboson production at the LHC.  Since the $\chi^2$ function is not gaussian, these correlations are not simply related to a covariance matrix. Instead, the 95\% CL regions for each pair of parameters with all others profiled are reported. See section~\ref{sec:corr} for comments.
\medskip

\noindent\textbf{Correlations for the current LHC data}
	
	\newpage
		
\noindent\textbf{Correlations expected at HL-LHC}
	
	\newpage

\section{Summary of LEP-1 bounds}
\label{app:lepconstraints}
	
	In this appendix, we present the LEP-1 constrains obtained by profiling the $\chi^2$ function obtained by Ref.~\cite{Efrati:2015eaa}.
	
\subsubsection*{Minimal Flavour Violation}
	In MVF scenarios, the vertex corrections have the following form:
	\be
	[\delta g_{L,R}^{Zu,d}]_{ij} \simeq \left( A_{L,R}^{u,d}+B_{L,R}^{u,d} \ \frac{m_i^2}{m_3^2} \right)\delta_{ij}\,,
	\label{eq:MFVdef}
	\ee
	where $i,j=1,2,3$ stand for the family index. We are only interested in the constraints on the light quarks $u,\,d$ that control the diboson production. Using the results in Ref.~\cite{Efrati:2015eaa} and after profiling over all other parameters related to the electron and neutrino couplings, we arrive at
	\be
	\begin{array}{ccrcl}
		\left[\delta g_L^{Zu}\right]_{11} & \, = & -0.002 & \pm & 0.003 \\
		\left[ \delta g_R^{Zu} \right]_{11} &  \, = & -0.003 & \pm & 0.005 \\
		\left[\delta g_L^{Zd} \right]_{11}&  \, = & \,0.002 & \pm & 0.005 \\
		\left[\delta g_R^{Zd} \right]_{11} &  \, = & \,0.016 & \pm & 0.027 
	\end{array}
	\,,\quad
	\rho = \begin{pmatrix}
		1 & 0.43 & 0.52 & 0.23 \\
		& 1 & 0.19 & 0.36 \\
		& & 1 & 0.90\\
		& & & 1
	\end{pmatrix},
	\label{eq:mfvbounds}
	\ee
	In this flavour scenario, the vertex corrections are mostly sensitive to the $A$ coefficient in \eq{eq:MFVdef}, while the contribution from $B$, being suppressed by $m_{u,d}/m_{t,b}$, is negligible. The same bounds will also apply to the $c$ and $s$ quarks since the $B$ contribution remains negligible for the second family. 
	
\subsubsection*{Flavour Universality}
	
	In FU scenarios, all the vertex corrections have the same value irrespective of their family index, i.e.
	\be
	[\delta g_{L,R}^{Zu,d}]_{ij} = A_{L,R}^{u,d} \  \delta_{ij}\,.
	\ee
	In this case the bounds for the light quarks and heavy quarks coming from LEP-1 are the same.  Using the results of Ref.~\cite{Efrati:2015eaa} and after profiling over all other parameters, the LEP-1 bounds on the vertex corrections are found to be
	\be
	\begin{array}{ccrcl}
		\delta g_L^{Zu} & \, = & -0.0017 & \pm & 0.002 \\
		\delta g_R^{Zu} &  \, = & -0.0023 & \pm & 0.005 \\
		\delta g_L^{Zd} &  \, = & \,0.0028 & \pm & 0.001 \\
		\delta g_R^{Zd} &  \, = & \,0.019 & \pm & 0.008
	\end{array}
	\,,\quad
	\rho = \begin{pmatrix}
		1 & 0.83 & 0.04 & -0.11 \\
		& 1 & -0.13 & -0.05 \\
		& & 1 & 0.89\\
		& & & 1
	\end{pmatrix}.
	\label{eq:fubounds}
	\ee
	In this case, diboson production will set bounds on all of  $Z\bar{q}q$ from just measuring the vertices for $u$ and $d$.  It should be noted that, while the bounds on the $Z\bar{u}u$ couplings are rather similar in the two MFV and FU cases, the  bounds on the $Z\bar{d}d$ couplings are about 4 times more stringent in the FU case compared to the MFV case. This is a result of the fact that the $b$ quark can be efficiently tagged and  better discriminated than the light quarks. On the other hand, for the case of MFV, the $Z\bar{b}b$ vertex correction   gives a good constraint to the parameters $A+B$ in \eq{eq:MFVdef}, while $\left[\delta g_{L,R}^{Zd}\right]_{11}$ is only sensitive to $A$ and has a much lower precision from the Z-pole observables.
	
\subsubsection*{Universal theories}
	
	For universal theories where new physics coupled to the SM degrees of freedom via the SM currents only, the vertex corrections obey the relations~\footnote{This can be checked explicitly by for example writing the Higgs basis coefficients in terms of only bosonic operators in the SILH basis, see also Ref.~\cite{Wells:2015uba}.} 
	\be
	\delta g_R^{Zu} = 2 (\delta g_L^{Zu} + \delta g_L^{Zd})\,, \hspace{2cm} \delta g_R^{Zd} = - (\delta g_L^{Zu} + \delta g_L^{Zd}) \,,
	\label{eq:UniRel}
	\ee
and only two $Zqq$ couplings are independent. We can choose them to be $\delta g_L^{Zu}$ and $\delta g_L^{Zd}$. From the $\chi^2$ function corresponding to FU theories, one can derive the bounds on these two independent couplings
	\be
	\begin{array}{ccrcl}
		\delta g_L^{Zu} & \, = & -0.00010 & \pm & 0.00019 \\
		\delta g_L^{Zd} &  \, = & \,0.00008 & \pm & 0.00018
	\end{array}
	\,,\quad
	\rho = \begin{pmatrix}
		1 & -0.93\\
		& 1
	\end{pmatrix}.
	\label{eq:Bosonicbounds}
	\ee
	In this scenario,  the current diboson data do not set competitive bounds on the $Z\bar{q}q$ couplings.
	
For completeness we show in the following the  connection between $\delta g_{L,R}^{Zu,d}$, $\dgz$ and the oblique parameters when considering universal theories. For  $\delta g_{L,R}^{Zu,d}$ one finds:
	\be
	\delta g_{L,R}^{Zq}\,=\, \frac{1}{2}T^3_q\, \left(\hat{T}-W-Y \, \tan^2 \theta_W \right) \,+\, \frac{1}{2}Q_q\frac{\sin^2\theta_W}{\cos 2\theta_W}\left( W+\hat{T}-2 \hat{S} - Y (-2 + \tan^2 \theta_W) \right),
	\label{eq:Udirect}
	\ee 
which actually holds for any SM fermion. The two relations~\ref{eq:UniRel} are trivially satisfied. In addition, $\dgz$ can be written as:
\be
\delta g_{1z}\,=\, \frac{1}{2\cos 2\theta_W} \left( \hat{T} - 2\, S_\perp + W+ Y\,  \tan^2 \theta_W - \frac{\hat{S}-S_\perp}{\cos^2 \theta_W}  \right),
\ee
where the oblique parameters are obtained from the coefficients of the $d=6$ operators in the SILH basis:   $\hat{T}=c_T$, $\hat{S}=c_W+c_B$, $W = -2 c_{2W}$ and $Y= - 2 c_{2B}$. And  $S_{\perp}$ corresponds to $S_\perp=c_W-c_B$.
	
\section{Cross checks of the aTGC bounds}
	\label{sec:xchech}

\begin{figure}[h!!]\centering
		\includegraphics[scale=0.25]{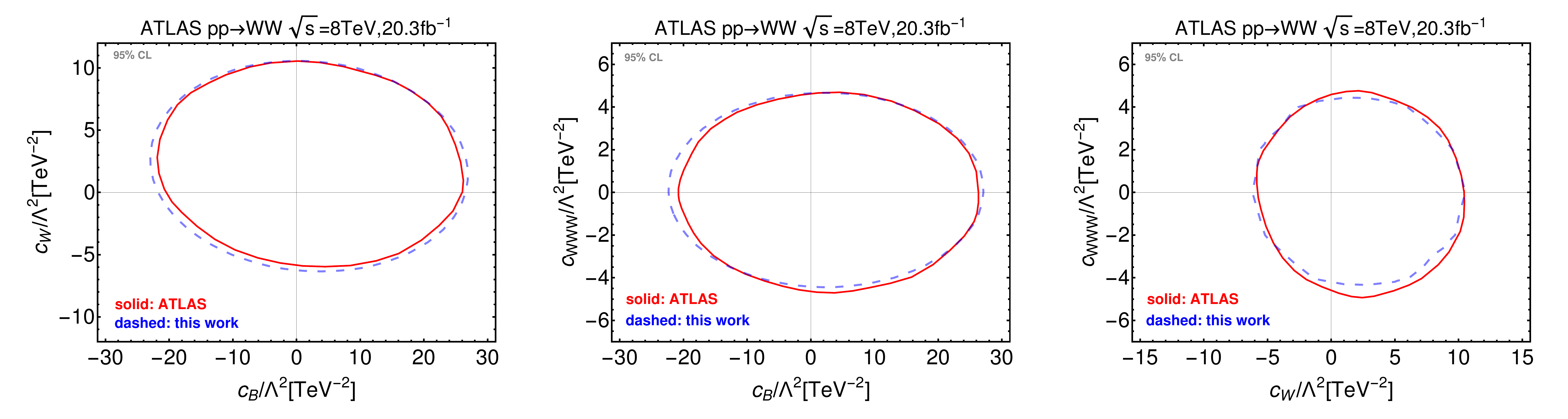} 
		\caption{Comparison between the 95\% CL contours obtained by the ATLAS collaboration~\cite{Aad:2016wpd}, and the results we obtained recasting their data.}
		\label{fig:xcheckATLAS}
	\end{figure}

As a cross check of our methodology and our assumptions, we compared the results of our fit with the ones presented by the experimental collaborations. For the $pp\to WW$ channel at 8\,TeV, Fig.~\ref{fig:xcheckATLAS} shows the comparison between the fit of Ref.~\cite{Aad:2016wpd} by the ATLAS collaboration and the results we obtained recasting the publicly available data. There is a good agreement. To compare with ATLAS results, we performed a change of basis and set bounds on the coefficients $c_{WWW}, \, c_{W}$ and $c_{B}$ corresponding to the following three operators which appear for instance in the HISZ basis, see Refs.~\cite{Hagiwara:1993ck, Falkowski:LHCHXSWG-INT-2015-001}:
\bea
{\cal O}_{WWW} &=& \text{Tr}[W_{\mu\nu} W^{\nu \rho} W_{\rho}^\mu]\,, \\
{\cal O}_{W} &=& (D_\mu H)^\dagger W^{\mu \nu} (D_\nu H)\,, \\
{\cal O}_{B} &=& (D_\mu H)^\dagger B^{\mu \nu} (D_\nu H) \,.
\eea 
The Wilson coefficients of the HISZ operators entering the aTGC are related to Higgs' basis coefficients as follows~\cite{Falkowski:LHCHXSWG-INT-2015-001}:
\bea
\dgz &=&  \frac{c_W}{\Lambda^2} \, \frac{g^2 +g'^2}{8} v^2\,, \\
\dka &=& \frac{c_W+c_B}{\Lambda^2} \, \frac{g^2}{8}  v^2\,, \\
\lambda_z &=&  \frac{c_{WWW}}{\Lambda^2} \, \frac{3 g^4}{8} v^2\,.
\eea

In the $pp\to WZ$ channel at 8\,TeV, we could not reach a similar agreement with the ATLAS results reported in Ref.~\cite{Aad:2016ett}, but we do agree with previous phenomenological studies~\cite{Butter:2016cvz, Falkowski:2016cxu}.

\begin{figure}[h!!]
		\begin{center}
			\includegraphics[scale=0.35]{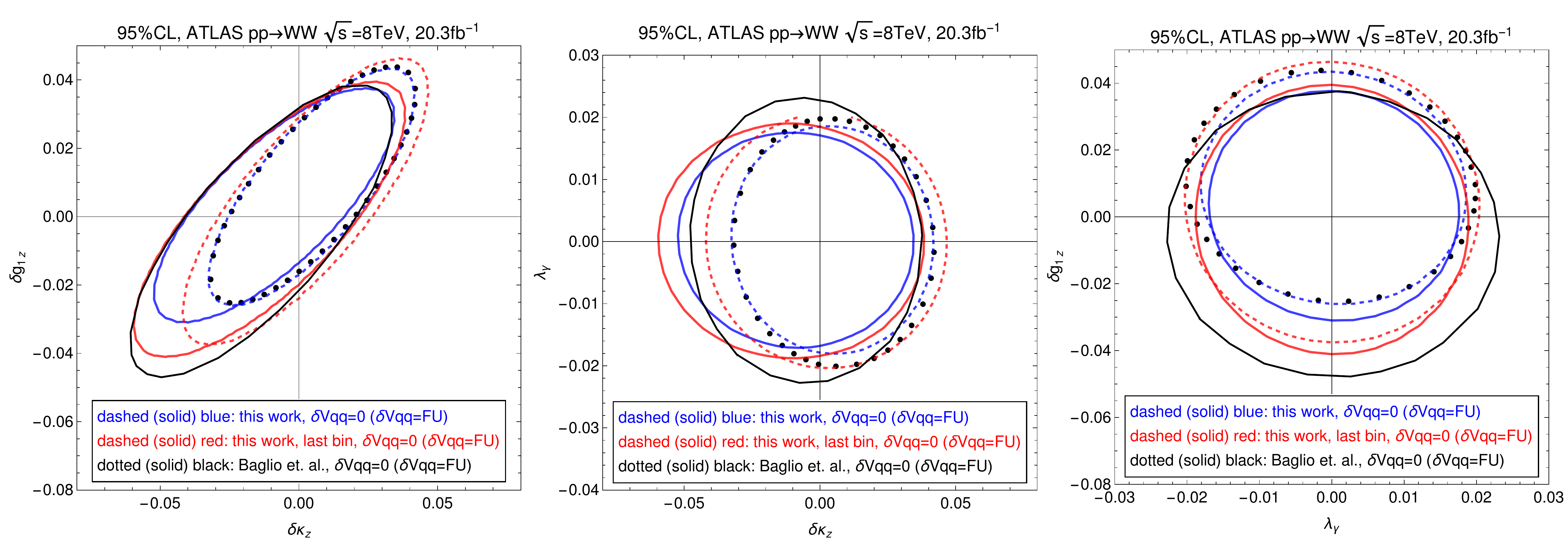} 
		\end{center}
		\caption{95\% CL contours for the aTGCs marginalizing over all other parameters. A three parameter fit (dashed or dotted) with 
			$\delta V \bar{q}q=0$ and a seven parameter fit (solid lines) combined LHC diboson data and LEP-1 data are performed.	For the later fit, a FU setup is considered. The blue lines are our results, in black the results obtained in Ref.~\cite{Baglio:2017bfe}, and in red, our results taking into account only the last bin.}
		\label{fig:xcheckSD}
	\end{figure}

In Fig.~\ref{fig:xcheckSD} we compare the fit on the aTGCs using the LHC diboson data reported in Ref.~\cite{Aad:2016wpd}, after profiling over the $\delta Vqq$ couplings, with the results in Ref.\cite{Baglio:2017bfe}. In our work, the aTGC bounds are derived from a global fit to the LHC diboson data and the $\chi^2$ extracted from Ref.~\cite{Efrati:2015eaa} for LEP-1. In dashed blue, we show our three parameter fit, which agrees well with the experimental results, as shown  in Fig.~\ref{fig:xcheckATLAS} already. Our results and those from Ref.~\cite{Baglio:2017bfe}, shown in botted black, are very similar. When the parameters $\delta V \bar{q}q$ are profiled using the LEP-1 constraints under the FU assumption, our results (solid blue) show deviations with respect those from Ref.~\cite{Baglio:2017bfe} (solid black). The slight differences with the fit from Ref.~\cite{Baglio:2017bfe} are  due to the following: {\it i)}  only the last bin of the experimental distribution is used in Ref.~\cite{Baglio:2017bfe} while we use all of them, {\it ii)} the procedure itself to set the bounds for the aTGC in Ref.~\cite{Baglio:2017bfe} is different which could also create some discrepancy with our results. To asses the first point, we show in red our fit  after profiling over the quark couplings when only the last bin is used. We find that this has a better agreement with Ref.~\cite{Baglio:2017bfe}, nonetheless not taking into account the subleading bins spoils our agreement with the ATLAS result.

\section{Comparison of HEP parameter bounds}
\label{sec:Comp}
	
\begin{figure}[t]
		\begin{center}
			\includegraphics[scale=0.5]{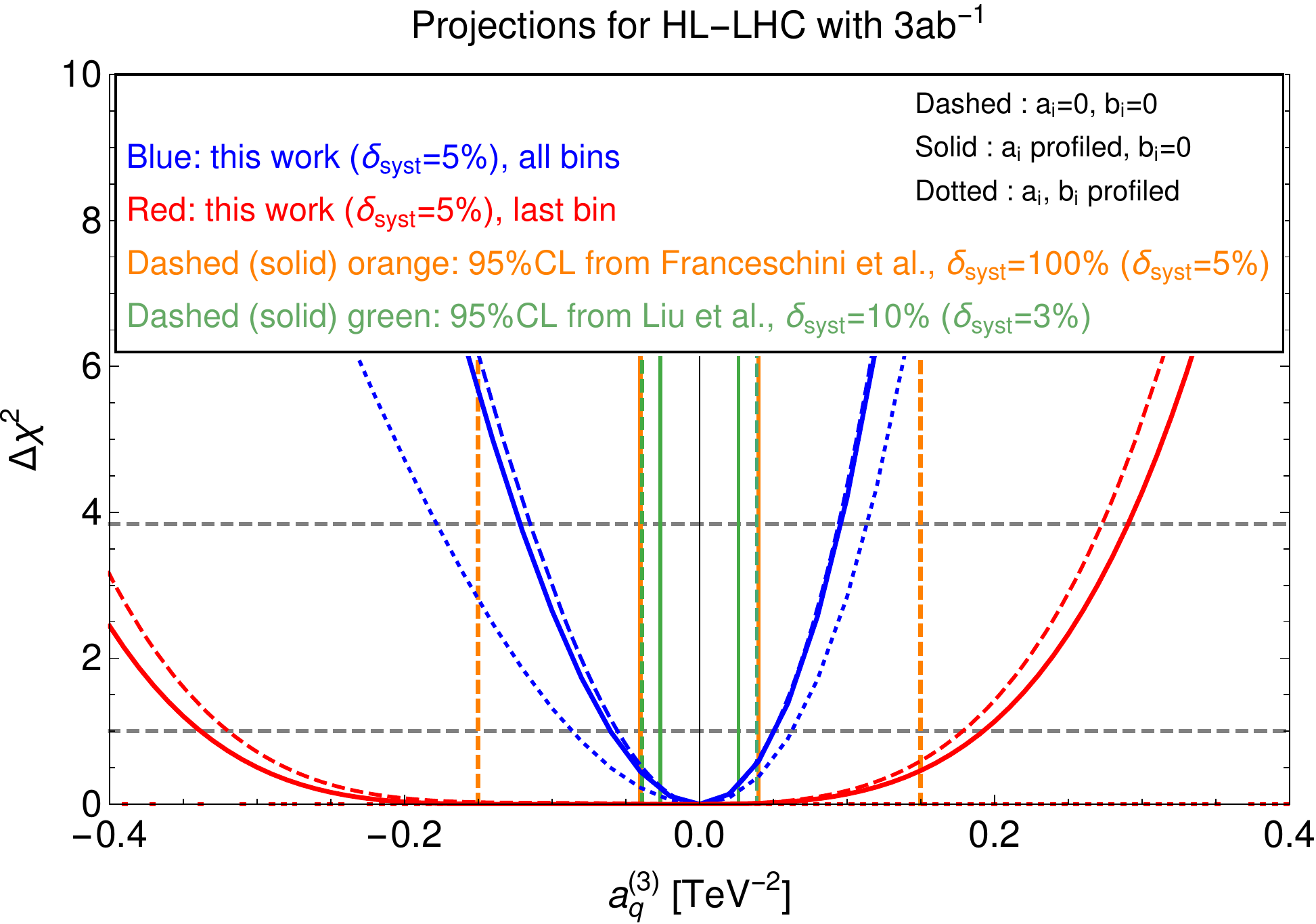} 
		\end{center}
		\caption{$\chi^2$ for $a_q^{(3)}$ ffrom the HL-LHC projections and comparison with the 95\% bounds obtained in Refs.~\cite{Franceschini:2017xkh, Liu:2018pkg}.}
		\label{fig:aq3}
	\end{figure}

To compare with previous works studying diboson production, in Fig.~\ref{fig:aq3} we present the HL-LHC bounds for the high energy parameters (HEP) defined in Ref.~\cite{Franceschini:2017xkh}. These HEP appear in the helicity amplitudes of Eqs.~(\ref{eq:AmpWW}), (\ref{eq:AmpWZ}).  In order to rewrite the Higgs basis in terms of the HEP, we perform a change of basis to in $\chi^2$ function  inverting the following relations:
\bea
a_q^{(3)} &=& \frac{g^2}{m_W^2} \big[\delta g_L^{Zu}-\delta g_L^{Zd} - c_W^2 \, \dgz \big], \, \\
a_q^{(1)} &=& - \frac{g^2}{3\, m_W^2} \big[3\, (\delta g_L^{Zu}+\delta g_L^{Zd}) + (\dka \,  t_W^2-\dgz \, s_W^2) \, \big],\, \\
a_u &=& - \frac{4\, g^2}{3\, m_W^2} \big[ \frac{3}{2} \,  \delta g_R^{Zu} + (\dka \,  t_W^2-\dgz \, s_W^2) \big], \, \\
a_d &=&  \frac{2\, g^2}{3\, m_W^2} \big[- 3 \,  \delta g_R^{Zd} + (\dka \,  t_W^2-\dgz \, s_W^2) \big].
\label{eq:HEPparameters}
\eea

The $\chi^2$ function then becomes a function of the four HEP, and $\lambda_\gamma$, and two other orthogonal combinations which we call $b_1, \, b_2$. These orthogonal combinations appear in the subleading amplitudes shown in Eqs.~(\ref{eq:subWW}), (\ref{eq:subWZ}). Figure~\ref{fig:aq3}  shows in red and blue the derived $\chi^2$ function for $a_q^{(3)}$ assuming $\delta_{syst} = 5\%$.~\footnote{See section~\ref{sec:HL-LHC} for the definition of $\delta_{syst}$.} The blue and red colours correspond to the case where all the bins of the differential distributions are used (blue) and the one where only the last bin is used (red).  Clearly, the actual bound is not entirely dominated by the most energetic bin and all the bins do contribute to setting the bound.  We explicitly studied three different cases: {\it i)} in dashed, we set  $\lambda_{\gamma}$, the three remaining HEP and the orthogonal directions $b_1, \, b_2$ to zero, {\it ii)}  in solid we set $b_1, \, b_2$ and $\lambda_\gamma$ to zero but profile over the three remaining HEP,  {\it iii)} in dotted we profile over all the parameters. We find that, as expected, the four HEP parameters are not very correlated among them, and therefore the solid and dashed lines differ by a small amount. On the other hand,  including or not the subleading terms $b_1, \, b_2$, which appear in the amplitudes shown in Eqs.~(\ref{eq:subWW}), (\ref{eq:subWZ}), makes a significant change. In general, we expect the subleading terms to be relatively important when the quadratic dimension-six amplitudes dominate the interference with the SM. On the other hand if the interference with the SM dominates, we expect these pieces to have an extra suppression coming from the SM amplitudes. In vertical orange and green lines we present the HL-LHC prospects of the leptonic $pp \to WZ$ and semi-leptonic $pp \to WV$ obtained in Refs.~\cite{Franceschini:2017xkh, Liu:2018pkg}; we differentiate in their case with solid and dashed lines two different assumptions on the systematic errors. If the new observables proposed by Refs.~\cite{Franceschini:2017xkh, Liu:2018pkg} are implemented, they will be able to set stronger bounds on $a_q^{(3)}$ by at least a factor 2.
	
The 95\% CL bounds for the HEP that we get when marginalizing over all the other parameters, i.e. three HEP, $b_1, b_2$ and $\lambda_\gamma$ are the following:
\bea
\Delta a_q^{(3)} &=& ^{+0.11}_{-0.18}\,, \hspace{2cm} \Delta a_q^{(1)} \ = \, ^{+0.34}_{-0.32} \,, \\
\Delta a_u &=& ^{+0.36}_{-0.47}\,, \hspace{2cm} \Delta a_d \ = \ ^{+0.60}_{-0.55} \,.
\eea
On the other hand, Ref.~\cite{Franceschini:2017xkh} gets $\Delta a_q^{(3)} \simeq \pm 0.04\, (0.15)$ for systematics of 5\% (100\%) and Ref.~\cite{Liu:2018pkg} finds $\Delta a_q^{(3)} \simeq \pm 0.03 \, (0.04)$ for systematics of 3\% (10\%).  
\medskip

\begin{figure}[h!!]
		\begin{center}
			\includegraphics[scale=0.478]{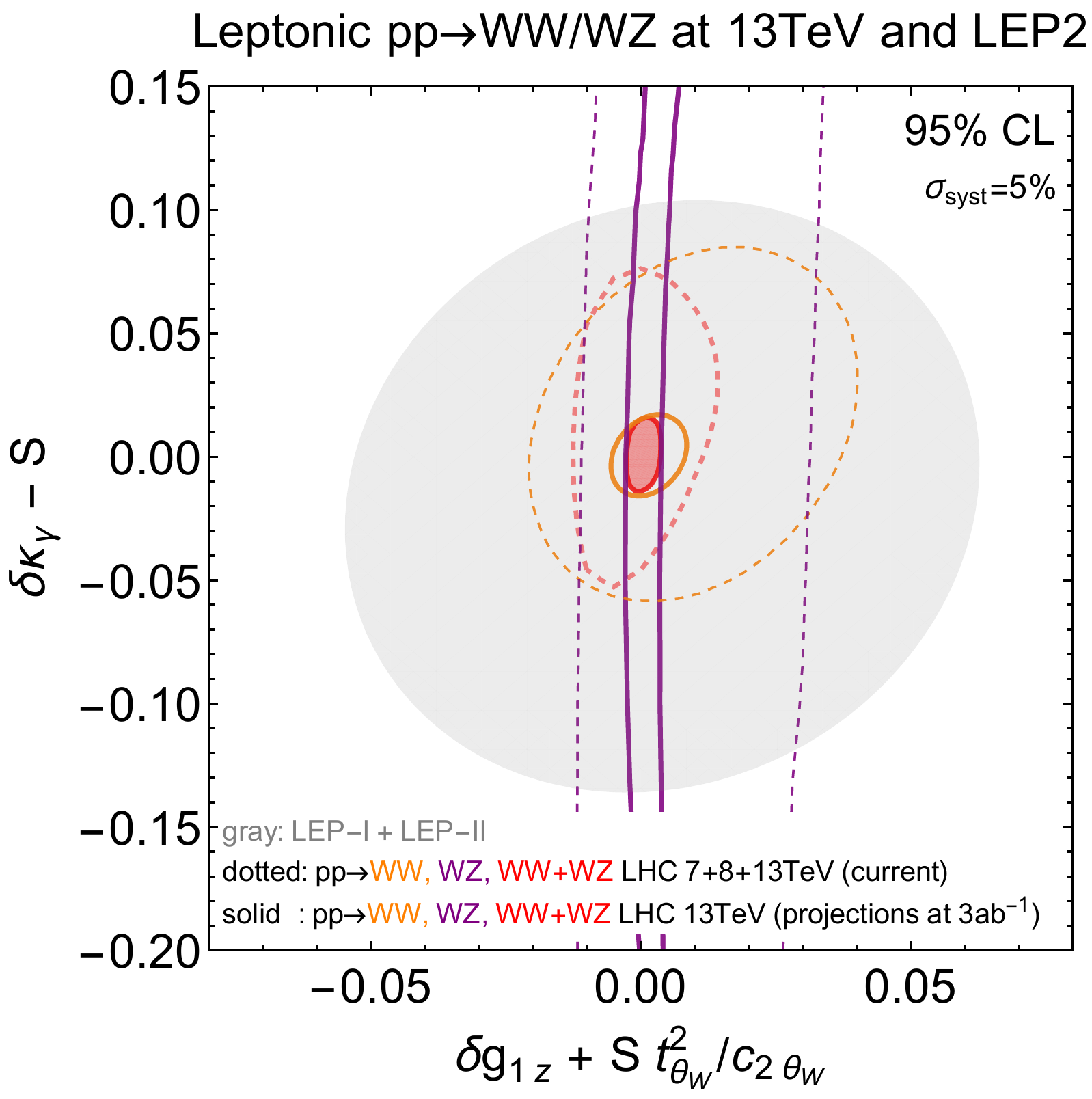} 
			\includegraphics[scale=0.5]{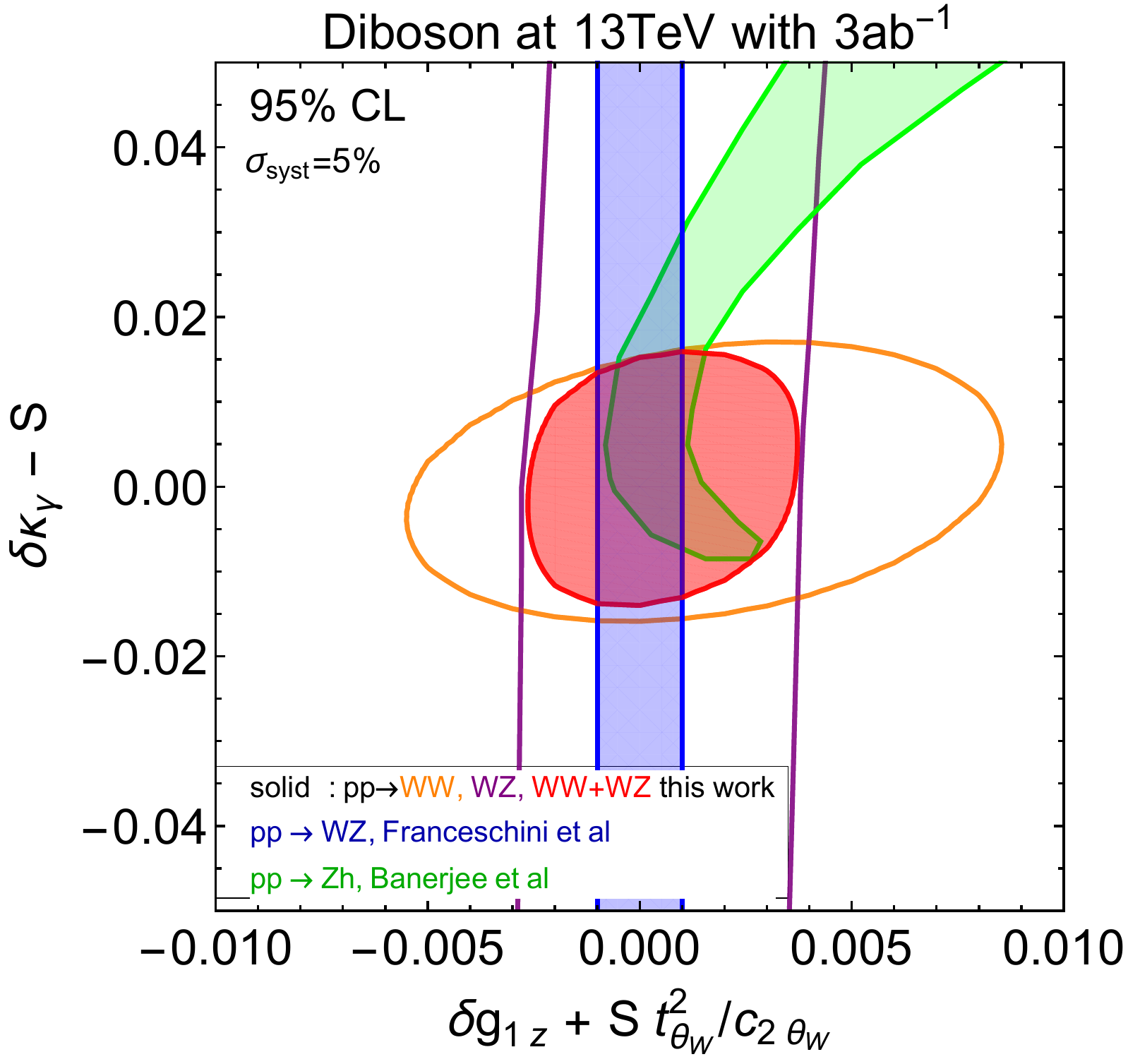} 
		\end{center}
		\caption{95\%CL constraints on universal theories with $W,Y\ll 1$. Left: Comparison of the LEP constraints with the ones extracted from the current  LHC diboson data (dotted) and the HL-LHC projections (solid). Right: Constraints of this work compared with the ones in Refs.~\cite{Franceschini:2017xkh,Banerjee:2018bio}. This fit for universal theories agrees with the 3-parameter fit of Fig.~\ref{fig:aTGCs_wVertex_HLLHC} when only the aTGC couplings are considered and the $\delta V\bar{q}q$ deviations are set to zero.}
		\label{fig:compPlot}
	\end{figure}

Another way to compare with previous works is to set bounds on the $a_q^{(3)}, \, a_q^{(1)}$ plane but considering only universal theories where the oblique parameters $W, \,  Y$ are negligible, see Refs.~\cite{Franceschini:2017xkh, Banerjee:2018bio}. In the SILH basis using the conventions in Ref.~\cite{Falkowski:LHCHXSWG-INT-2015-001}, one obtains
\bea
a_q^{(3)} &=& \frac{g^2}{m_W^2} \, ( c_{HW} + c_W - 2 c_{2W} )\, ,  \hspace{1cm} a_u = \frac{3 \, g^2}{m_W^2} c_{Hu} + \frac{4 \,g'^2}{3\, m_W^2} (c_B+c_{HB} + 2 c_{2B}) \,, \\ \nn
a_q^{(1)} &=& - \frac{g'^2}{3 m_W^2} \, (c_B + c_{HB} + 2 c_{2B})\,, \hspace{0.73cm} a_d =  \frac{g^2}{m_W^2} c_{Hd} - \frac{2 \,g'^2}{3\, m_W^2} (c_B+c_{HB} + 2 c_{2B})\,,
\eea
therefore, for universal theories and neglecting $W = -2 c_{2W}$ and $Y=- 2c_{2B}$ the HEP are:
\bea
a_q^{(3)} &=& \frac{g^2}{m_W^2} \, ( c_{HW} + c_W)\, ,  \hspace{1cm} a_q^{(1)} = -\frac{1}{4} a_u= \frac{1}{2} a_d =- \frac{g'^2}{3 m_W^2} \, (c_B + c_{HB}) \,,
\eea
which in terms of $\hat{S}$, $\dgz$ and $\dka$, using the conventions in Ref.~\cite{Falkowski:LHCHXSWG-INT-2015-001}, can be written as:
\bea
a_q^{(3)} &=&-\frac{g^2}{m_W^2} (\dgz \, c_W^2 + \frac{g'^2}{g^2-g'^2} \hat{S}) ,  \hspace{1cm} a_q^{(1)} = - a_{q}^{(3)} \, \frac{t_W^2}{3} - \frac{g'^2}{3 m_W^2} (\dka - \hat{S}) \,.
\eea
In Refs.~\cite{Franceschini:2017xkh, Banerjee:2018bio} for convenience they choose as independent directions $\dgz \, c_W^2 + \frac{g'^2}{g'^2-g^2} \hat{S}$ and $\dka - \hat{S}$. We show our bounds in this plane in Fig.~\ref{fig:compPlot}.

\section{High energy amplitudes in the Warsaw basis}
\label{sec:Warsaw}

Using the dictionary of Ref.~\cite{Falkowski:LHCHXSWG-INT-2015-001} we express the amplitude shown in Eqs. (\ref{eq:AmpWW})--(\ref{eq:AmpWZ}) in  the Warsaw basis. The high energy amplitudes for $pp \to WW$ are given by:
\bea
	\mathcal{M}(u_L u_L \rightarrow LL;00) &=& i  \frac{\hat{s}}{m_W^2} \frac{e^2}{4  \, s_W^2} \sin \theta \, \left[\frac{v^2}{\Lambda^2}(\omega_{\phi q}^{(1)} + \omega_{\phi q}^{(3)}) \right] \, , \\\nonumber
	\mathcal{M}(d_L d_L \rightarrow LL;00) &=& i  \frac{\hat{s}}{m_W^2} \frac{e^2}{4 \, s_W^2}  \sin \theta \, \left[\frac{v^2}{\Lambda^2}(\omega_{\phi q}^{(1)} - \omega_{\phi q}^{(3)}) \right] \, , \\\nonumber
	\mathcal{M}(u_R u_R \rightarrow RR;00) &=& i  \frac{\hat{s}}{m_W^2}\frac{e^2}{4   \, s_W^2} \sin \theta  \, \left[-\frac{v^2}{\Lambda^2} \omega_{\phi u} \right] \, , \\\nonumber
	\mathcal{M}(d_R d_R \rightarrow RR;00) &=&  i  \frac{\hat{s}}{m_W^2} \frac{e^2}{4  \, s_W^2} \sin \theta  \, \left[-\frac{v^2}{\Lambda^2} \omega_{\phi d} \right] \, , \\\nonumber
	\mathcal{M}(LL;\pm\pm)  &=&  i \frac{\hat{s}}{m_W^2}\frac{3e^3 }{4 \, t_W s_W}  \, T_q^3 \, \sin\theta\, \left[ -\frac{v^2}{\Lambda^2} \, \omega_W \right] \,,
	\label{eq:AmpWW_Warsaw}
	\eea
while for $pp \to W Z$, these are:
	\bea\nonumber
	\mathcal{M}(LL;00) &=& i \frac{\hat{s}}{m_W^2} \frac{e^2}{2\sqrt{2}s_W^2c_W} \sin\theta\left[- \frac{v^2}{\Lambda^2} \omega_{\phi q}^{(3)} \right] \, , \\
	\mathcal{M}(LL;\pm\pm) &=& i\frac{\hat{s}}{m_W^2}\frac{3 e^3}{4\sqrt{2}s_W^2}\, \sin\theta \left[ \frac{v^2}{\Lambda^2} \omega_W \right] .
	\label{eq:AmpWZ_Warsaw}
\eea

\newpage

\section{Summary tables}
\label{sec:summarytables}	
	
In this appendix, we report the results of the various fits performed in this paper.
	
\subsection*{Constraints on $\delta V \bar{q}q$}
	
   We first report the bounds on  $\delta V \bar{q}q$ under various assumptions on the aTGCs. See Tables \ref{tab:allthreetgcprofiled}, \ref{tab:nolooptgcs}, \ref{tab:notgcs},  \ref{tab:projvertexs}.

	\begin{table}[h!!]
		\begin{center}
			\begin{tabular}{|c||c||c|c||c|c|}   
				\multicolumn{6}{c}{1$\sigma$ bounds on $\delta V\bar{q}q$ from current LHC data (three aTGCs profiled)}\\
				\hline
				$\times 10^3$ & Diboson  & LEP (MFV)&LEP (FU) & Comb. (MFV) & Comb. (FU)\\ 
				\hline
				\hline
				\rule{0pt}{3ex}  
				$[\delta g_L^{Zu}]_{11}$ & $-7.9\pm 10 $  &$ -1.9\pm 3.1$ &  $-1.7\pm 2.1$ & $-1.9\pm 2.8$ & $ -0.9\pm 2.0 $ \\ 
				\rule{0pt}{3ex}  
				$[\delta g_R^{Zu}]_{11}$ & $-40\pm 24 $  &$ -2.6\pm 5$ &  $-2.3\pm 4.6$ & $-1.3\pm 4.4$ & $ -0.9\pm 4.3 $ \\ 
				\rule{0pt}{3ex}  
				$[\delta g_L^{Zd}]_{11}$ & $-18  \pm 7 $  &$ 2  \pm  5.4$ & $ 2.8 \pm 1.5$ & $-2\pm 2.5$  & $0.8 \pm 1.2$ \\ 
				\rule{0pt}{3ex} 
				$[\delta g_R^{Zd}]_{11}$ & $20.3  \pm 14.2 $  &$ 16  \pm 27$ & $ 20 \pm 7.7$ & $-1.8\pm 7.9$  & $8.6 \pm 5.6$ \\ 
				\hline
			\end{tabular}
			\caption{Constraints $(\times 10^{3})$ on the $\delta V\bar{q}q$ vertex corrections from a seven parameter global fit combining LHC diboson data and LEP-1 measurements. The first column gives the bounds using the LHC diboson data alone. The second and third columns report the LEP-1 bounds derived in Ref.~\cite{Efrati:2015eaa} under the MFV and FU assumptions respectively. Finally, the last two columns show the combination of the current LHC and LEP-1 data for the two flavour assumptions.} 
			\label{tab:allthreetgcprofiled}
		\end{center}
	\end{table}

	\begin{table}[h!!]
		\begin{center}
			\begin{tabular}{|c||c||c|c||c|c|}   
				\multicolumn{6}{c}{1$\sigma$ bounds on $\delta V\bar{q}q$ from current LHC data ($\delta\kappa_\gamma=\lambda_\gamma=0$)}\\
				\hline
				$\times 10^3$ & Diboson  & LEP (MFV)&LEP (FU) & Comb. (MFV) & Comb. (FU)\\ 
				\hline
				\hline
				\rule{0pt}{3ex}  
				$[\delta g_L^{Zu}]_{11}$ & $6.4\pm 6.4 $  &$ -1.9\pm 3.1$ &  $-1.7\pm 2.1$ & $-1.5\pm 2.7$ & $ -0.3\pm 1.8 $ \\ 
				\rule{0pt}{3ex}  
				$[\delta g_R^{Zu}]_{11}$ & $4.3\pm 6.4 $  &$ -2.6\pm 5$ &  $-2.3\pm 4.6$ & $-0.3\pm 3.9$ & $ 0.6\pm 3.8 $ \\ 
				\rule{0pt}{3ex}  
				$[\delta g_L^{Zd}]_{11}$ & $-8.7  \pm 5.2 $  &$ 2  \pm  5.4$ & $ 2.8 \pm 1.5$ & $-2\pm 2.5$  & $0.6 \pm 1.1$ \\ 
				\rule{0pt}{3ex} 
				$[\delta g_R^{Zd}]_{11}$ & $-2.1  \pm 8 $  &$ 16  \pm 27$ & $ 20 \pm 7.7$ & $-2.7\pm 7.7$  & $7.7 \pm 5.4$ \\ 
				\hline
			\end{tabular}
			\caption{Constraints $(\times 10^{3})$ on the $\delta V\bar{q}q$ vertex corrections from a five parameter global fit combining LHC diboson data and LEP-1 measurements, setting $\delta\kappa_\gamma=\lambda_\gamma=0$. The first column gives the bounds using the LHC diboson data from Table~\ref{tab:experimental} setting $\delta\kappa_\gamma=\lambda_\gamma=0$. The first column gives the bounds using the LHC diboson data alone. The second and third columns report the LEP-1 bounds derived in Ref.~\cite{Efrati:2015eaa} under the MFV and FU assumptions respectively. Finally, the last two columns show the combination of the current LHC and LEP-1 data for the two flavour assumptions.} 
			\label{tab:nolooptgcs}
		\end{center}
	\end{table}
	
	\begin{table}[h!!]
		\begin{center}
			\begin{tabular}{|c||c||c|c||c|c|}   
				\multicolumn{6}{c}{1$\sigma$ bounds on $\delta V\bar{q}q$ from current LHC data ($\delta g_{1z}=\delta\kappa_\gamma=\lambda_\gamma=0$)}\\
				\hline
				$\times 10^3$ & Diboson  & LEP (MFV)&LEP (FU) & Comb. (MFV) & Comb. (FU)\\ 
				\hline
				\hline
				\rule{0pt}{3ex}  
				$[\delta g_L^{Zu}]_{11}$ & $0.6\pm 5.0 $  &$ -1.9\pm 3.1$ &  $-1.7\pm 2.1$ & $-0.7\pm 2.6$ & $ 1.1\pm 1.6 $ \\ 
				\rule{0pt}{3ex}  
				$[\delta g_R^{Zu}]_{11}$ & $1.5\pm 6.1 $  &$ -2.6\pm 5$ &  $-2.3\pm 4.6$ & $1.5\pm 3.6$ & $ 3.5\pm 3.4 $ \\ 
				\rule{0pt}{3ex}  
				$[\delta g_L^{Zd}]_{11}$ & $-5.3  \pm 4.6 $  &$ 2  \pm  5.4$ & $ 2.8 \pm 1.5$ & $-2.8\pm 2.4$  & $0.1 \pm 1.1$ \\ 
				\rule{0pt}{3ex} 
				$[\delta g_R^{Zd}]_{11}$ & $-0.71  \pm 8 $  &$ 16  \pm 27$ & $ 20 \pm 7.7$ & $-6.2\pm 6.8$   & $5.0 \pm 5.4$ \\ 
				\hline
			\end{tabular}
			\caption{Constraints $(\times 10^{3})$ on the $\delta V\bar{q}q$ vertex corrections from a five parameter global fit combining LHC diboson data and LEP-1 measurements, setting $\delta g_{1z}=\delta\kappa_\gamma=\lambda_\gamma=0$. The first column gives the bounds using the LHC diboson data alone. The second and third columns report the LEP-1 bounds derived in Ref.~\cite{Efrati:2015eaa} under the MFV and FU assumptions respectively. Finally, the last two columns show the combination of the current LHC and LEP-1 data for the two flavour assumptions.} 
			\label{tab:notgcs}
		\end{center}
	\end{table}

		\begin{table}[h!!]
			\begin{center}
				\begin{tabular}{|c||c||c|c||c|}   
					\multicolumn{5}{c}{1$\sigma$ bounds on $\delta V\bar{q}q$ expected at HL-LHC}\\
					\hline
					$\times 10^3$ & aTGCs profiled  & no loop $(\delta\kappa_\gamma=\lambda_\gamma=0)$ & no aTGCs & exclusive fit \\ 
					\hline
					\hline
					\rule{0pt}{3ex}  
					$[\delta g_L^{Zu}]_{11}$ & $\pm 2.5 $  &$ \pm 1.5$ &  $ \pm 1.4$ & $\pm 0.5$ \\ 
					\rule{0pt}{3ex}  
					$[\delta g_R^{Zu}]_{11}$ & $\pm 4.0 $  &$  \pm 1.7$ &  $ \pm 1.5$ & $ \pm 1.1$\\ 
					\rule{0pt}{3ex}  
					$[\delta g_L^{Zd}]_{11}$ & $  \pm 1.5 $  &$   \pm  1.2$ & $ \pm 1.2$ & $ \pm 0.45$ \\ 
					\rule{0pt}{3ex} 
					$[\delta g_R^{Zd}]_{11}$ & $  \pm 4.0 $  &$  \pm 2.5$ & $ \pm 2.5$ & $ \pm 1.8$ \\ 
					\hline
				\end{tabular}
				\caption{Expected constraints ($\times 10^{3}$) at HL-LHC on the $\delta V\bar{q}q$ vertex corrections. 
				The constraints are obtained from the projections at HL-LHC for the $pp\to W^+W^-\to\ell\nu\ell\nu$ channel combined with the LEP-1 constraints for a MFV setup. The first column gives the constraints resulting from a seven parameter fit. In the second, the two aTGCs usually generated at the loop level are set to zero. In the third column, all the three aTGCs are set to zero. Finally, the last column reports the constraints obtained from an exclusive fit with only one parameter considered at a time.} 
				\label{tab:projvertexs}
			\end{center}
		\end{table}
\bigskip

\newpage
\
\newpage

\subsection*{Constraints on the aTGCs}

We now report the bounds on  the aTGCs under various assumptions on  $\delta V \bar{q}q$. These are shown in Tables \ref{tab:tgcconstraints} and \ref{tab:tgcconstraintsproj}.
	
	\begin{table}[h!!]
		\begin{center}
			\begin{tabular}{|c||c||c|c|}   
			\multicolumn{4}{c}{1$\sigma$ bounds on aTGC from current LHC data}\\
				\hline
				{\vrule height 14pt depth 8pt width 0pt}
				$\times 10^3$ & $\delta g_{L,R}^{Zu,d}=0$  & $\delta g_{L,R}^{Zu,d}=$MFV &$\delta g_{L,R}^{Zu,d}=$FU \\ 
				\hline
				\hline
				\rule{0pt}{3ex}  
				$\delta \kappa_\gamma$ & $12\pm 31$  &$ 18\pm 35$ &  $24\pm 35$\\ 
				\rule{0pt}{3ex}  
				$\delta g_{1z}$ & $-7 \pm 4$  &$ -7\pm 5 $ &  $ -9 \pm 5$\\
				\rule{0pt}{3ex}  
				$\lambda_\gamma$ & $0\pm 6$  &$ 0\pm 6$ &  $0\pm 6$\\
				\hline
			\end{tabular}
			\caption{Constraints $(\times 10^{3})$ on the anomalous triple gauge couplings from the current LHC diboson data. The first column corresponds to the traditional diboson analysis that considers only aTGCs and sets to zero all anomalous fermion-gauge vertices $\delta V\bar{q}q=0$. The next two columns show the effect of letting these anomalous fermion-gauge vertices float, assuming either a MFV or a FU setup respectively.} 
		\label{tab:tgcconstraints}
		\end{center}
	\end{table}

	\begin{table}[h!!]
		\begin{center}
			\begin{tabular}{|c||c||c|c|}   
				\multicolumn{4}{c}{1$\sigma$ bounds on aTGC expected at HL-LHC}\\
				\hline
				{\vrule height 14pt depth 8pt width 0pt}
				$\times 10^3$ & $\delta g_{L,R}^{Zu,d}=0$  & $\delta g_{L,R}^{Zu,d}=$MFV &$\delta g_{L,R}^{Zu,d}=$FU \\ 
				\hline
				\hline
				\rule{0pt}{3ex}  
				$\delta \kappa_\gamma$ & $ \pm 10$  &$  \pm 22$ &  $ \pm 20$\\ 
				\rule{0pt}{3ex}  
				$\delta g_{1z}$ & $  \pm 1.5$  &$  \pm 3.5 $ &  $ \pm 3.0$\\
				\rule{0pt}{3ex}  
				$\lambda_\gamma$ & $ \pm 2.2$  &$ \pm 2.3$ &  $ \pm 2.2$\\
				\hline
			\end{tabular}
			\caption{Constraints ($\times 10^{3}$) on the anomalous triple gauge couplings using the projections at HL-LHC of the $pp\to W^+W^-\to\ell\nu\ell\nu$ channel. The first column corresponds to the traditional diboson analysis that considers only aTGCs and sets to zero all anomalous fermion-gauge vertices $\delta V\bar{q}q=0$. The next two columns show the effect of letting these anomalou fermion-gauge vertices float, assuming either a MFV or a FU setup respectively.} 
		\label{tab:tgcconstraintsproj}
		\end{center}
	\end{table}

\end{document}